\documentclass[journal]{new-aiaa}
\usepackage[utf8]{inputenc}
\usepackage{textcomp}

\usepackage{graphicx}
\usepackage{float}
\usepackage{amsmath}
\usepackage[version=4]{mhchem}
\usepackage{siunitx}
\usepackage{longtable,tabularx}
\usepackage{multirow}
\usepackage{multicol}
\usepackage{dcolumn}
  \newcolumntype{d}{D{.}{.}{-1}}
\usepackage{color}

\title{Aeroacoustics of Twin Rectangular Jets Including Screech: Large-eddy Simulations with Experimental Validation}

\author{%
Jinah Jeun\footnote{Postdoctoral Fellow, Center for Turbulence Research; jjeun@stanford.edu. Member AIAA.}}
\affil{Stanford University, Stanford, CA 94305}
\author{%
Aatresh Karnam\footnote{PhD student, Department of Aerospace Engineering and Engineering Mechanics, Student Member AIAA.}}
\affil{University of Cincinnati, Cincinnati, OH 45221}
\author{%
Gao Jun Wu\footnote{PhD Student, Department of Aeronautics and Astronautics, Student Member AIAA.}, \ and \ Sanjiva K. Lele\footnote{Professor, Department of Aeronautics and Astronautics and Department of Mechanical Engineering, Associate Fellow AIAA.}}
\affil{Stanford University, Stanford, CA 94305}
\author{%
Florian Baier\footnote{PhD student, Department of Aerospace Engineering and Engineering Mechanics, Student Member AIAA.}, \ and \ Ephraim J. Gutmark\footnote{Distinguished Professor, Department of Aerospace Engineering and Engineering Mechanics, Fellow AIAA.}}
\affil{University of Cincinnati, Cincinnati, OH 45221}

\begin{document}

\maketitle

\begin{abstract}
High-fidelity large-eddy simulations (LES) are performed to investigate aeroacoustic characteristics of jets issuing from twin rectangular nozzles with an aspect ratio of 2:1 at two over-expanded conditions and the design condition. For all three jet conditions simulated, LES predicts qualitatively similar near-field flow statistics to those measured at the University of Cincinnati. Using the Ffowcs Williams-Hawkings method, LES captures the fundamental screech tone and its harmonics fairly well at multiple observer locations in the far-field. Intense jet flapping motions in the near-field along the minor axis, which are influenced by jet-to-jet interactions, are found to correspond to those frequencies. Moreover, the predicted overall sound pressure levels are within 1-2 dB of the experimental measurements. However, the screech tones appear to be intermittent, as the twin-jet interaction pattern varies irregularly. To extract dominant flow structures at the screech frequencies and identify the twin-jet coupling modes, spectral proper orthogonal decomposition (SPOD) analysis is used. SPOD analysis recovers energetic peaks at the screech frequencies, and the corresponding leading modes indicate strong upstream radiation originating from the fifth/sixth shock-cells. For the two over-expanded conditions, the leading modes show anti-symmetric coupling in the minor axis at the fundamental screech frequencies. In contrast, the two jets behave symmetrically with respect to each other in the major axis, in line with the absence of jet flapping in this direction. Furthermore, the leading SPOD eigenvalues turn out to be, at least, two orders of magnitude larger than higher-order eigenvalues, suggesting potential of reduced-order models for the twin-jet screech.
\end{abstract}

\section*{Nomenclature}
{\renewcommand\arraystretch{1.0}
\noindent\begin{longtable*}{@{}l @{\quad=\quad} l@{}}
$b$ & Width at the nozzle exit \\
$c$ & Speed of sound \\
$D_e$ & Equivalent nozzle diameter \\
$dt$ & Simulation time step \\
$f$ & Frequency \\
$f_{s}$ & Screech frequency \\
$h$ & Height at the nozzle exit \\
$\lambda$ & SPOD eigenvalues \\
$L_c$ & Jet potential core length \\
$L_s$ & Shock-cell spacing \\
$M_a$ & Acoustic Mach number, $U_j / c_\infty$ \\
$M_j$ & Fully-expanded jet Mach number, $U_j / c_j$ \\
NPR & Nozzle pressure ratio as computed by total pressure over ambient static pressure \\
$\mu$ & Dynamic viscosity \\
$p$ & Pressure \\
$Re_j$ & Reynolds number based on the nozzle exit height and the fully expanded jet velocity, $\rho_j U_j h / \mu_j$ \\
$\rho$ & Density \\
$St$ & Strouhal number, $f D_e / U_j$ \\
$St_g$ & Grid cut-off Strouhal number \\
$T$ & Temperature \\
TR & Temperature ratio as computed by total temperature over ambient static temperature \\
$t_{sim}$ & Total simulation time \\
$u,v,w$ & Velocity components in $x,y,z$ \\
$u_c$ & jet convection velocity \\
$U_j$ & Mean streamwise jet velocity \\
$\Delta$ & (Voronoi) Cell spacing \\
$\Delta t$ & Sampling period \\
$\phi$ & Inlet angle \\
\multicolumn{2}{@{}l}{Subscripts}\\
$\infty$ & Ambient properties \\
$j$ & Fully-expanded jet properties\\
\end{longtable*}}

\footnotetext{Presented as Paper 2021-1290 at the AIAA Scitech 2021 Forum, Virtual Event, 11--15 \& 19--21 January 2021}

\section{Introduction}
\label{sec:intro}
\lettrine{M}{ulti-jet} configurations are common in modern aircraft as a means to protect against emergencies such as the failure of one jet engine. When jet engines are closely placed, however, interactions between two jet plumes produce complex flow-fields around them that could be significantly different from those induced by a single, isolated jet. Moreover, when the nozzle exit pressure is inconsistent with the ambient condition, sometimes an aeroacoustic resonance, also known as screech, is generated with distinct tones and severely elevated noise levels. If this is the case, the proximity of the jet engines triggers coupling between them, which consequently alters their screech characteristics to a great extent. A series of previous studies provided abundant experimental evidence to support that, even though their frequencies are left unchanged, screech amplitudes may be drastically amplified or reduced by introducing additional jets~\cite{bozak2011,bozak2014a,bridges2014,kuo2017}. Such jet-to-jet interactions are identified as a function of several parameters such as the inter-nozzle spacing ($s$), nozzle pressure ratio (NPR), nozzle temperature ratio (TR), jet Mach number, and so forth. The aeroacoustics of twin jets is comprehensively reviewed by~\citet{raman2012}, and an overview of jet screech in general is given by~\citet{raman1999} and with more recent progress by~\citet{edgington2019}.    

Rectangular exit nozzles are widely used in advanced tactical aircraft because they allow variable aspect ratios and thrust-vectoring capabilities. Compared to the traditional axisymmetric jets, rectangular nozzles offer reduced complexity in terms of structural integration. Such design benefits have attracted researchers for the past few decades. For various nozzle exit aspect ratios and jet operating conditions, aeroacoustics of a single rectangular jet has been well documented experimentally~\cite{krothapalli1981,krothapalli1986,gutmark1990,quinn1992,zaman1999,alkislar2003,goss2009,heeb2013,valentich2016,karnam2019}. Numerical simulations have met some success in predicting the screech frequency~\cite{nichols2011,bres2017a,viswanath2017,gojon2019,chakrabarti2020,wu2020a,wu2021}, but fully recovering screech amplitudes were in many cases challenging. Noting that screech tones can be produced by a feedback loop associated with the interaction of instability waves and shock-cell structures and the receptivity at the nozzle lip,~\citet{powell1953} first derived screech frequency prediction formula for an axisymmetric jet using an averaged shock-cell spacing. Later,~\citet{tam1986} suggested a formula based on the wavenumber of the shock-cell systems. \citet{tam1988} further extended it for non-axisymmetric jets and found a close agreement with the experiments. In fact, his formula was obtained by a first-order approximation of the shock-cell systems bounded by thin vortex sheets. Success of his models for rectangular jets therefore supports a linear screech generation mechanism in such jets. On the other hand, \citet{panda1997} pointed out that the wavelength of standing waves produced by oppositely-travelling hydro-acoustic waves near the shear layer would be a more appropriate length scale to estimate the screech frequency, but recent success of a multi-mode screech prediction formula based on the shock-cell spacing~\cite{gao2010} indicates that it is a still relevant length scale for screech prediction. 

Complex geometries and highly sensitive screech phenomenon involved in twin rectangular jets require huge amount of computational resources to simulate flow-field around them. For this reason, most studies on twin rectangular jets to date have relied on experimental measurements~\cite{seiner1987,walker1990,zilz1990,raman1998,bozak2014a,bozak2014b,bridges2014,karnam2020,karnam2021,esfahani2021}. Amongst many efforts, in their seminal study \citet{raman1998} thoroughly investigated the coupling of twin rectangular jets of an aspect ratio of 4. The coherence between the two jets was quantified, and a simple test to identify their coupling modes was introduced. Based on the inter-nozzle spacing $(s)$ and the radius of the `null' region $(n)$, over which the phase of acoustic wavefronts arriving from downstream remained unchanged, a parameter $(\alpha)$ was defined such that $\alpha = \frac{s-2n}{h}$. Here, the reference length scale was given by the nozzle exit height ($h$). The sign of $\alpha$ then determined whether the `null' phase regions of the two jets overlapped or not; when they were separated ($\alpha > 0$), anti-symmetric coupling occurred. In contrast, when they were overlapped ($\alpha < 0$), the two jets were coupled symmetrically. \citet{raman1998} further found that the switch of coupling modes was associated with the transition of effective source locations, leading to the growth of such null regions. In this study the anti-symmetric coupling mode reduced the sound pressure level in the inter-nozzle region, while the other case amplified it.

An empirical model to estimate variation of far-field acoustic radiation ($\Delta$SPL) due to twin effects was suggested by~\citet{bozak2014b}. The effects of inter-nozzle spacing, temperature ratio, and jet Mach number as well as the dependence on frequency and observer angles were included in this model. To account for the effects of nozzle aspect ratio, the model was applied to two different nozzle geometries with aspect ratios of 2 and 8. The easiest estimation of twin-jet noise can be achieved by simply doubling sound pressure of a single jet with the assumption of uncorrelated twins. Compared to this scenario, the empirical model reduced errors in $\Delta$SPL slightly by considering the shielding effects of the twin jet. Nevertheless, the data set used to derive this empirical model covered subsonic jets only.

Leveraging on the recent progress in massively parallel implementation of large-eddy simulations (LES), numerically simulating twin rectangular jets became feasible. For twin rectangular jets with an aspect ratio of 2:1, which were designed and tested at the University of Cincinnati~\cite{karnam2020}, a first numerical campaign was launched recently~\cite{jeun2020a,viswanath2020}. These studies were able to provide initial assessment of flow-fields produced by the twin jets, including those inside the nozzle and in the very vicinity of the nozzle exits that were inaccessible in the experiments. Nevertheless, due to the complexities associated with the sharp converging-diverging sections of the nozzles and interaction between the twin jets, there still existed some deviations from the experiments.

This paper aims to obtain an experimentally-validated LES database for twin rectangular jet screech and characterize the dominant coherence associated with screech generation and twin-jet coupling. Firstly, high-fidelity LES data of the twin-jet flow dynamics in both near- and far-fields are compared with experimental measurements, including major improvements on the shock systems near the nozzle compared to our previously published results~\cite{jeun2020a,jeun2020b,jeun2021}. Next, the continuous wavelet transform (CWT) of the numerical data is used to reveal the intermittent nature of the screech tones. Lastly, Dominant coherent structures at screech frequencies are extracted by applying spectral proper orthogonal decomposition (SPOD) analysis~\cite{schmidt2018,towne2018}. The leading SPOD modes at screech frequencies are utilized to determine the coupling modes of the two jets.

The remainder of this paper is organized as follows. Section~\ref{sec:methods} details the experimental and numerical setups, including nozzle geometry and flow configurations. Section~\ref{sec:results} presents detailed results and discussions in three parts. Section~\ref{subsec:near_field} compares near-field flow statistics from  the LES with experiments. Characteristic scales of shock-cell systems and convection velocities are also measured. In Sec.~\ref{subsec:far_field} far-field sound predicted by LES is assessed with microphone measurements. The radiation patterns and intermittency of screech tones are discussed. Section~\ref{subsec:spod} provides SPOD analysis of the twin jets at the maximum screech condition and identifies the jet coupling modes at screech frequencies. Lastly, Sec.~\ref{sec:conclusion} concludes this paper with a summary of key observations and future work.

\section{Methodologies}
\label{sec:methods}
\subsection{Nozzle geometry and flow parameters}
\begin{figure}
\centering
\begin{tabular}{cc}      
  \includegraphics[width=.35\textwidth]{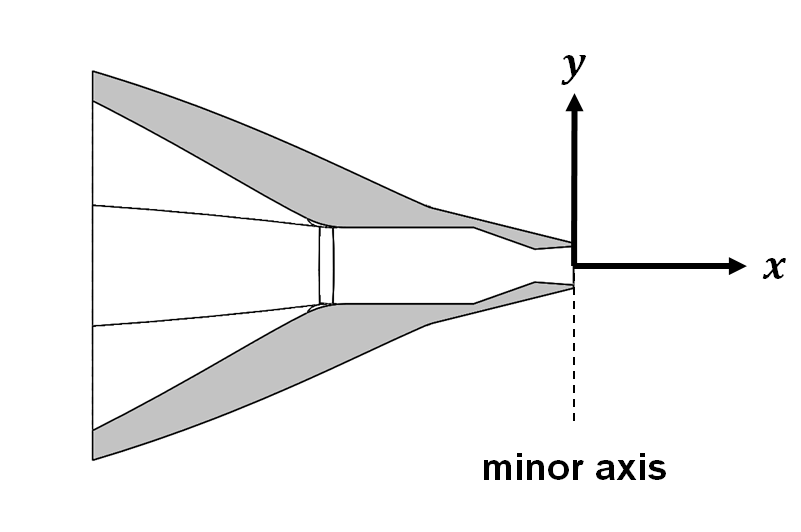} & \includegraphics[width=.35\textwidth]{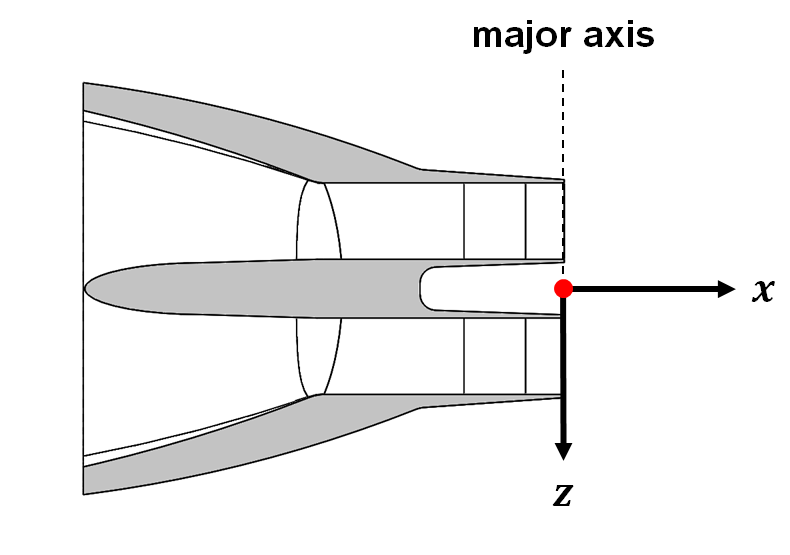} \\
  a) & b)\\ 
  \multicolumn{2}{c}{
    \includegraphics[width=.35\textwidth]{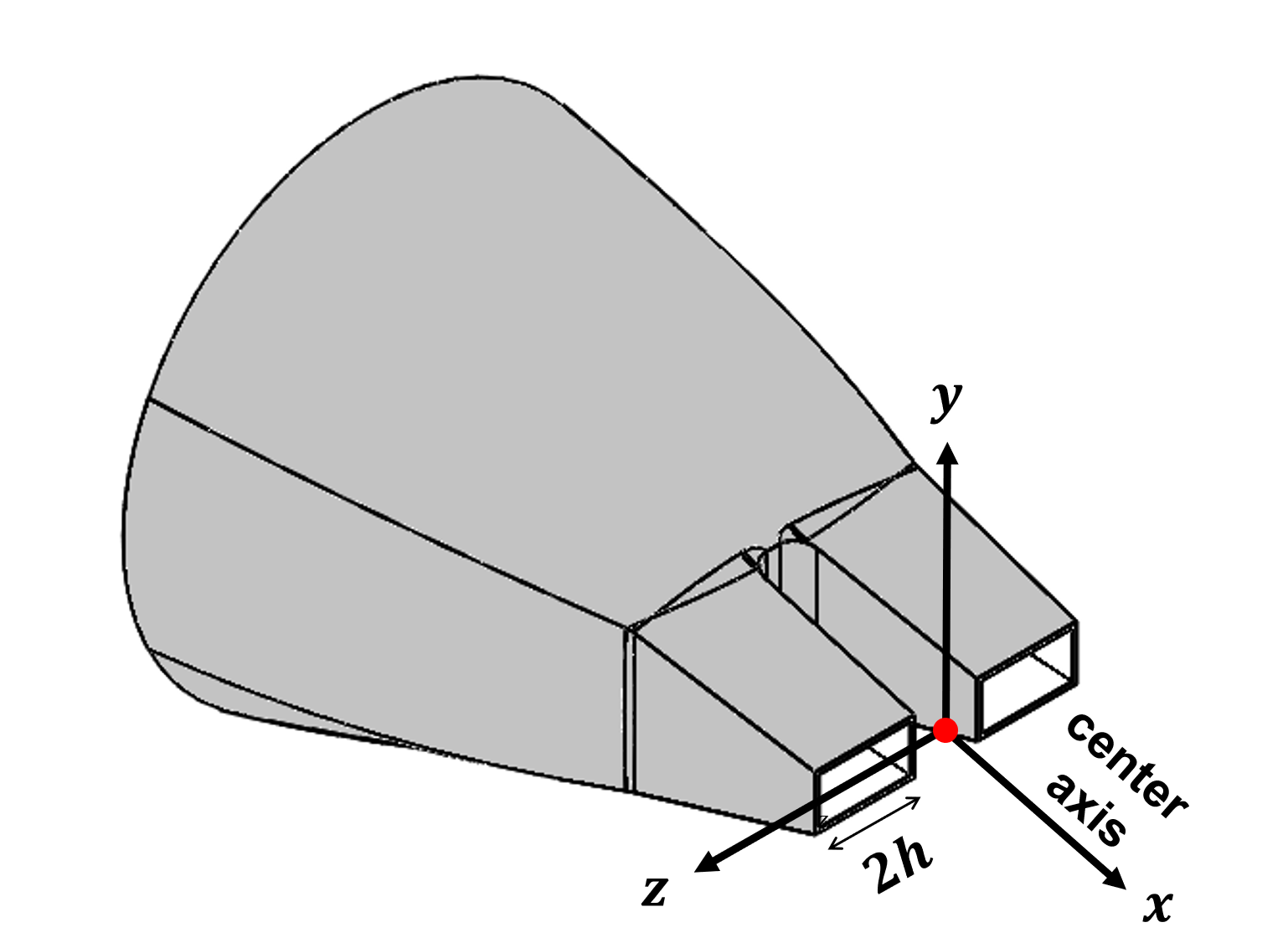} 
	} \\
	\multicolumn{2}{c}{
	c)
  } \\ 
\end{tabular}
\caption{Nozzle geometry: a) mid-plane cross-section along the minor axis, b) mid-plane cross-section along the major axis, and c) nozzle exterior (Figures drawn to scale).}
\label{fig:nozzle_geom}
\end{figure}

We consider twin rectangular nozzles each with an aspect ratio of 2:1, designed and tested at the University of Cincinnati~\cite{karnam2020} as shown in Fig.\ref{fig:nozzle_geom}. The origin of the coordinate system is located mid-way between two nozzle exits and denoted by a red dot in this figure. With respect to this, the $+x$-axis is defined along the streamwise direction, and the $y$- and $z$-axes are in the minor and major axes of the nozzles, respectively. In other words, the minor axis is defined along the converging-diverging (CD) side, and the major axis is in the flat side of the bi-conic nozzles. Throughout this study, we use the nozzle exit height ($h$) as a reference length scale, unless otherwise stated. As seen in this figure, the nozzle configuration initially has a circular cross-section, which is split in half and transitions into two 12.19 mm by 24.13 mm rectangular exits by a separator at $x/h = -12.4$. The equivalent diameter $(D_e)$ is defined as $D_e = 1.30 (b h)^{0.625} / (b + h)^{0.25}$ so as to yield the same pressure drop between a round nozzle and a rectangular nozzle with the exit width ($b$) and height ($h$)~\cite{huebscher1948}. For our twin nozzles, $D_e = 1.52 h$.

The inter-nozzle spacing ($s$) is given by $3.5 h$. It is well known that for twin jets, this distance hugely impacts interactions between nozzles and consequently screech characteristics~\cite{raman1998,raman2012,bozak2014a,bridges2014}. However, a parametric study on the effect of nozzle spacing for our low aspect ratio rectangular nozzles is beyond the scope of this paper. Between the two jets, one may notice cavity-like region, extending up to $x/h$ = -3.73. By approximating it as a rectangular cavity that is confined in the major axis only and open to the ambient air, we find that the resulting natural frequencies do not correspond to screech frequencies of the jets we consider in the present paper so any resonant effects that may result from this region are neglected.

Mimicking tactical aircraft, the twin nozzles have sharp converging-diverging sections in the minor axis, but the nozzle geometry remains flat along the major axis. Such bi-conic nozzle design, together with the rectangular exits under non-ideal expansion conditions, yields intricate flow fields inside the nozzles and following jet plumes, making simulating them challenging. Furthermore, the large inlet-to-outlet ratio ($\approx$ 33.7) of this nozzle configuration introduces strong flow acceleration, which causes any numerically triggered turbulence from the upstream to be significantly damped downstream. Therefore, special efforts need to be undertaken to achieve a turbulent nozzle-exit boundary layer with fluctuation levels similar to experimental conditions~\cite{bres2018a}. In the experiments, the nozzles were 3D printed in plastic and inherently had some level of surface roughness, leading to fully turbulent boundary layer state. The results produced by LES are, however, based on a laminar boundary layer near the nozzle exits but still show favorable agreement with the far-field noise and twin-jet coupling observed in the experiments, as will be discussed in later sections. 

For this nozzle configuration, two over-expanded conditions (NPR = 2.5 and 3) and the design condition (NPR = 3.67) are investigated. Here, the nozzle pressure ratio (NPR) is calculated as the ratio between total pressure and ambient pressure. The fully expanded jet velocity at the design condition $(M_d)$ is 1.5. In all cases, the nozzle temperature ratio (TR), which is defined as the ratio of total temperature to ambient static temperature, remains at $1$ so jets are considered to be cold. We chose these jet operating conditions for a direct comparison with the experiments conducted at the University of Cincinnati. The summary of three different jet operating conditions is given in Table~\ref{tab:operating_conditions}. In this table, $M_j$ represents the fully expanded jet Mach number computed from the isentropic relations, and $M_a$ means the acoustic Mach number, respectively. The Reynolds number ($Re_j$) is defined based on the nozzle exit height and the fully expanded jet velocity at each condition.

\begin{table}[t]
\centering
\caption{Summary of jet operating conditions}
\begin{tabular}{ccccccc}
\hline\hline
  Description & NPR & TR & $M_j$ & $M_a$ & $T_j / T_\infty$ & $Re_j$ \\ 
\hline
  Cold over-expanded supersonic jet & 2.5 & 1.0 & 1.22 & 1.07 & 0.769 & $4.70 \times 10^5$ \\ 
  Cold over-expanded supersonic jet & 3 & 1.0 & 1.36 & 1.16 & 0.730 & $5.60 \times 10^5$ \\ 
  Cold ideally-expanded supersonic jet & 3.67 & 1.0 & 1.50 & 1.25 & 0.689 & $6.68 \times 10^5$ \\ 
\hline\hline \\
\end{tabular}
\label{tab:operating_conditions}
\end{table}

\subsection{Experimental setup}
\label{sec:exp}
Flow field measurements using particle image velocimetry (PIV) and acoustic measurements using a far field microphone array were conducted in the Gas Dynamics and Propulsion Laboratory at the University of Cincinnati~\cite{karnam2020}. The nozzle was installed slightly off center in a 10 ft $\times$ 7 ft $\times$ 10 ft fully anechoic chamber, and air is supplied from a 22,000 lb high pressure (1800 psig) storage tank filled by a 360 SCFM compressor. The experimental setup allowed to capture a flow field up to 16$h$ downstream from the nozzle exit. The data captured from the cameras were then processed through a correlation technique to extract velocity vectors in the streamwise and cross-streamwise directions ($v$ or $w$, depending on the orientation). Utilizing the fact that the two cross-stream velocity fluctuations were found to be almost identical, the turbulent kinetic energy (TKE) was computed as:
\begin{equation}
    TKE = \frac{1}{2}(\overline{u^{'}u^{'}} + 2\overline{v^{'}v^{'}}) \quad or \quad \frac{1}{2}(\overline{u^{'}u^{'}} + 2\overline{w^{'}w^{'}}),
\label{eq:tke}
\end{equation}
depending on the nozzle orientation.

\subsubsection{Acoustic measurements}
\begin{figure}
  \centering
  \begin{tabular}{cc}      
    \includegraphics[width=.4\textwidth]{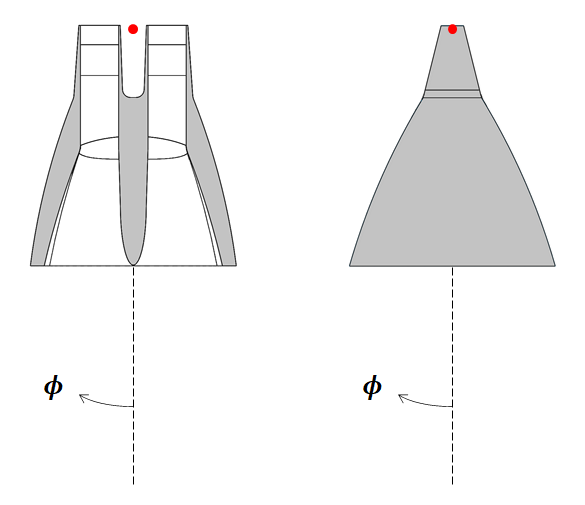} \\    
  \end{tabular}
  \caption{Far field microphone array setup: (left) major axis and (right) minor axis. Red dot represents the origin of the coordinate system.}
  \label{fig:mic}
\end{figure}

Acoustic measurements were taken using a far field microphone array fixed at 61.5$D_e$ from the system origin. The array consisted of 16 microphones ranging from $\phi$ = 45$^\circ$ to 152$^\circ$ in both the major and minor directions, where the polar angles ($\phi$) were measured from far upstream of the jet as shown in Fig.~\ref{fig:mic}. The sampling rate for the data acquisition was set at 204,800 Hz with a duration of 2 seconds. Each data set was collected multiple times on different days to achieve repeatability of the experiments, ensuring the overall variation between runs to be less than 1 dB. This yielded the raw data of 409,600 samples that were sequentially divided into 100 blocks. The acoustic spectra were then calculated by taking fast Fourier transform over each block and averaging all segments. The overall sound pressure level (OASPL) was calculated as an integral of the frequency spectrum over a relevant frequency range (from 300 Hz to 100,000 Hz) due to the anechoic chamber threshold. More details on the acoustic measurements can be found in~\citep{karnam2020}.

\subsubsection{Particle image velocimetry}
PIV images were acquired using a pair of LaVision Imager Intense CCD cameras, each with a 1376 $\times$ 1040 pixel array (pixel pitch: 6.45$\mathrm{\mu}$m) stacked vertically as seen below in Fig.~\ref{fig:piv_setup}. An Evergreen Dual Pulse Nd: Yag laser operating at a peak frequency of 5 Hz at 532 nm (170 mJ per pulse) was used for illumination. The laser beam was passed through an iris to minimize distortion effects due to non-circular beam cross-section, then through a focusing lens and finally a cylindrical lens to produce a sheet $\approx$1 mm in thickness. The two cameras were paired with Nikon (NIKKOR) 50 mm lenses at f/16 aperture to maximize the measurement area. An optical bandpass filter (Wavelength: 532 nm; Optical Density: 6; Full-Width Half Max (FWHM): $\pm$10 nm) was used with each lens to avoid capturing light from surface/stray reflections without affecting the light reflected off the seed particles. A knife edge was also used to limit the spread of the laser sheet to prevent strong reflections from the nozzle lip bleeding into the camera frame. The choice of the lens was made to accommodate both the jets in the twin-jet configuration. The interval between successive frames was set to 0.5 $\mu$s to minimize measurement lag given the large field of view (FOV) provided by the lens. For each test condition a series of 1200 images are recorded for vector field and flow property computation. The camera-lens combination resulted in a magnification of 12.55 pixels (px) /mm with a pixel shift of 2.33 px at 370 m/s (NPR 2.5) and 2.7 px at 430 m/s (NPR 3.67). Table~\ref{tab:piv_parameters} shows some of the crucial PIV parameters.

\begin{figure}
  \centering
  \begin{tabular}{cc}      
    \includegraphics[width=.2347\textwidth]{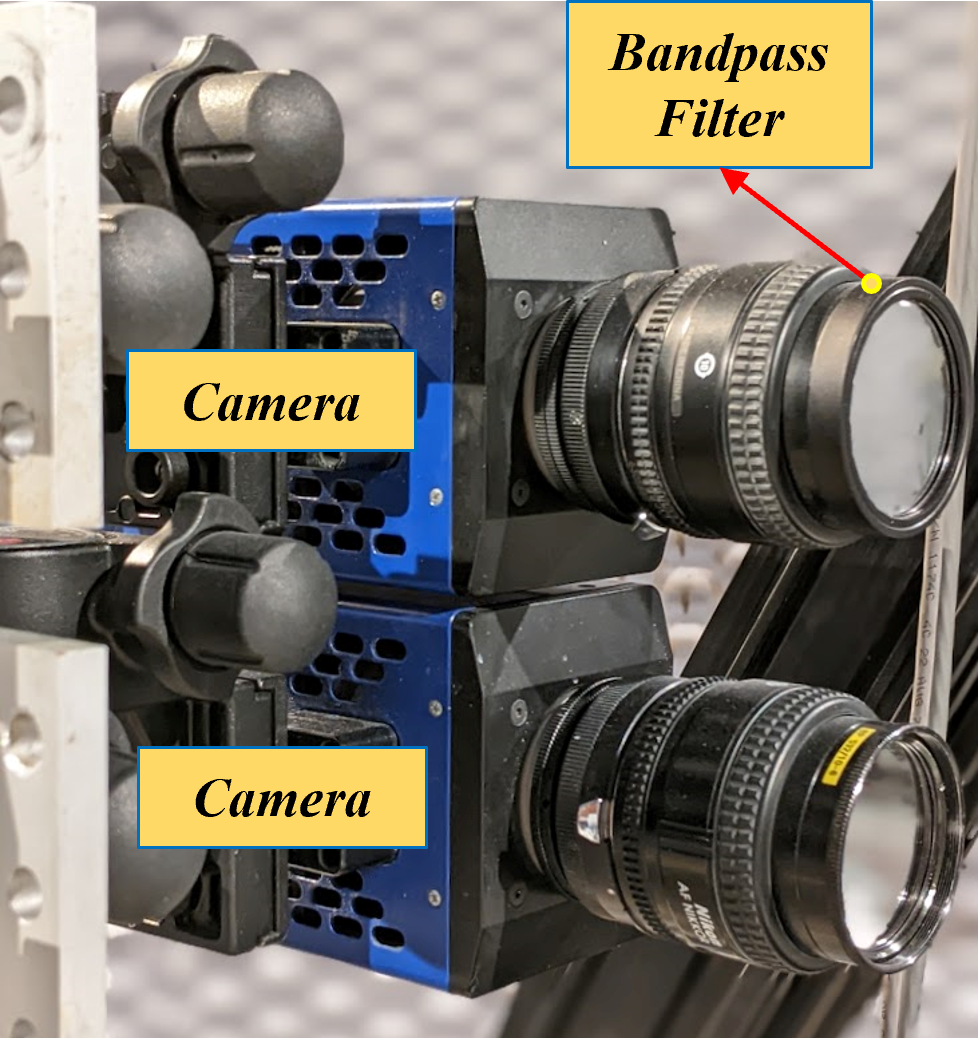} & \includegraphics[width=.57\textwidth]{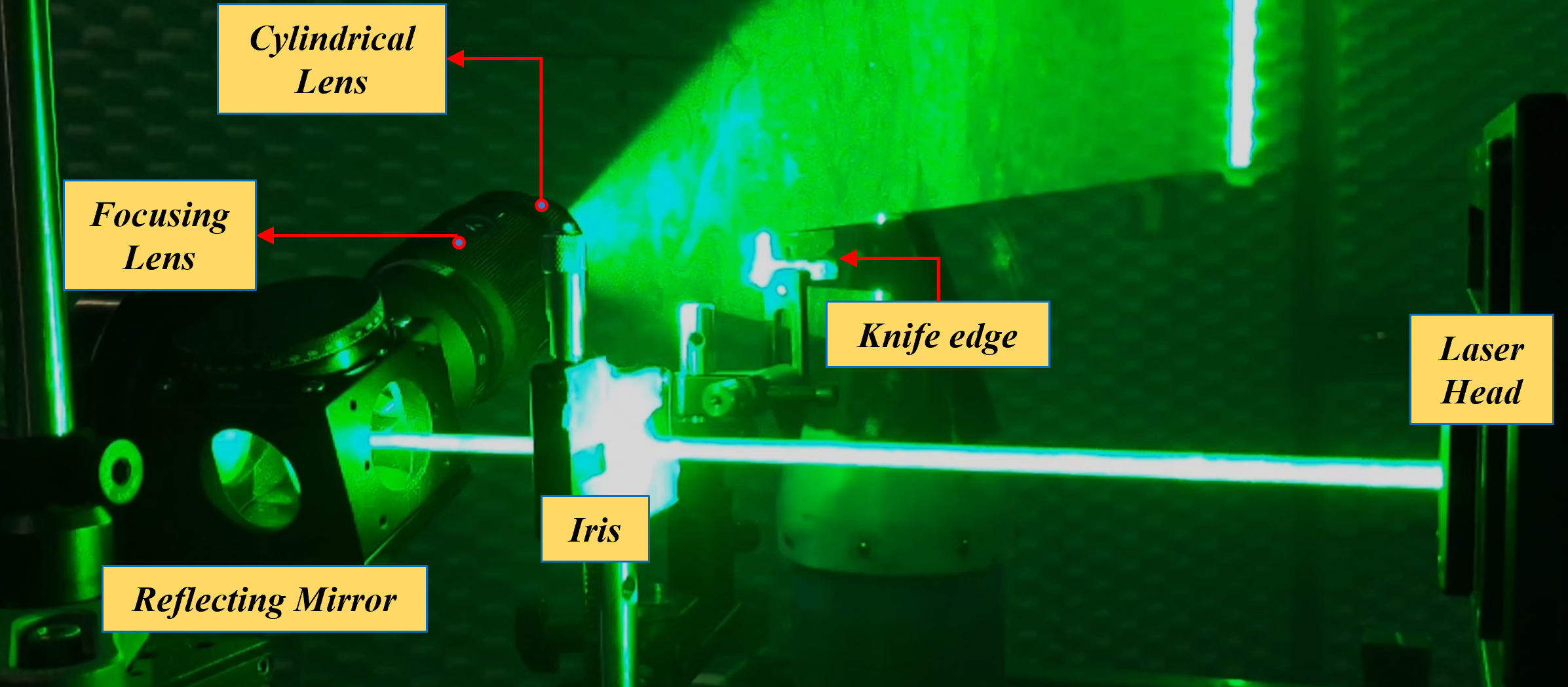} \\
  \end{tabular}
  \caption{PIV setup for image capture. (Left) Camera set up and orientation. (Right) Laser optics for illumination.}
  \label{fig:piv_setup}
\end{figure}

\begin{table}[t]
\begin{center}
\caption{PIV parameters.}
\begin{tabular}{cc}
  \hline\hline
  Parameter & Value \\ 
  \hline
  Interrogation window -- Initial pass, px & 128 $\times$ 128 \\
  Interrogation window -- Final pass, px & 16 $\times$ 16 \\
  Interrogation window -- Final, $D_{e}$ & 0.07 $\times$ 0.07 \\
  Overlap, $\%$ & 50 \\
  Frame separation, $\mu$s & 0.5 \\
  Exposure time, ns & 500 \\
  Field of view (FOV), $D_{e}$ & 5.92 $\times$ 4.48 \\
  Digital resolution, px/$D_{e}$ & 231.6 \\
  \hline\hline
\end{tabular}
\label{tab:piv_parameters}
\end{center}
\end{table}

\subsubsection{Seed particle choice and particle lag study}
Previous studies on single rectangular jets~\cite{gojon2019,baier2020,chakrabarti2021} conducted at the current facility showed aluminium oxide as a suitable candidate for flow seeding for quantification of flow properties in supersonic jets and for validation of LES results. The seed particles used in the current study were sourced from Buehler Inc. with a manufacturer specified size of 0.3 $\mu$m. The ambient was seeded using fog from a water-based fog generator with a manufacturer specified size of 1 $\mu$m.

\begin{figure}
  \centering
  \begin{tabular}{c}      
    \includegraphics[width=.45\textwidth]{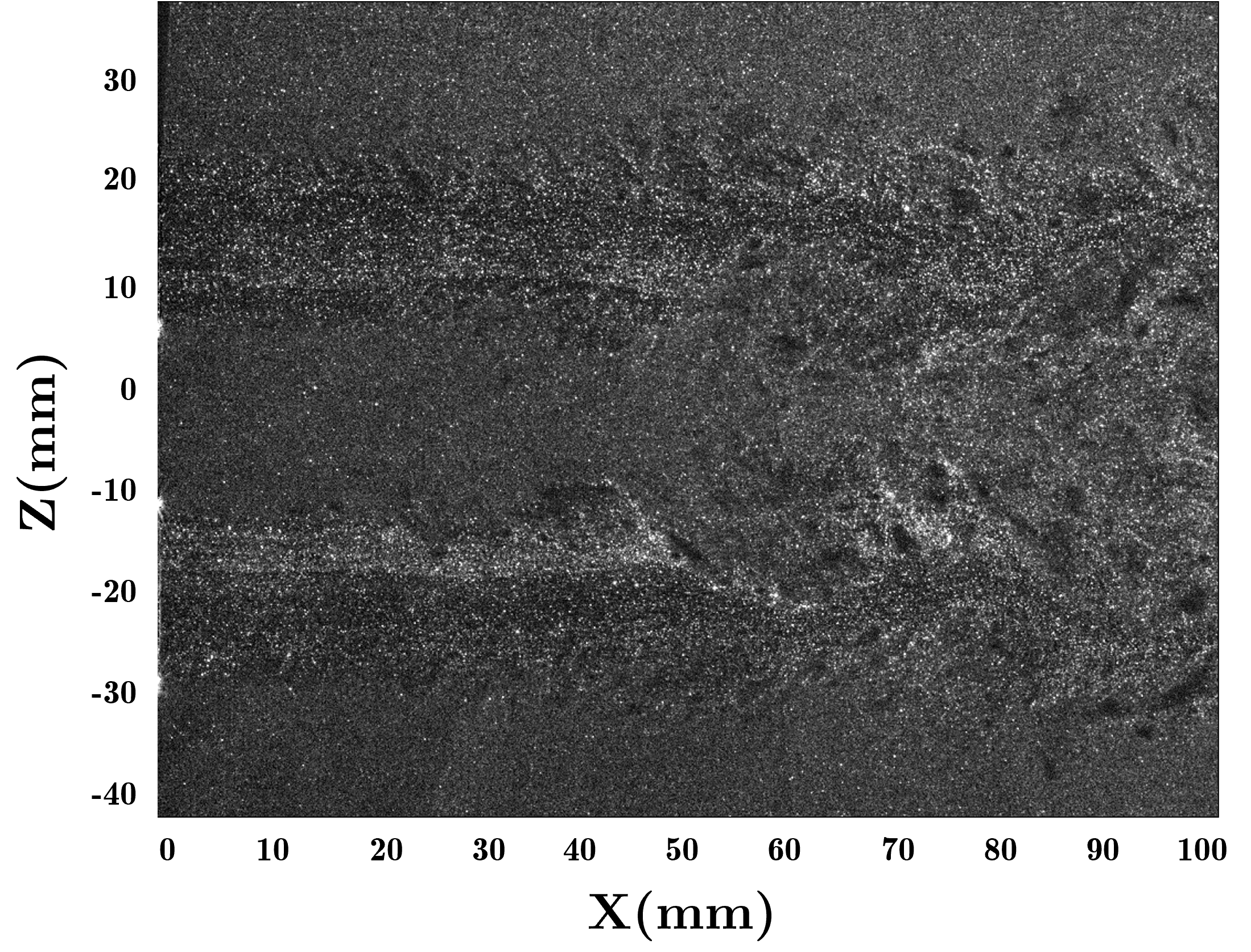} \\
  \end{tabular}
  \caption{Combined seed distribution achieved from the core seed and the ambient seed (Flow direction: left to right).}
  \label{fig:seed_dist}
\end{figure}

\begin{figure}[H]
  \centering
  \begin{tabular}{ccc}      
    \includegraphics[width=.4\textwidth]{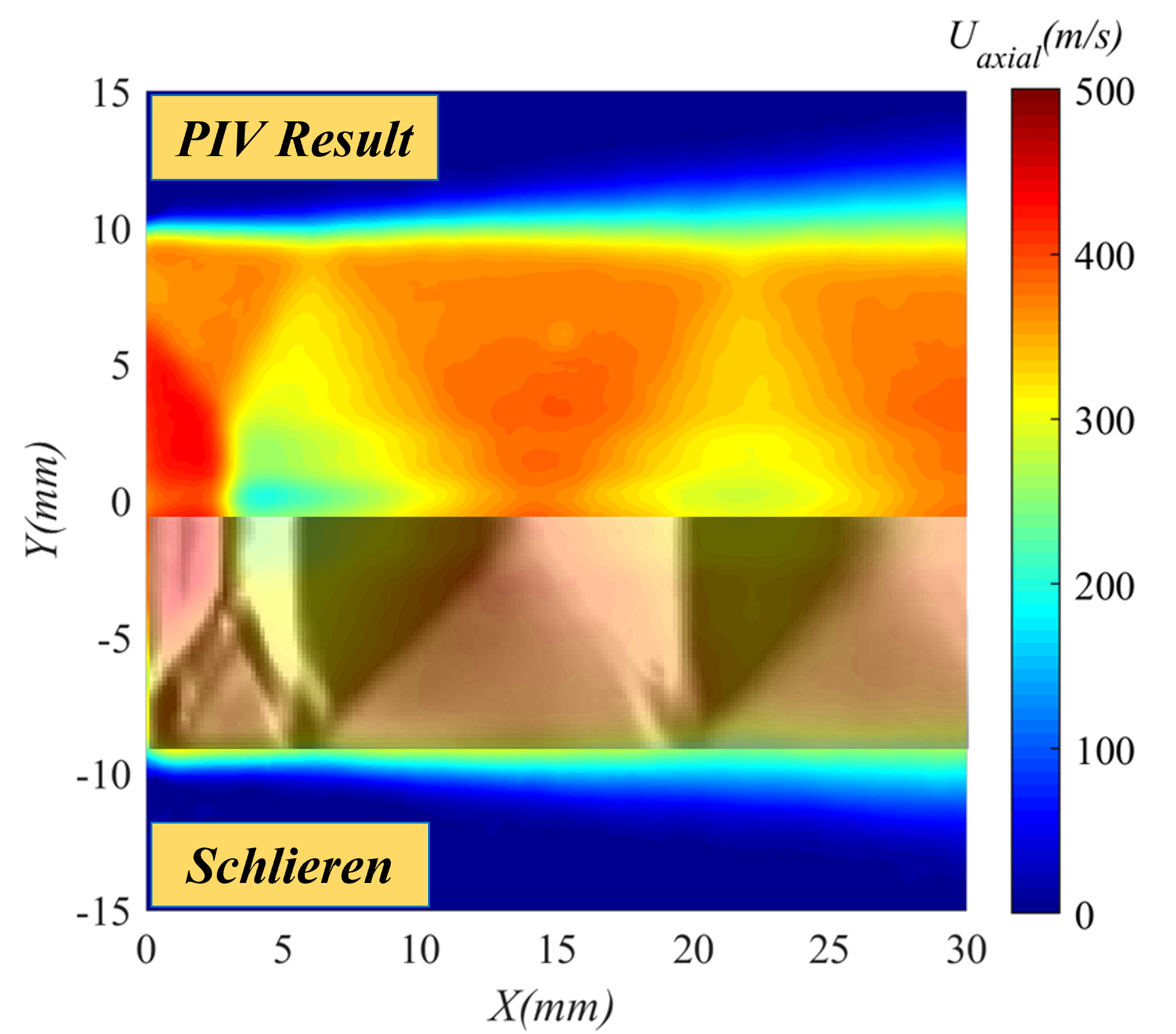} & \includegraphics[width=.3\textwidth]{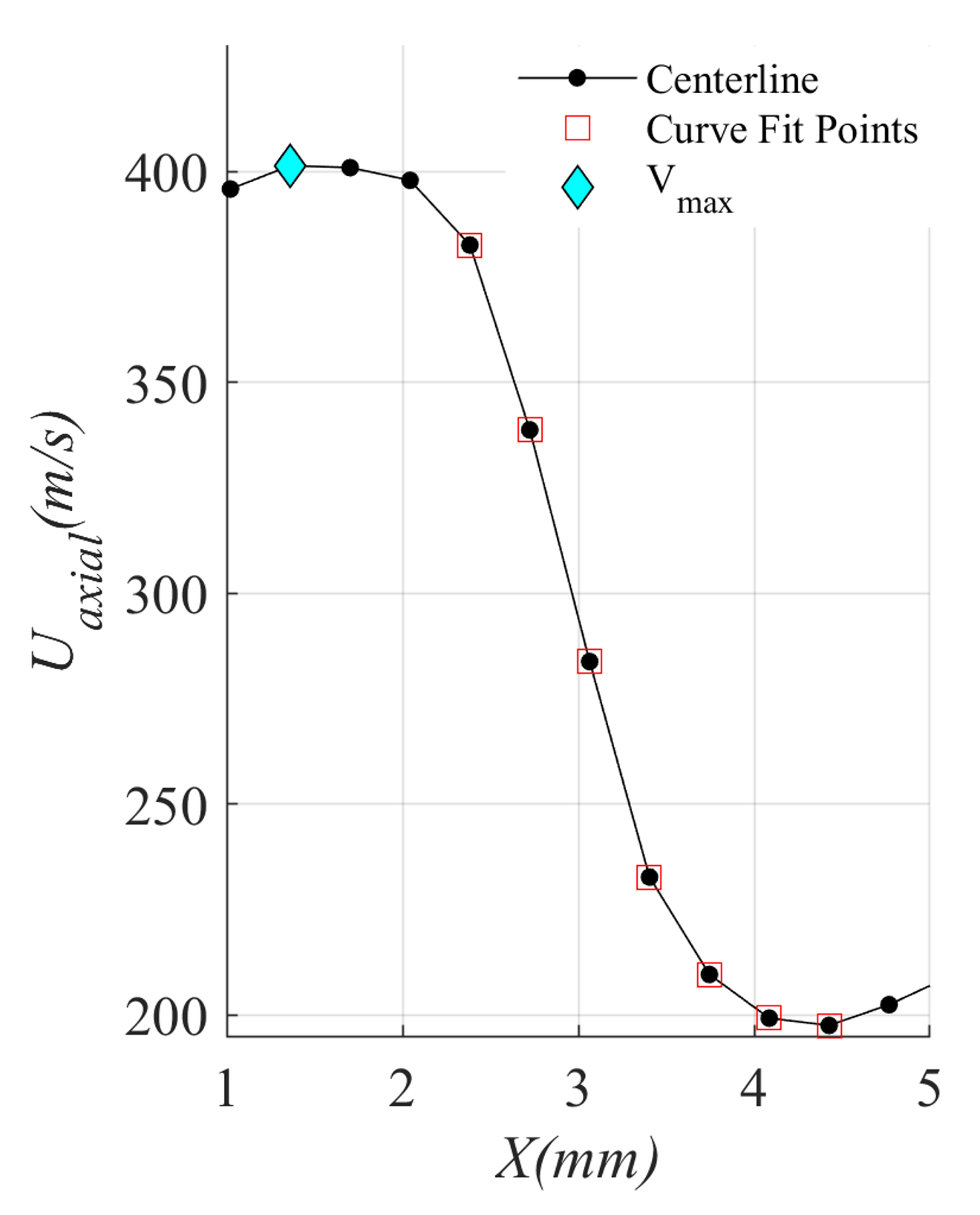} & \includegraphics[width=.22\textwidth]{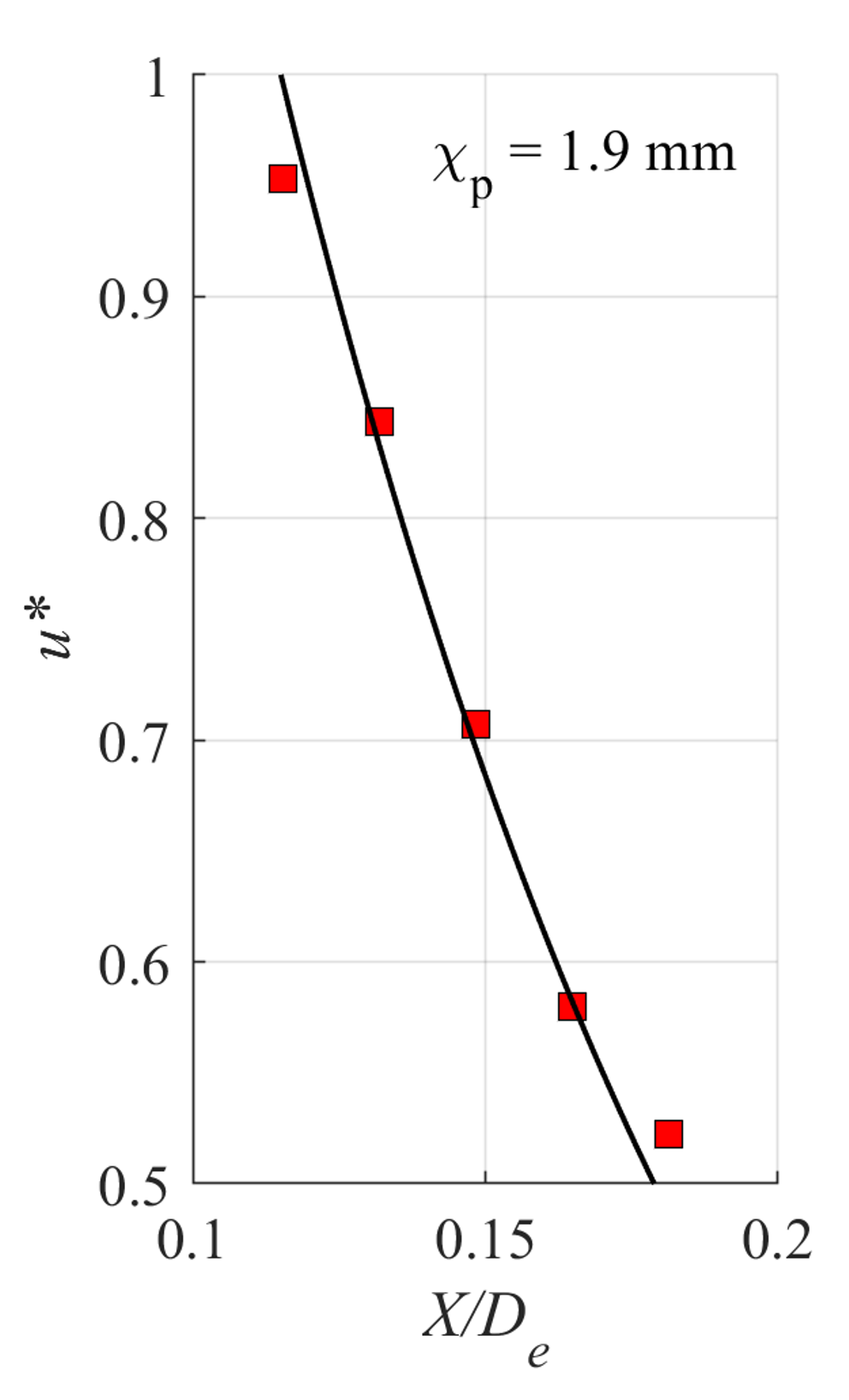} \\
  \end{tabular}
  \caption{Results from a circular jet at NPR = 2.5: (left) schlieren image overlaid on PIV contour showing the normal shock at jet exit, (middle) peak velocity and points chosen for exponential fit, and (right) particle relaxation distance computed from the curve fit.}
  \label{fig:particle_lag_study}
\end{figure}

The distribution of ambient seed and core seed is shown in Fig.~\ref{fig:seed_dist} in an instantaneous image from the measured data set for the condition of NPR = 2.5 looking at the flat side of the jet. To quantify the particle lag effects, the methodology used by~\citet{haghdoost2020} was adopted for the current study. The particle relaxation time was computed based on Melling’s~\cite{melling1997} exponential velocity decay criteria for seed particles using a circular CD nozzle with an exit diameter $D$ = 20.65 mm with the same design conditions ($M_d$ = 1.5 and NPR = 3.67) as the twin jets. The over-expanded condition of NPR = 2.5 was chosen for characterization due to the formation of a discernible normal shock seen from the schlieren image in Fig.~\ref{fig:particle_lag_study}(a). The results obtained from the PIV test are shown in Fig.~\ref{fig:particle_lag_study}(b), where the points marked in the velocity profile were used for the exponential fit for relaxation time computation. The particle relaxation length ($\upchi_p$) for the current setup was found to be 1.9 mm which results in a particle relaxation time of 5.5 $\mu$s. The flow relaxation time, computed based on the maximum exit velocity of the flow and the nozzle exit diameter~\cite{edgington2014}, was found to be 47 $\mu$s which results in a Strouhal number $St \approx$ 0.11 for this setup which satisfies the commonly established criteria of $St \ll$ 1. Although the Strouhal number for the current set up was not as low as that seen in other studies~\cite{haghdoost2020,edgington2014}, the reader must bear in mind that the FOV used in the current study is much larger in relation to the camera sensor when compared to studies of similar nature. The choice of FOV was based on the focus of this study which is centered on understanding the interactions between the twin jets requiring the need for a larger measurement window to capture sufficient flow dynamics while not sacrificing measurement accuracy.

\subsubsection{Notes regarding the updates on results}
Substantial changes have been made to the experimental setup for flow acquisition after initial results were published in~\cite{karnam2020,jeun2021}. Overall, the qualitative comparisons between the LES and experiments remain unchanged. The updated experiments however offer improved quantification of the flow field, particularly concerning the shock-cell system.

\subsection{Numerical setup}
\label{sec:les}
\subsubsection{Large-eddy simulations}
High-fidelity large-eddy simulations are performed using an unstructured compressible solver, CharLES, developed by Cascade Technologies. CharLES is equipped with a Voronoi-based mesh-generation framework~\cite{bres2018b} and a shock-capturing method based on kinetic energy and entropy preserving (KEEP)~\cite{tadmor2003,chandrashekar2013,fisher2013}. CharLES has demonstrated its predictive capabilities for jets issuing from various complex nozzle configurations~\cite{bres2014,bres2017a,bres2018b,bres2019,jeun2021,wu2021}. For more details on the flow solver, readers may refer to~\citet{bres2017b,bres2019}.

\begin{figure}
\centering
  \includegraphics[width=.6\textwidth]{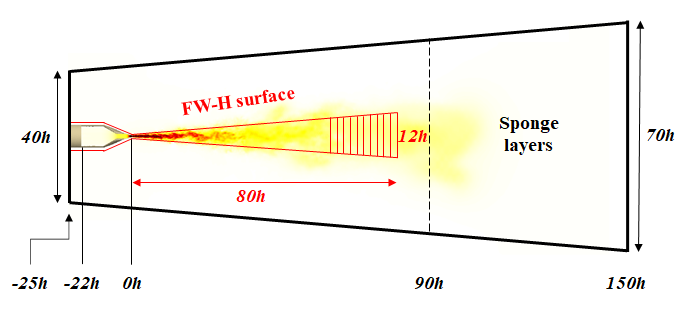}
\caption{A schematic representation of the computational domain and numerical setup.}
\label{fig:les_domain}
\end{figure}

As visualized in Fig.~\ref{fig:les_domain} the numerical domain extends from -25$h$ to 150$h$ in the streamwise direction and flares slightly in $y$- and $z$-directions from $\pm 20h$ to $\pm 35h$. Along the nozzle wall boundaries, the equilibrium wall-model is applied. At the nozzle inlet at $x/h$ = -22, the pressure, velocity, and temperature are specified such that the desired flow conditions are met at the exits. In the experiments, the total conditions were in fact measured at a location much upstream ($x/h = -31$), but pressure loss introduced by truncating the upstream domain boundary is predicted to be negligible. Furthermore, LES uses plug flow-like boundary conditions, while the actual flow appeared to be already fully developed pipe flow in the experiments. Again, pressure loss across the pipe diameter at this location will be less than $0.1\%$ so the use of such boundary conditions can be justified. In addition, unphysical reflections of outgoing waves into the numerical domain are avoided by employing the downstream sponge layer at $x/h$ = 90 as denoted by a black dashed line in Fig.~\ref{fig:les_domain}, with appropriate sponge strength determined following the recommendations in the literature~\cite{mani2012,bres2017b}. To compute the far-field sound, we use a permeable formulation of the Ffowcs Williams-Hawkings (FW-H) equations~\cite{ffowcs1969,lockard2000,bres2017b}. The near-field flow data are extracted on the FW-H surfaces as schematically represented by red lines in Fig.~\ref{fig:les_domain}. The choice of FW-H surface is made after testing several candidates so that it can contain as much of the noise sources as possible while ensuring sufficient numerical resolution. The cross-sections of the FW-H surfaces are in forms of a rounded rectangle at the beginning and smoothly transformed through an elliptic to a circle, as the twin jets merge together at far downstream regions. To cancel spurious sound produced by convecting vortices, the outflow disks of the FW-H surfaces are closed with $11$ end-caps~\cite{shur2005}, which are equally distributed from $x/h$ = 60 to 80. 

\subsubsection{Mesh generation}
\label{subsubsec:mesh}
We follow similar mesh generation strategies discussed in a series of the authors' previous studies on single and twin rectangular jet simulations\cite{jeun2020a,jeun2020b,wu2021}. To account for oblique shocks formed downstream of the nozzle throats and upstream-propagating waves, the mesh is carefully refined along the nozzle walls and near the nozzle. The regions along the jet shear layers and in the potential cores also require high resolution grids to resolve the shock-cell systems in the jet plumes and instability waves along the jet boundaries. In addition, the experiments~\cite{karnam2020} showed that the two jets started merging around $x/h$ = 6 in the streamwise direction. The grids between the twin jets are therefore accordingly refined downstream of this location. Compared to the other two test cases, the stronger over-expanded condition with NPR = 2.5 formed Mach stems inside the jet potential core that might induce further velocity drops. Hence, the final mesh for this operating condition has higher grid resolutions in the initial potential core regions than those in the other cases, yielding the largest mesh size among the three conditions simulated in this paper. However, note that the additional refinement windows extend to only a short distance from the nozzle exits. Moreover, preliminary LES data revealed that the cases with higher NPRs registered longer potential cores and wider shear layers, requiring larger refined zones to resolve them. Despite such refinements, the mesh size changes by approximately 10\% only across all three test cases. In addition, the authors' previous studies~\cite{jeun2020b,wu2021} indicated that frequent mesh transitions led to inaccurate prediction of the far-field sound with missing screech tones especially at downstream radiation angles by progressively coarsening the grid resolutions of FW-H surfaces in the downstream. The cases of NPR = 3 and 3.67 still use meshes generated by the old mesh generation strategies resulting in slightly more number of mesh transitions, the 210M mesh for NPR = 2.5 undergoes less frequent transitions for better noise predictions. 

Table~\ref{tab:numerics} summarizes mesh resolutions and characteristic time parameters of the simulations for all cases. In this table, $dt$ denotes the simulation time step, $\Delta t$ is the FW-H data sampling period, and $t_{sim}$ represents the total simulation duration, respectively. Considering that the high-frequency noise components are mainly generated from sources residing in the region extending from the nozzle exit to the jet potential core and that 4-8 grid points per wavelength are required to sufficiently resolve an acoustic wave, the grid cut-off Strouahl number $St_g$ is estimated by $St_g$ = $D_e / (8 \Delta M_a)$ = 3 to 4 with $\Delta$ being the cell spacing based on the current mesh resolutions. A schematic of the meshes viewed in each direction is provided in~\cite{jeun2020b}.

\begin{table}
\begin{center}
\caption{Summary of mesh resolutions and characteristic time parameters of the simulation and post-processing of it.}
\begin{tabular}{cccccccc}
  \hline\hline
  NPR & \begin{tabular}[c]{@{}c@{}} Mesh \\ size\end{tabular} & \begin{tabular}[c]{@{}c@{}}Internal nozzle \\ wall resolution\end{tabular} & \begin{tabular}[c]{@{}c@{}}Potential core \\ resolution\end{tabular} & \begin{tabular}[c]{@{}c@{}}Jet plume \\ resolution \\\end{tabular} & $dt c_\infty / h$ & $\Delta t c_\infty / h$ & $t_{sim} c_\infty / h$ \\ 
  \hline
  2.5 & 210M & 0.005-0.01$h$ & \begin{tabular}[c]{@{}c@{}} 0.01-0.02$h$\vspace{-0mm} \\ $(0 \le x/h \le 24)$\end{tabular} & \begin{tabular}[c]{@{}c@{}}0.04-0.08$h$\vspace{-0mm} \\ $(24 \le x/h \le 50)$\end{tabular} & 0.001 & 0.05 & 1,400 \\
  3 & 190M & 0.005-0.01$h$ & \begin{tabular}[c]{@{}c@{}} 0.01-0.04$h$\vspace{-0mm} \\ $(0 \le x/h \le 24)$\end{tabular} & \begin{tabular}[c]{@{}c@{}}0.04-0.16$h$\vspace{-0mm} \\ $(24 \le x/h \le 50)$\end{tabular} & 0.001 & 0.05 & 1,400 \\
  3.67 & 200M & 0.005-0.01$h$ & \begin{tabular}[c]{@{}c@{}} 0.01-0.04$h$\vspace{-0mm} \\ $(0 \le x/h \le 24)$\end{tabular} & \begin{tabular}[c]{@{}c@{}}0.04-0.16$h$\vspace{-0mm} \\ $(24 \le x/h \le 50)$\end{tabular} & 0.001 & 0.05 & 1,000 \\
  \hline\hline
\end{tabular}
\label{tab:numerics}
\end{center}
\end{table}

\section{Results and Discussion}
\label{sec:results}
\subsection{Near-field aerodynamics}
\label{subsec:near_field}
\subsubsection{Instantaneous snapshots}
\begin{figure}
\centering
\begin{tabular}{c}
  \includegraphics[width=0.6\textwidth]{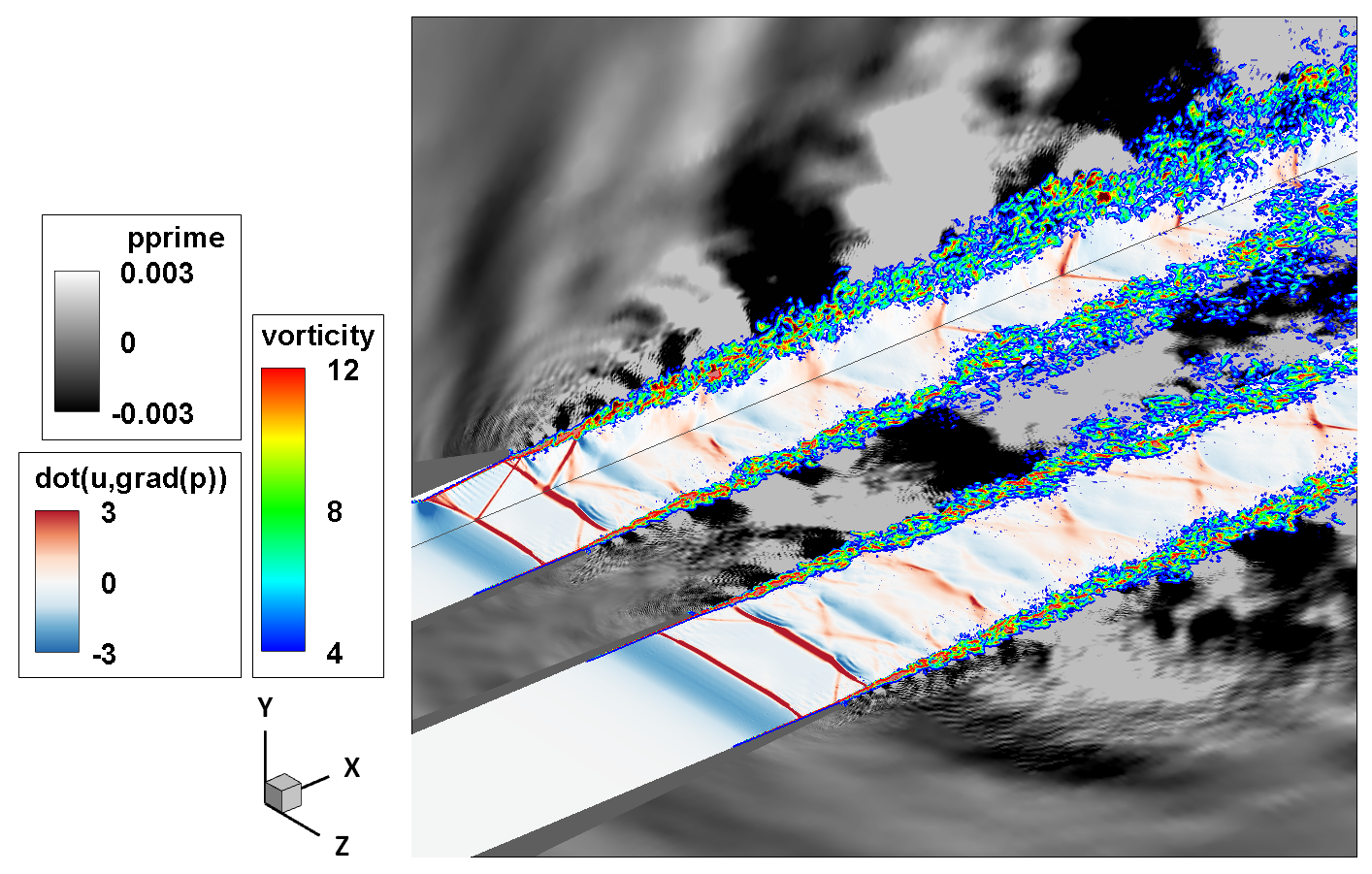} \\
\end{tabular}
\caption{Instantaneous snapshot of the twin rectangular jets with NPR = 3. A gray line is added to represent the border between the major and minor axis views of the jet on the left-side.}
\label{fig:snapshots}
\end{figure}

Figure~\ref{fig:snapshots} shows the instantaneous flow snapshots captured by LES, which visualize the overall flow dynamics in the near-field of the twin rectangular jets at NPR = 3. The snapshots are taken in the mid-plane of the twin nozzles along the major axis and also in that of one nozzle along the minor axis to highlight unique flow structures appearing in each axis. Red and blue contours of the variable $\textbf{u} \cdot \nabla{p}$ describe the instantaneous shock/expansion wave structures. Shock diamond patterns in the jet plumes are originated from the oblique shocks formed at the sharp throat and additional shocks generated just prior to the nozzle exits due to the non-ideal expansion operating condition. Since the internal nozzle wall remains flat in the major direction, the oblique shocks viewed in the major axis plane have normal shock fronts. Gray scale contours in the background visualize pressure fluctuations produced by the jet turbulence. In addition to the dominant downstream-travelling components resembling wavepackets, upstream-propagating waves associated with the jet screech are clearly visible in both axes. Finally, colored vorticity contours overlaid on top of this figure depict the shear layers.

\subsubsection{Mean field}
\begin{figure}
\centering
\begin{tabular}{c}
  \includegraphics[width=0.85\textwidth]{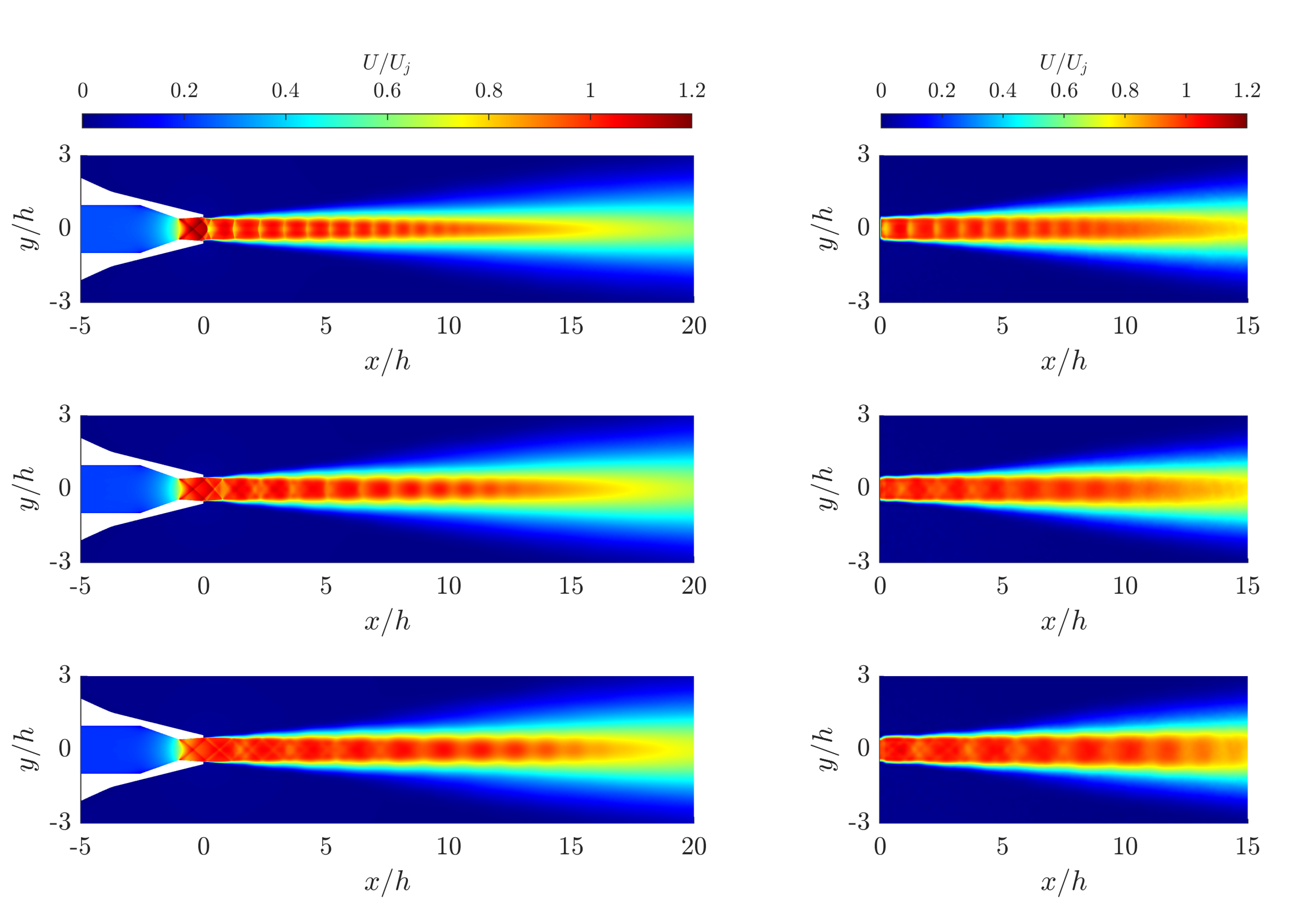} \\
\end{tabular}
\caption{Comparison of the mean streamwise velocity contours normalized by the fully expanded jet velocity between the (left) LES and (right) experiments in the minor axis plane: (top) NPR = 2.5, (middle) NPR = 3, and (bottom) NPR = 3.67.}
\label{fig:Uavg_contours_minor}
\end{figure}

\begin{figure}
\centering
\begin{tabular}{ccc}
  \includegraphics[width=0.3\textwidth]{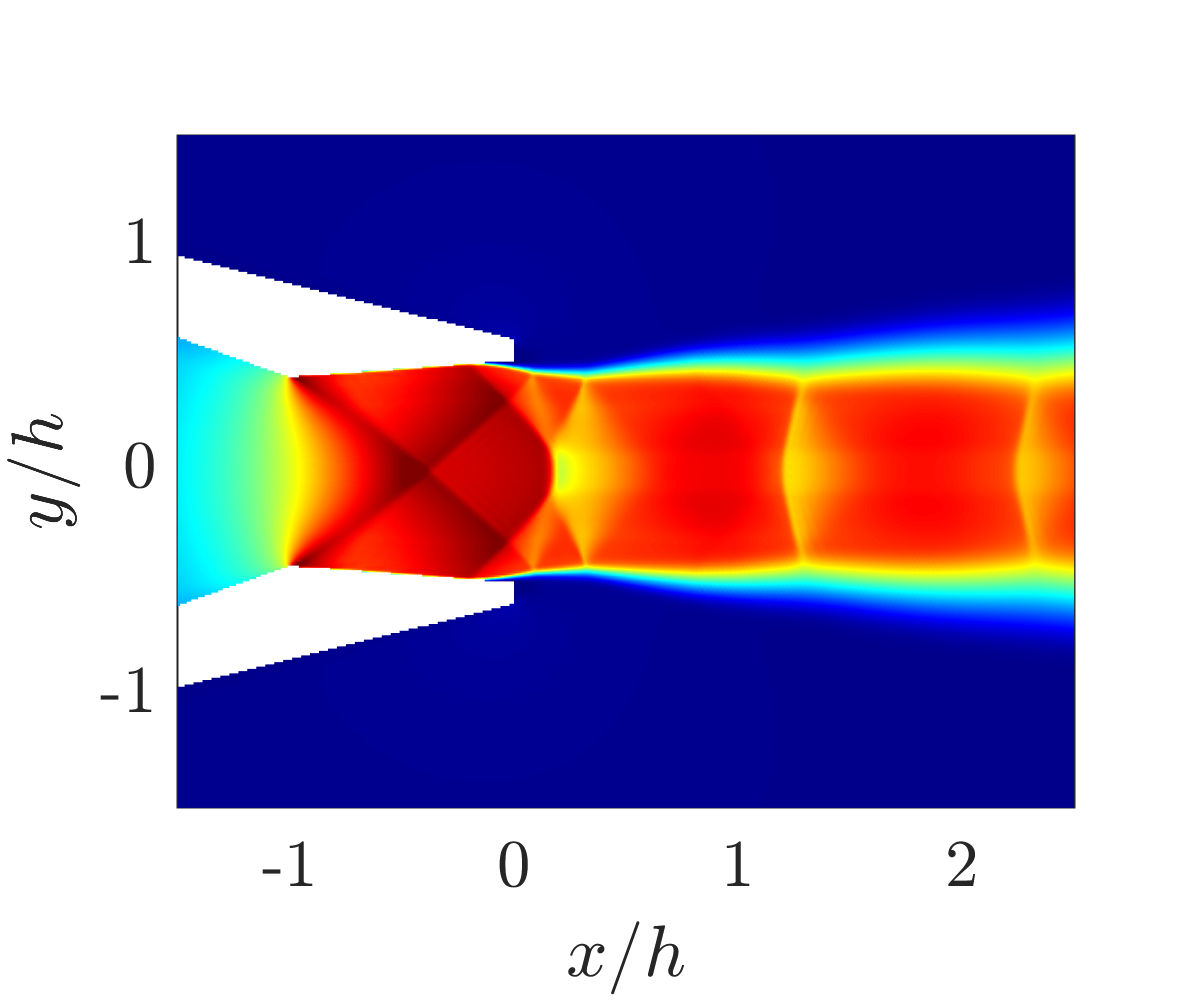} & \includegraphics[width=0.3\textwidth]{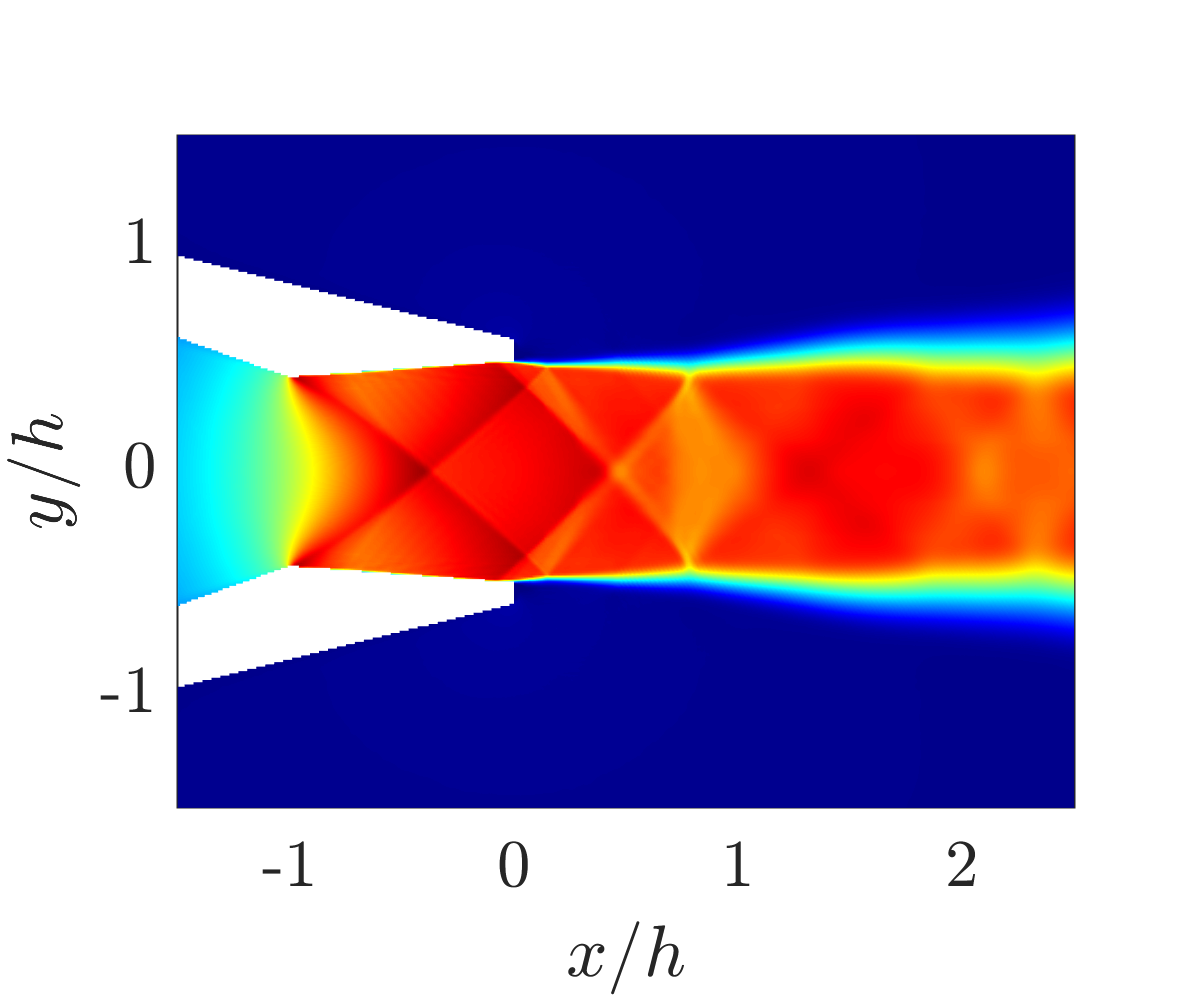} & \includegraphics[width=0.3\textwidth]{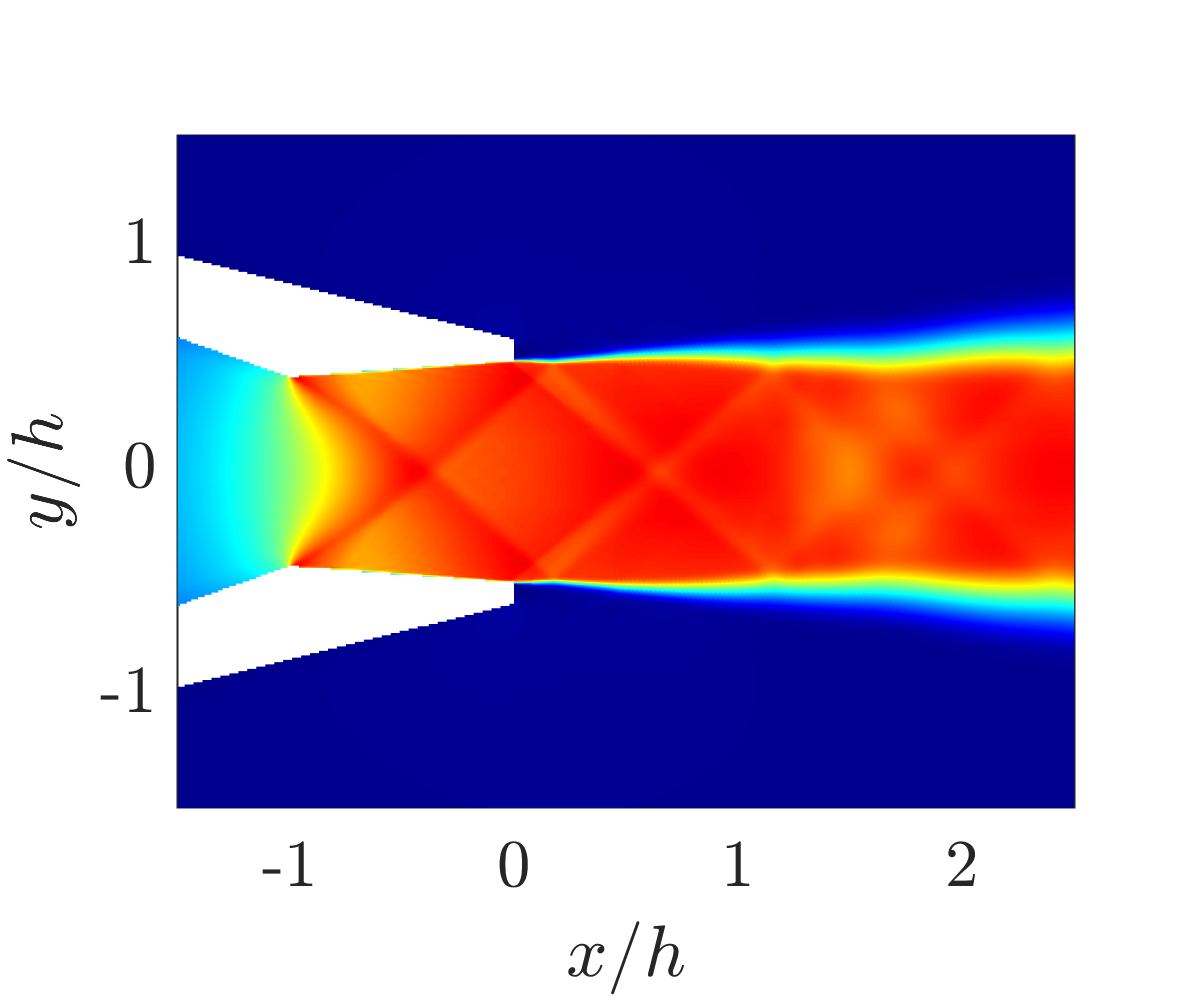} \\
  a) NPR = 2.5 & b) NPR = 3 & c) NPR = 3.67 \\
\end{tabular}
\caption{Mean streamwise velocity contours normalized by the fully expanded jet velocity zoomed-in near the nozzle exit. The colorscale ranges from 0 to 1.2.}
\label{fig:Uavg_internal_contours_minor}
\end{figure}

Figure~\ref{fig:Uavg_contours_minor} shows the normalized mean streamwise velocity contours in the minor axis plane predicted by LES, in comparison with the experimental measurements. By utilizing the fact that the twin jets share almost identical flow fields with each other, the LES contours are obtained by averaging the flow fields in the mid-plane cross-sections of the two nozzles along the minor axis, while the contours for the experimental results are obtained for one of the jets (centered at z/h = 1.75) only. The entire domain extends further downstream, but only a part of it is shown. The contours inside and in the vicinity of the nozzle are further zoomed in Fig.~\ref{fig:Uavg_internal_contours_minor} to highlight flow structures inaccessible in the experiments. As shown in the instantaneous snapshot, the oblique shocks initiated at the sharp throat propagate downstream, while interacting with additional shocks generated near the nozzle exit, ultimately shaping into a shock-cell system in the jet plume. In Fig.~\ref{fig:Uavg_internal_contours_minor}a the stronger over-expanded condition (NPR = 2.5) exhibits the formation of Mach stems, allowing additional velocity drops along the jet centerline. Their strength is measured to be weaker in the experiments as shown in Fig.~\ref{fig:Uavg_contours_minor}. It is also important to note that, as seen in Fig.~\ref{fig:Uavg_internal_contours_minor}c, the bi-conic converging-diverging sections and slight mismatch in the nozzle exit pressure to the ambient condition produce weak shock systems even at the design condition (NPR = 3.67). Overall, LES predicts qualitatively similar flow features to those captured in the experiments. However, whereas in the experiment the case at NPR = 3 registered the shortest jet potential core length ($L_c$ = 10.8$h$) and the maximum screech amplitude along with intense jet flapping motions in the minor axis due to strong twin-jet interactions, LES shows longer jet potential cores as NPR increases such that $L_c$ = 10.3$h$, 11.9$h$, and 14.1$h$ for NPR = 2.5, 3, and 3.67, respectively. Here, the jet potential core lengths ($L_c$) are estimated by $U(x = L_c)$ = 0.95$U_j$ along the jet centerlines. For the two high NPR cases, the experiments show that the potential cores have rounded tails~\cite{karnam2020}, but they remain sharp in LES across all three cases.

\begin{figure}[t]
\centering
\begin{tabular}{c}
  \includegraphics[width=0.85\textwidth]{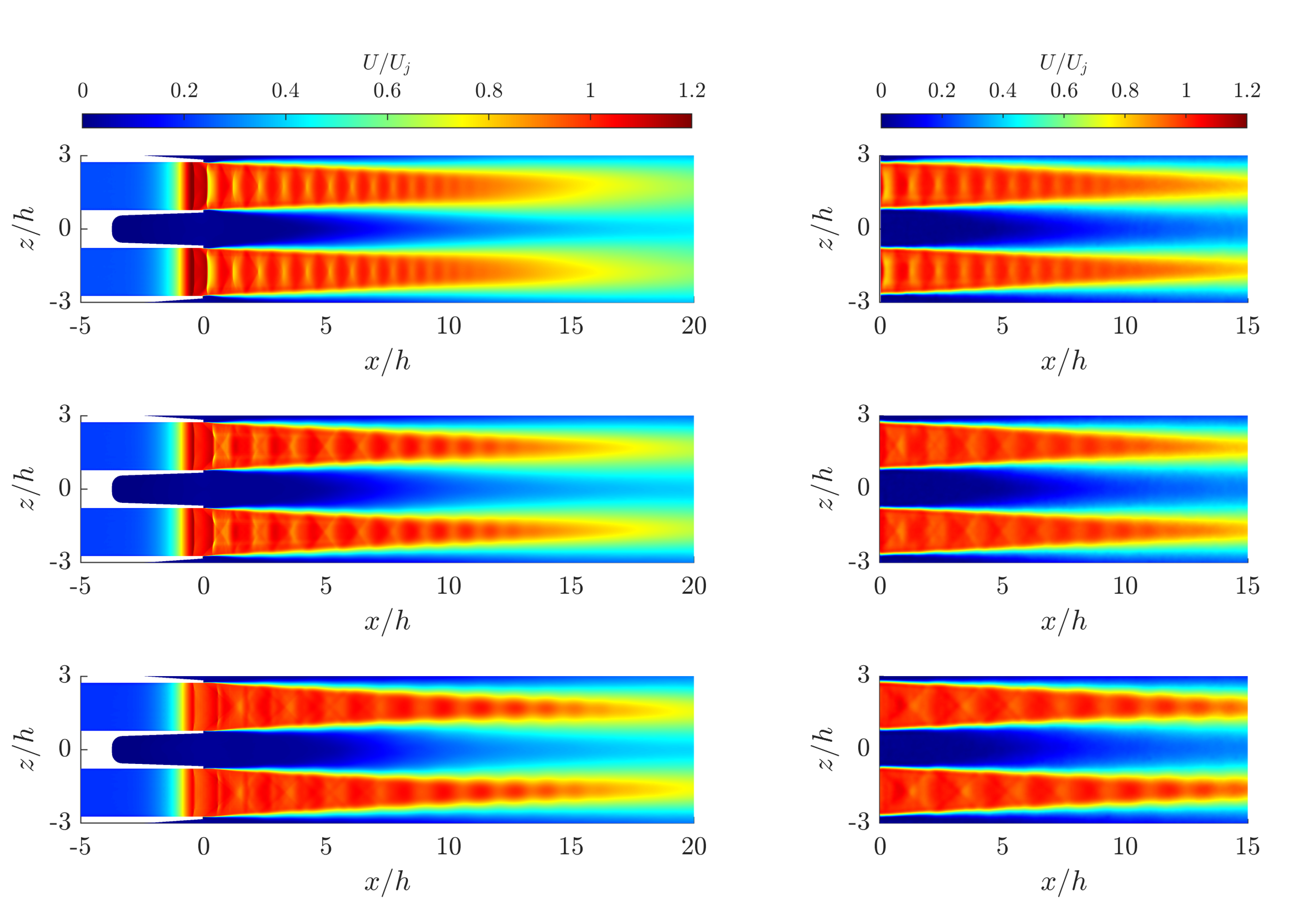} \\
\end{tabular}
\caption{Comparison of the mean streamwise velocity contours normalized by the fully expanded jet velocity between the (left) LES and (right) experiments in the major axis plane: (top) NPR = 2.5, (middle) NPR = 3, and (bottom) NPR = 3.67.}
\label{fig:Uavg_contours_major}
\end{figure}

The normalized mean streamwise velocity contours in the mid-plane cross-section along the major axis ($y/h$ = 0) are given in Fig.~\ref{fig:Uavg_contours_major}. Owing to the twin geometry, the flow field appears to be almost symmetric about $z/h$ = 0. Similarly to the experiments, LES captures the two jet plumes pulling towards each other as they spread in the downstream region for approximately $x/h > 6$. Regardless of NPR, the merging between the two jets seems to begin at almost the same streamwise location in both the LES and experiments. The differences in the potential core lengths and shock strengths between the LES and experiments are consistent to those in the minor axis plane.

Mean streamwise velocity contours in the center axis plane, i.e. in the mid-way cross-section between the two nozzles ($z/h$ = 0), were not available in the new experimental setup. Comparisons with the previous experimental data can be found in~\citet{jeun2021}. They are expected to remain qualitatively similar even in the new PIV setup. Contours in this direction visualize the merging of the two jets more directly.

\begin{figure}[h]
\centering
\begin{tabular}{cc}
  \includegraphics[width=0.6\textwidth]{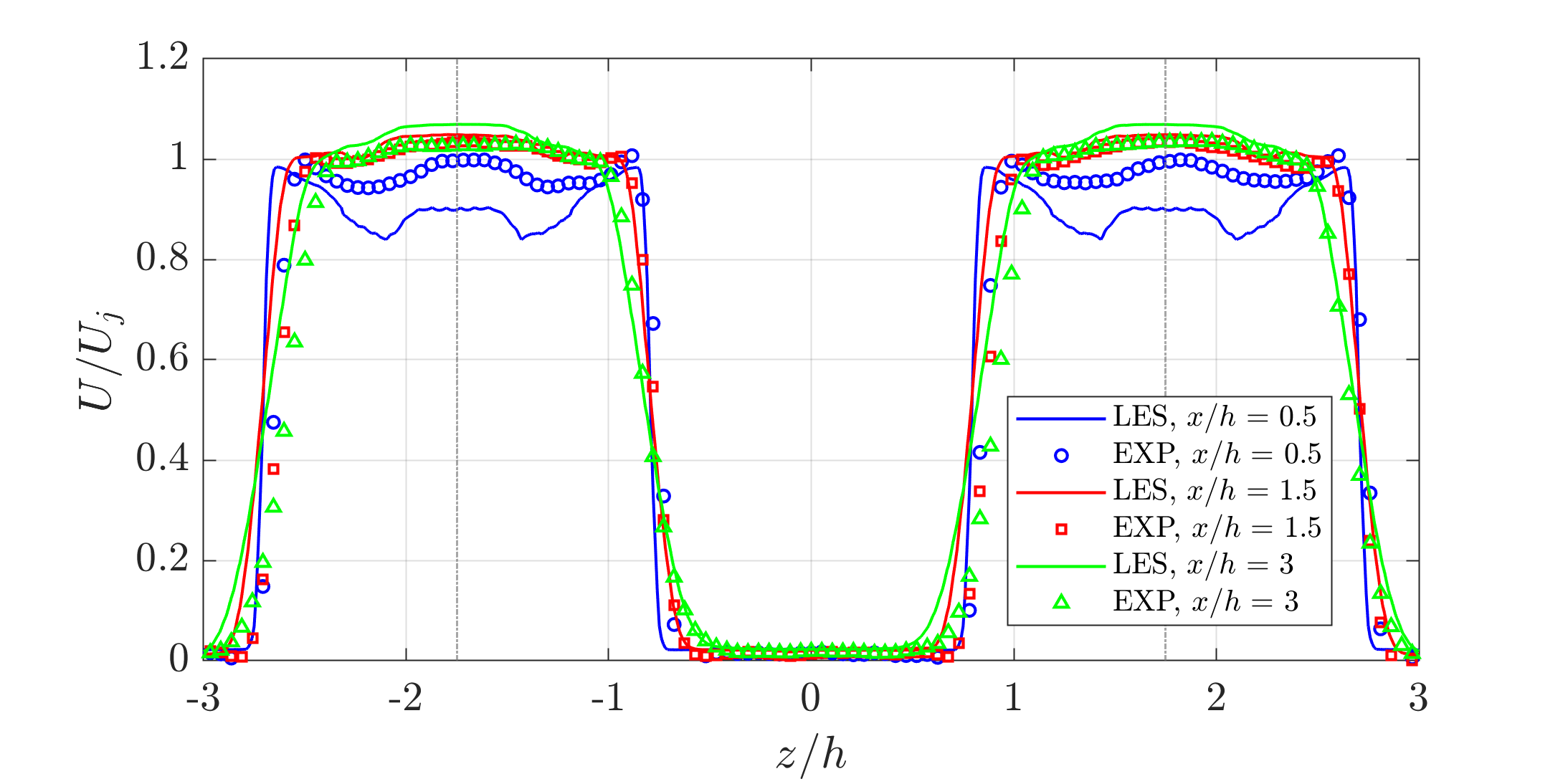} & \includegraphics[width=0.31286\textwidth]{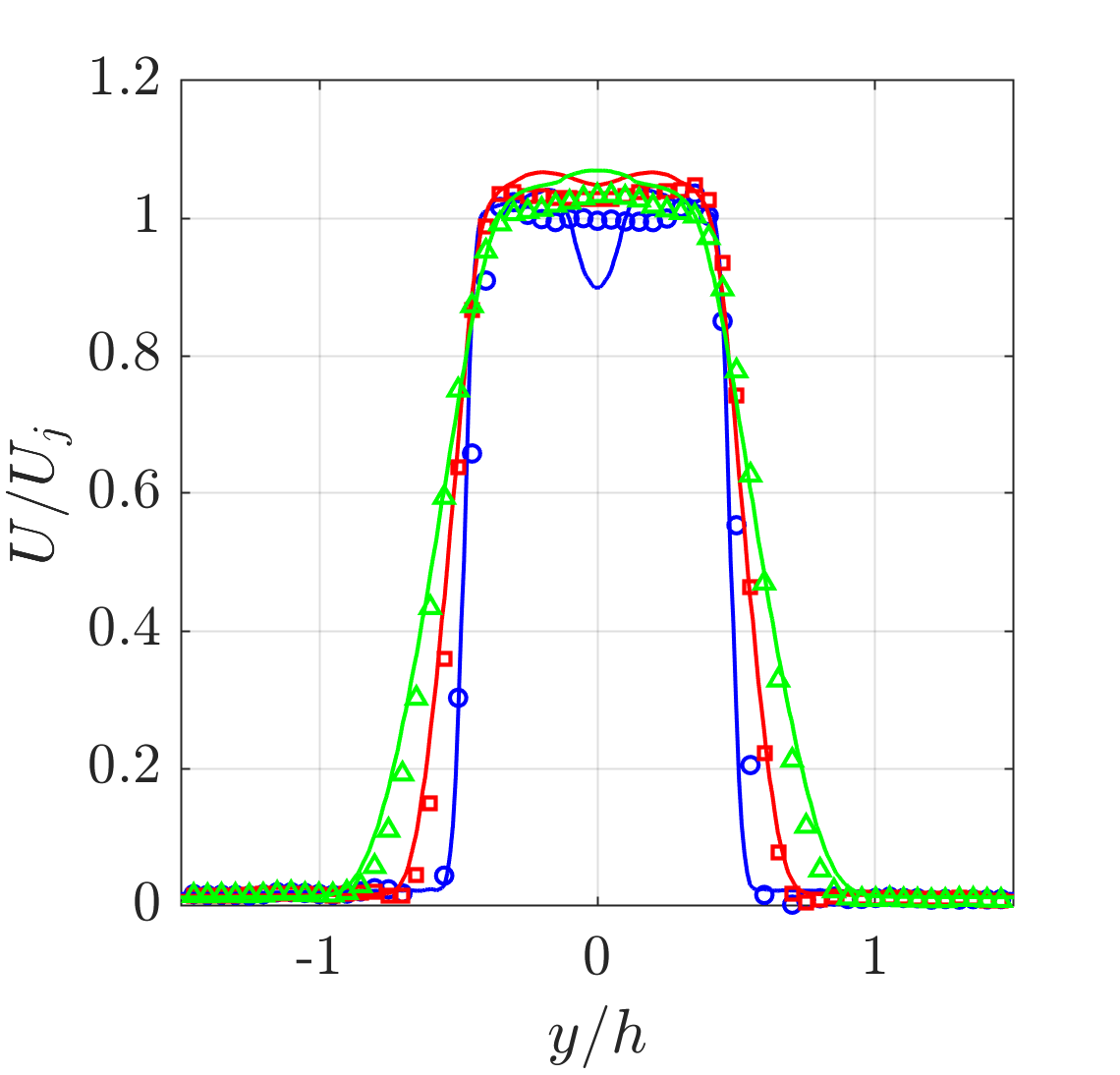} \\ 
  a) Major axis & b) Minor axis \\
\end{tabular}
\caption{Comparison between the LES and experimental mean streamwise velocities at various streamwise locations for NPR = 3. Dash-dot lines in a) represent the nozzle centerlines.}
\label{fig:Uavg_profiles_transverse}
\end{figure}

Quantitative comparisons of the mean streamwise velocity profiles are discussed next. First, Fig.~\ref{fig:Uavg_profiles_transverse} compares the experimental measurements of the streamwise velocity with the LES predictions in the major and minor axes at various streamwise locations near the nozzle exits. The experiments are denoted by symbols, and the LES results are represented by solid lines. For simplicity, profiles are provided for NPR = 3 only, but the two datasets show good agreement at all NPRs. The errors in the very vicinity of the nozzle exits ($x/h$ = 0.5) are relatively high compared to the other two locations (the shock system originating at the nozzle throat, and its subsequent reflections are presumably weaker in the experiment). In the major axis the LES and experiments show slightly different initial shear layer spreading rates. Apart from this, the overall agreement between the LES and experiments are fairly good in this near-nozzle region.

\begin{figure}[t]
\centering
\begin{tabular}{ccc}
  \includegraphics[width=0.3\textwidth]{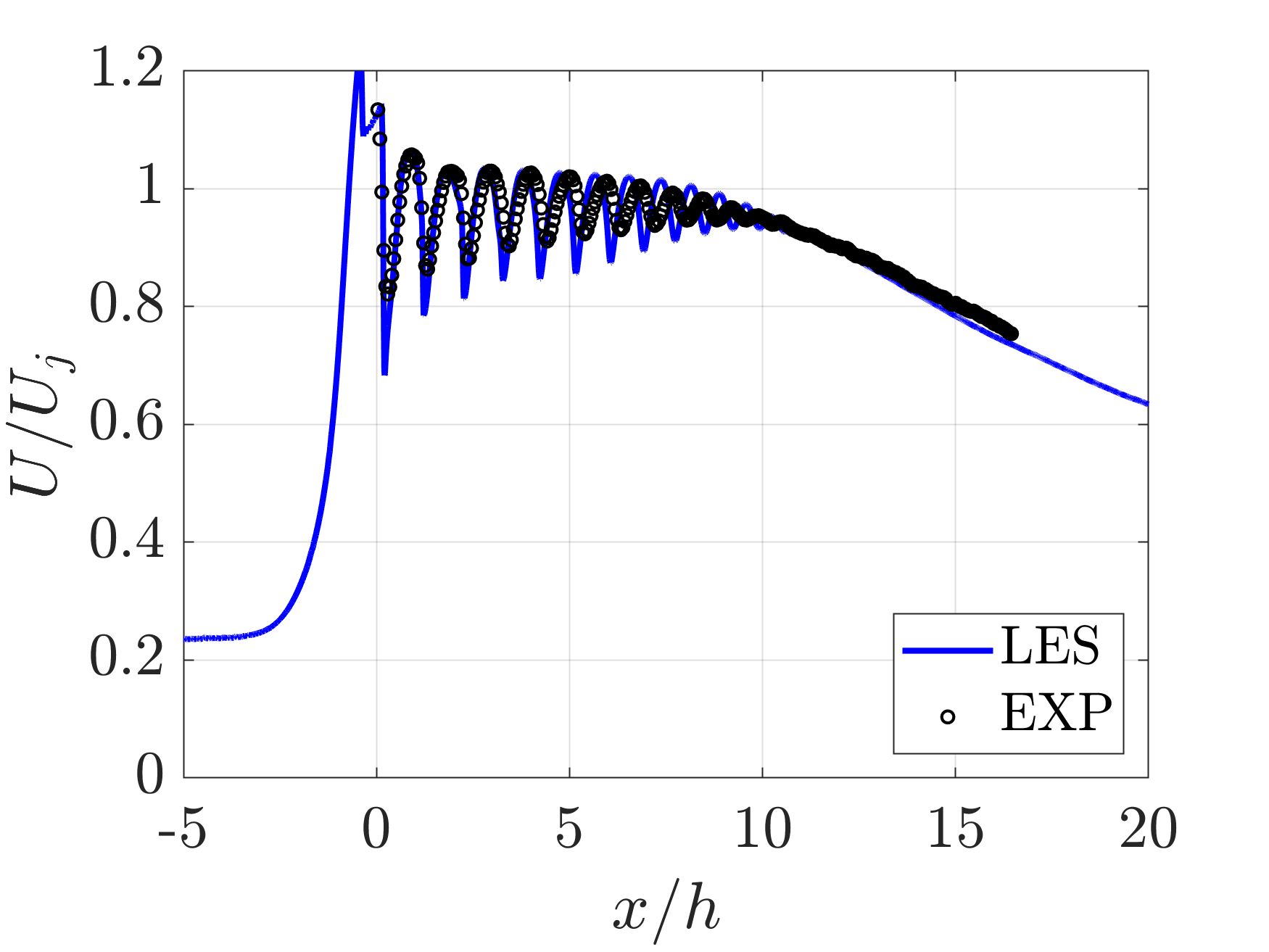} &   \includegraphics[width=0.3\textwidth]{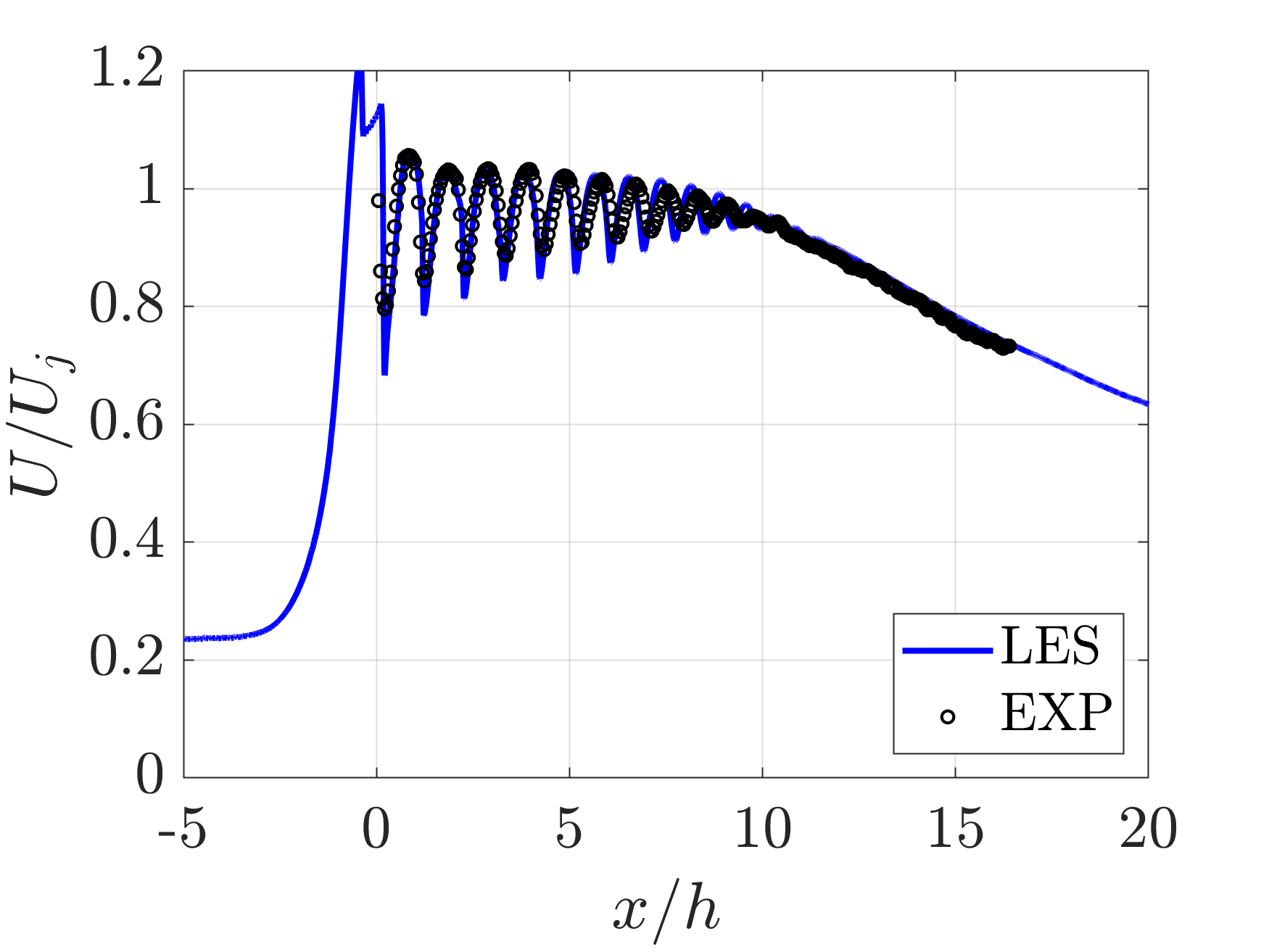} & \includegraphics[width=0.3\textwidth]{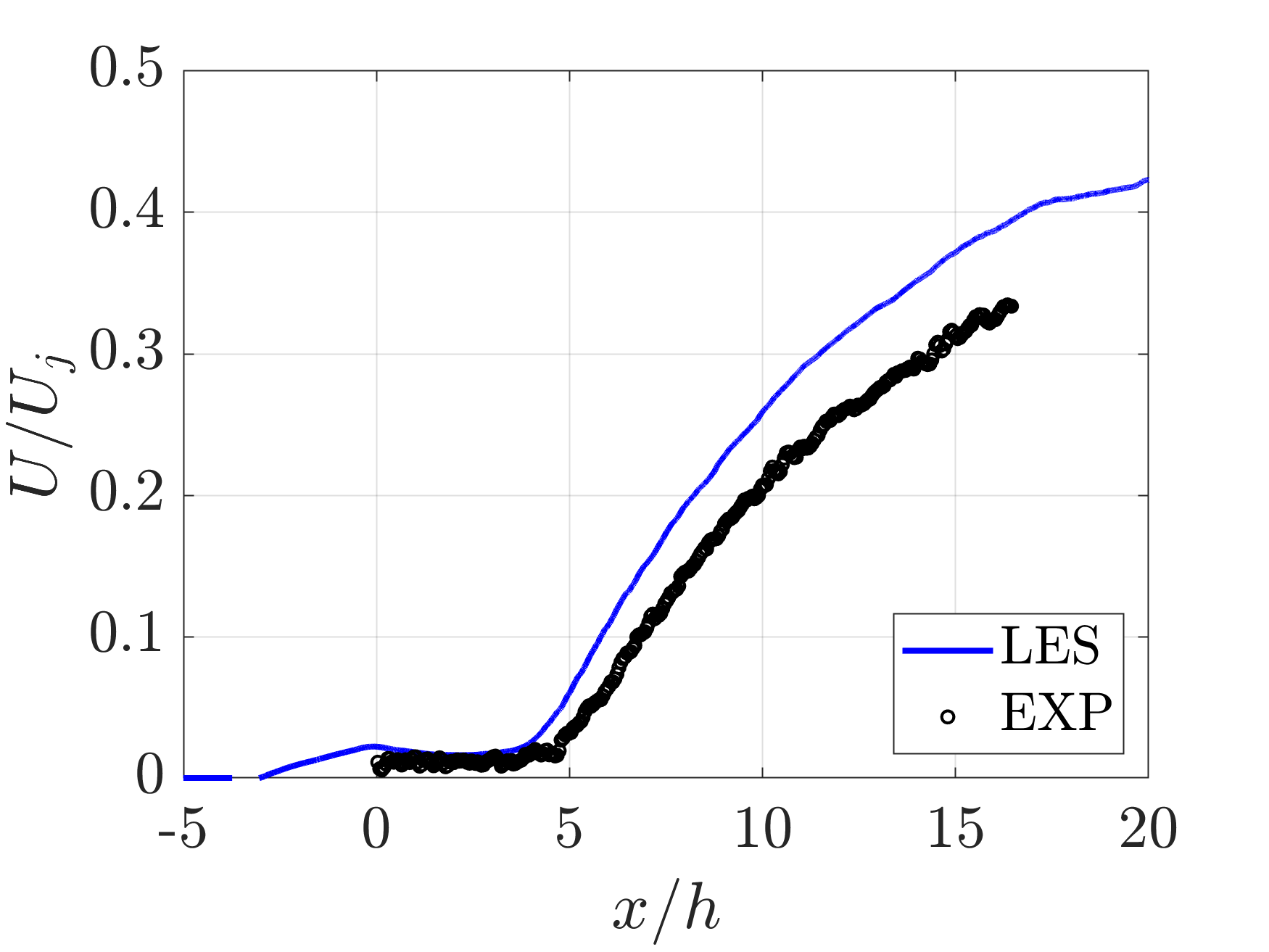} \\
  \includegraphics[width=0.3\textwidth]{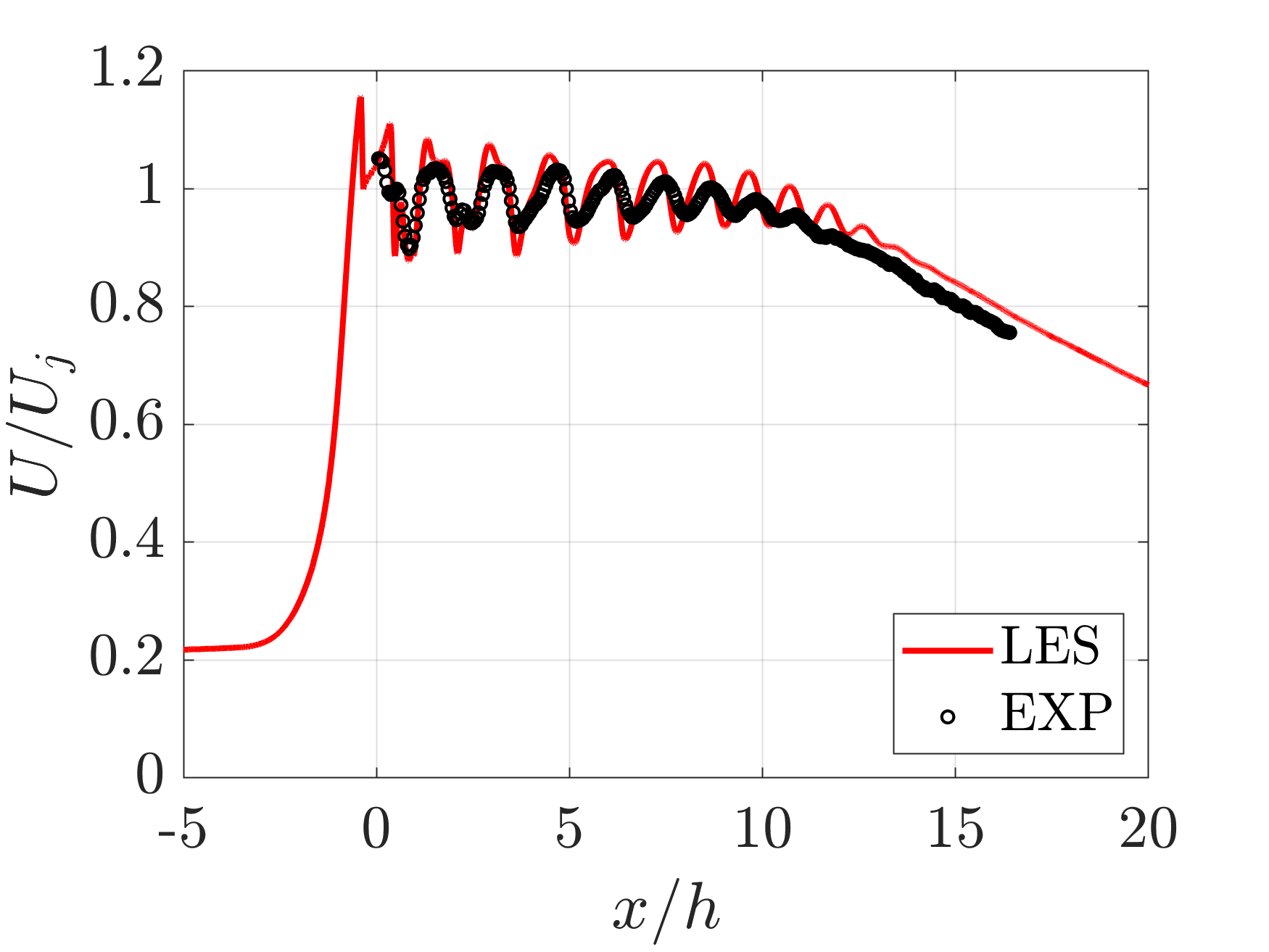} &   \includegraphics[width=0.3\textwidth]{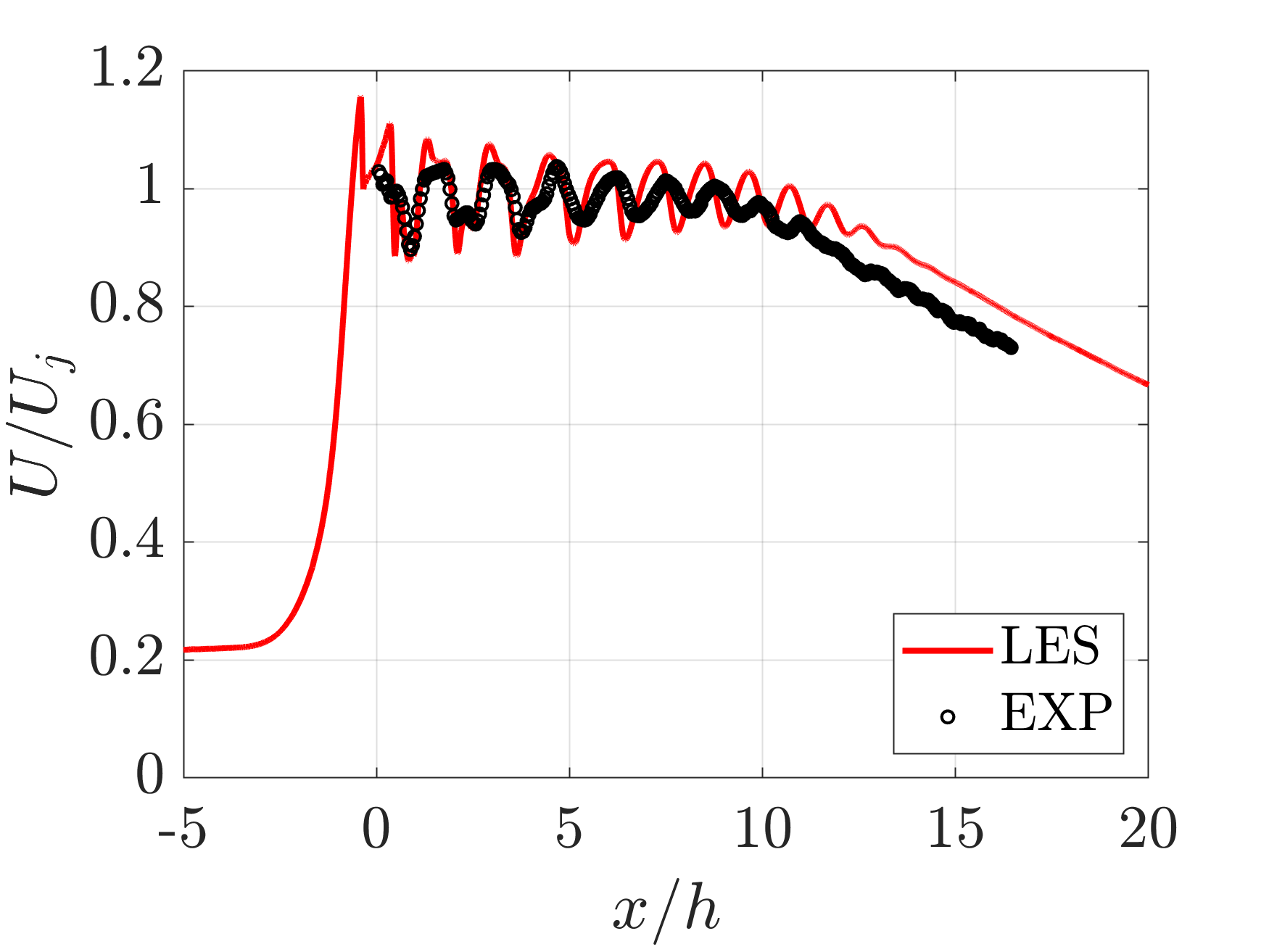} & \includegraphics[width=0.3\textwidth]{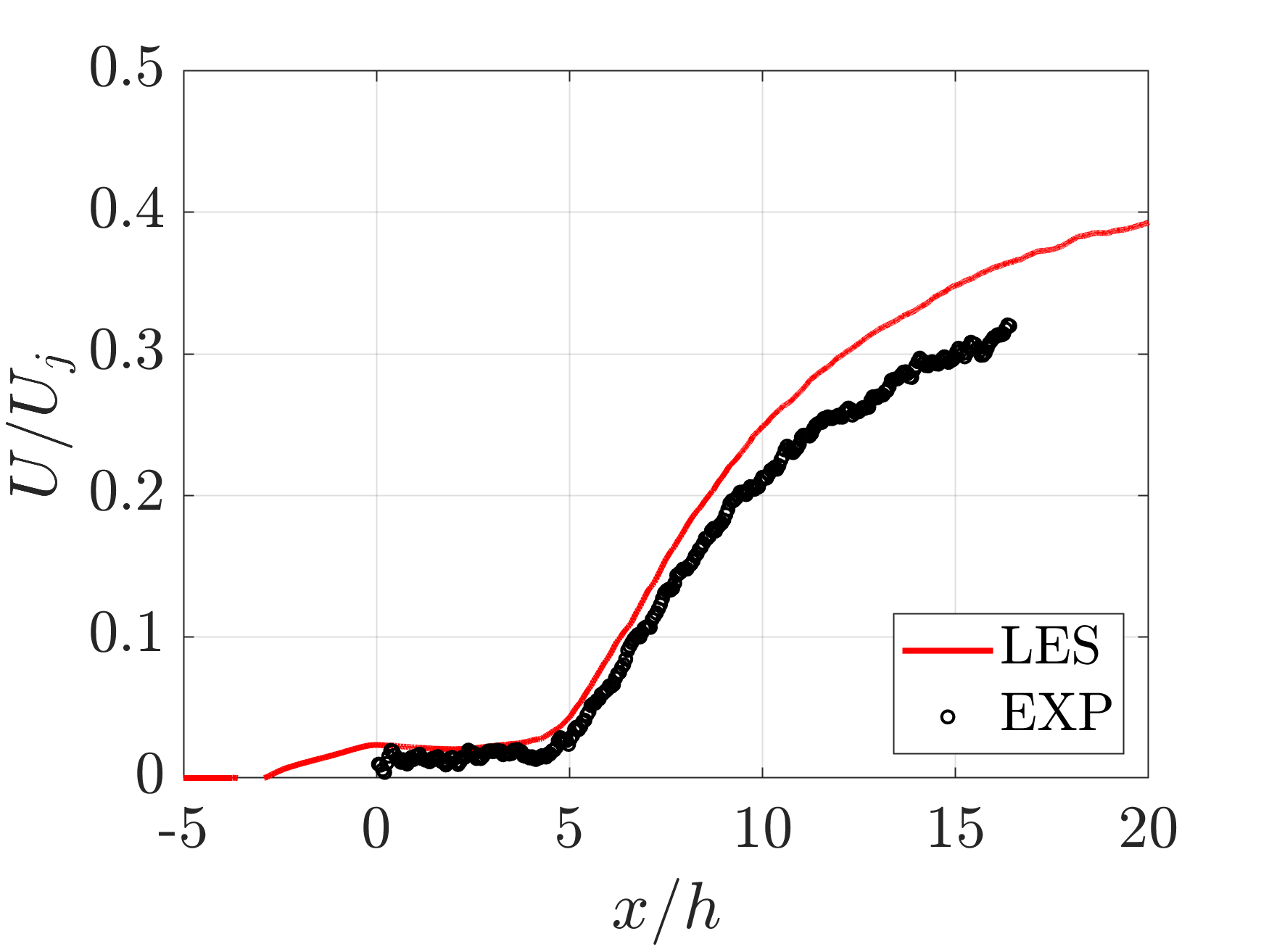} \\
  \includegraphics[width=0.3\textwidth]{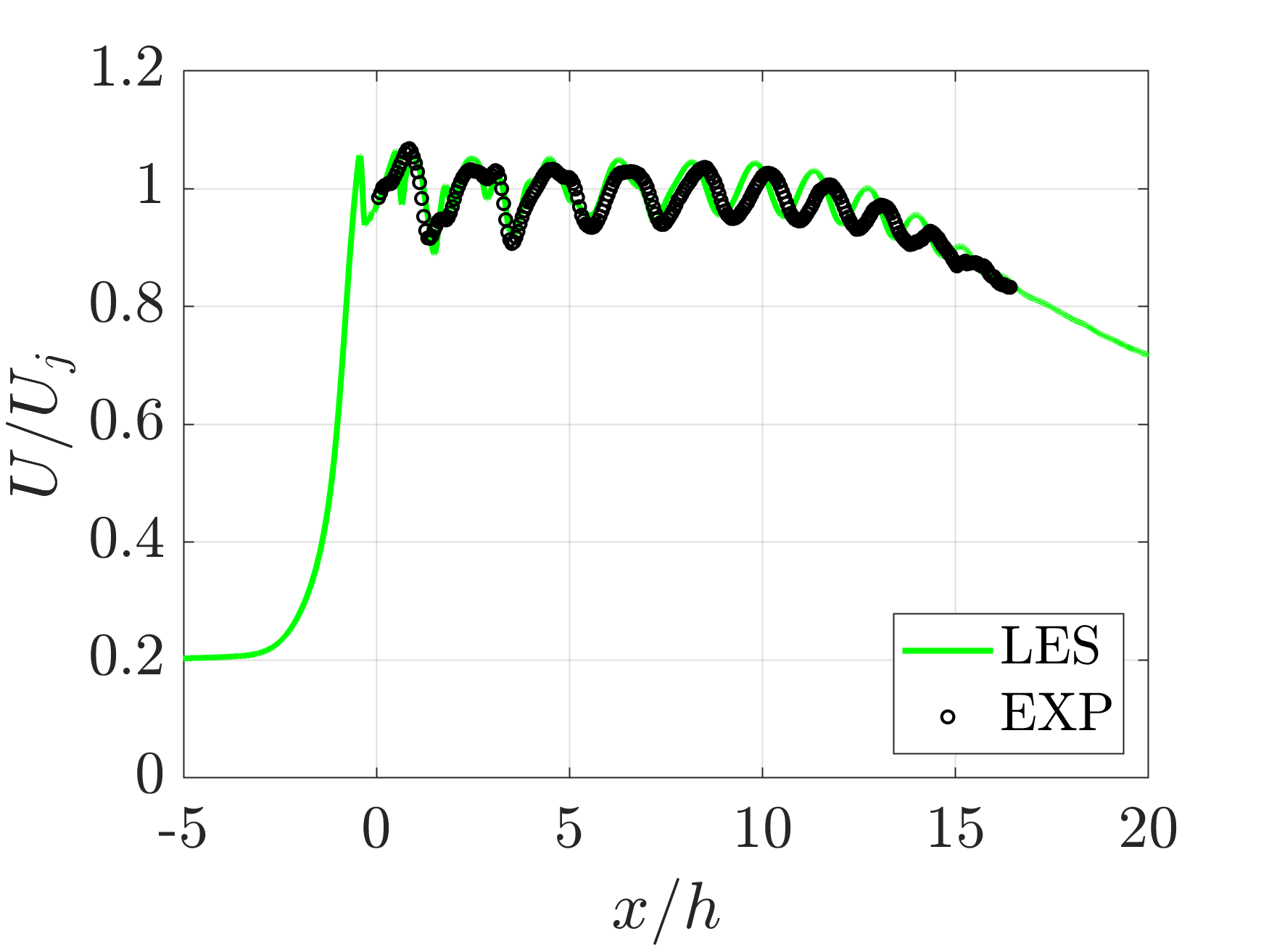} &   \includegraphics[width=0.3\textwidth]{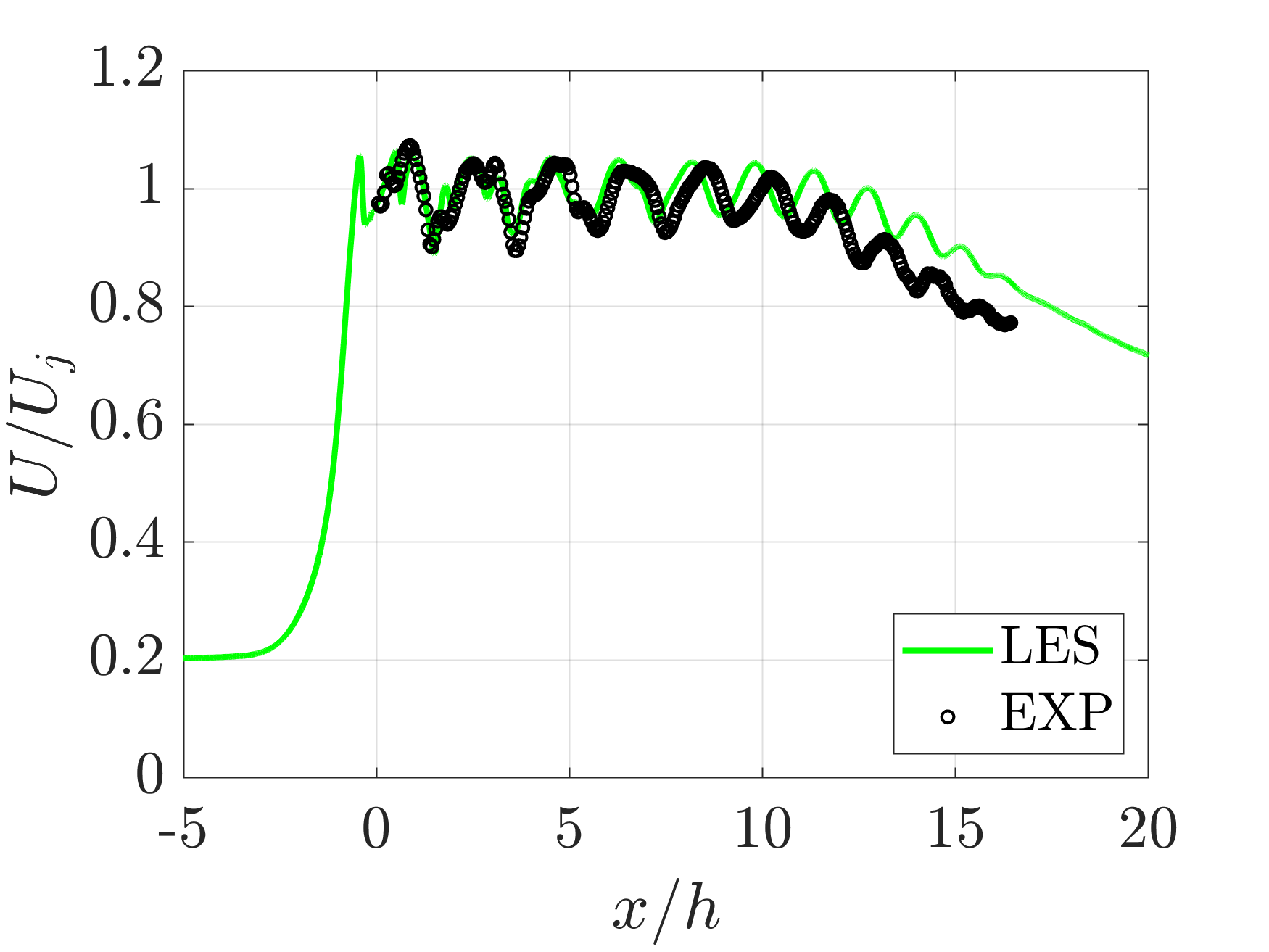} &  \includegraphics[width=0.3\textwidth]{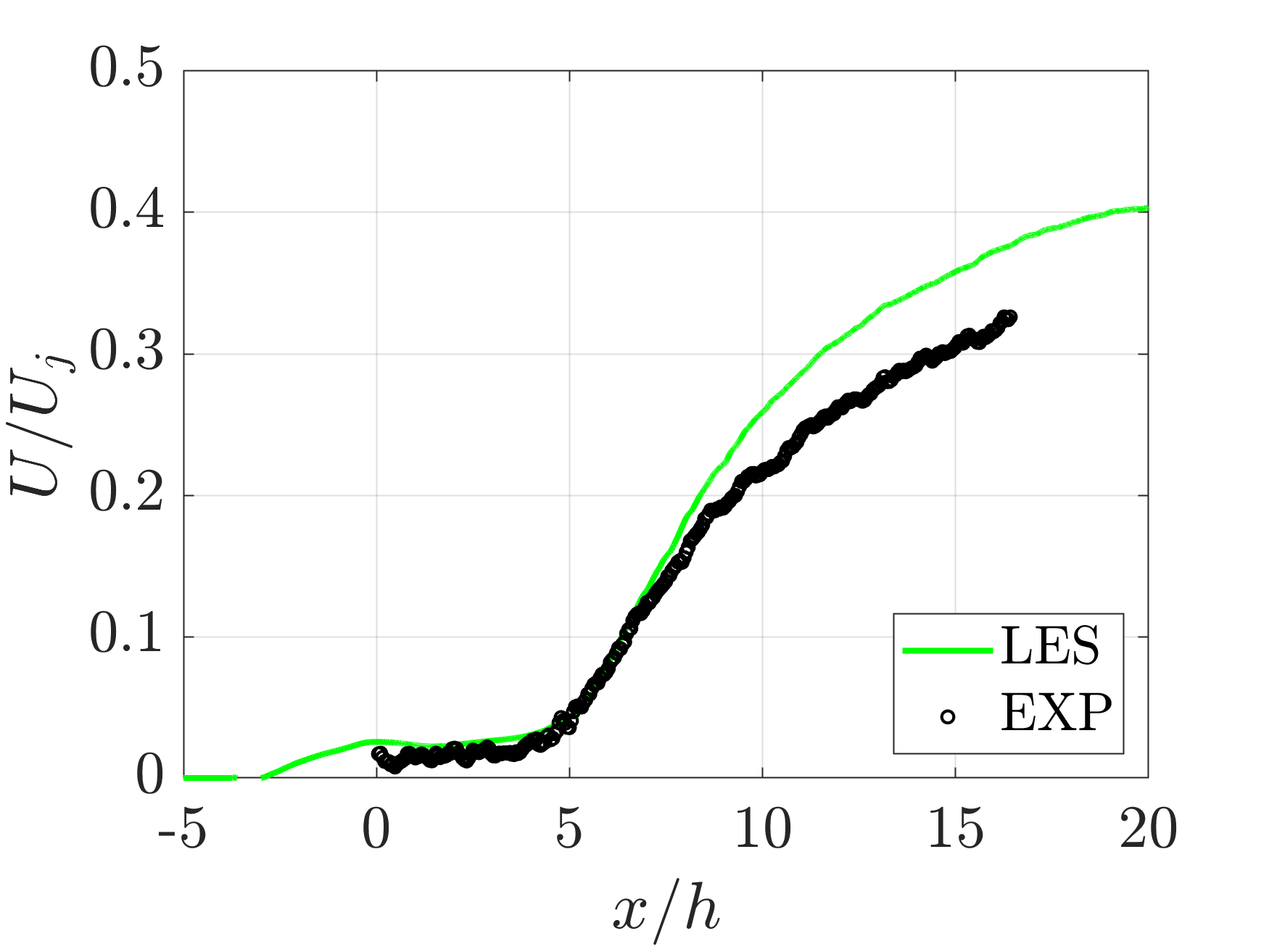} \\
  a) Major axis & b) Minor axis & c) Center axis \\
\end{tabular}
\caption{Comparison between the LES and experimental mean streamwise velocities along the centerlines of the a) major, b) minor, and c) center axes: (top) NPR = 2.5, (middle) NPR = 3, and (bottom) NPR = 3.67.}
\label{fig:Uavg_profiles}
\end{figure}

The mean streamwise velocity profiles are shown in Fig.~\ref{fig:Uavg_profiles} along the jet centerline in each nozzle orientation. In the major axis velocities are averaged between the two jets, but the minor axis velocity profile corresponds to the measurements of Jet 1 only. While profiles of the two jets are indistinguishable in LES, they show some deviations from each other in the experiment, particularly around the end of the potential core at NPR = 3. Overall, the LES and experiments show excellent agreement to each other for all test cases up to $x/h$ = 5--7, although the shock-cell strengths are slightly stronger in LES. By recalling the comparisons of the LES predictions with the previous experimental data~\cite{jeun2021}, the discrepancies in strength and phase of the shock-cell system are substantially reduced as a result of the updated experimental setup. Further downstream the shock-cell system predicted by LES deviates slightly from the experiments. Although the profiles are extracted along the same geometric centerline of the nozzles, the fact that the jets deflect inwards in the experimental results leads to some deviations from the LES predictions in the downstream locations. The transitional nozzle-exit boundary layer state, as indicated in the RMS velocity profiles along the lipline that will be discussed in the later section (see Fig.~\ref{fig:urms_profile}), could also cause some deviation. For turbulent boundary layers at the nozzle exit, the shocks may jiggle/oscillate more and lead to smoother variations in the centerline shock-structure. The potential core decays earlier in the experiments than in the LES, likely because of enhanced turbulent mixing in the experiment. This is more evident in the minor axis, but these discrepancies may be mitigated if average profiles of both jets are considered in this orientations as well. Although it is not shown in this figure, the experiments capture noticeable variations of the mean velocity profiles between the two jets along the major axis centerlines in the maximum screech case (NPR = 3). It may be the result of one of the jets carrying a dominant screech mode leading to larger fluctuations at this condition~\cite{prasad2022,jeun2021b}. Along the center axis, the twin configuration yields merging of inner shear layers of the two jets as they spread. The onset of the merging is marked by an increase in centerline velocity at about $5h$ downstream of the nozzle exits in both the experiments and LES except for NPR = 2.5, where the LES results spread slightly faster. The mean streamwise velocity profile from the LES has a larger spatial growth rate in the mid-stream region but ends up around the same value as the experimental measurements farther downstream. Different jet spreading rates in the experiments and LES can thus lead to discrepancy along the (geometric) centerline values.

\subsubsection{Twin-jet spreading/merging}
\begin{figure}[t]
\centering
\begin{tabular}{c}
  \includegraphics[width=0.5\textwidth]{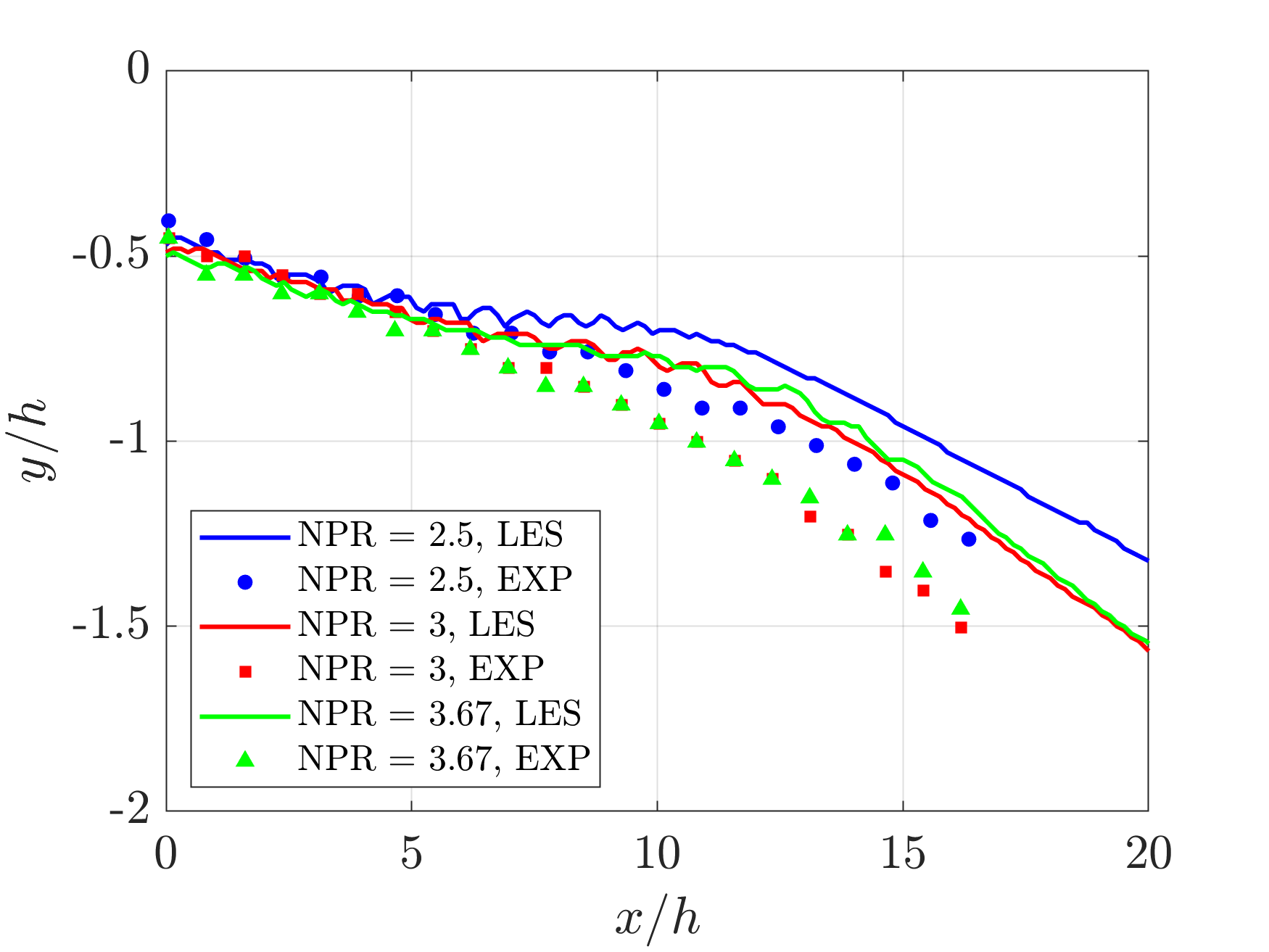} \\
\end{tabular}
\caption{Streamwise variation of the jet half-width based on the bottom shear layer in the minor axis view.}
\label{fig:half_width}
\end{figure}

Figure~\ref{fig:half_width} shows the streamwise development of the jet half-width (i.e. the radial location at which the velocity equals to half of the centerline value) as a proxy to estimate the jet spreading rates. We consider the bottom shear layer in the minor axis view to measure the jet half-width, following the experimental setup~\cite{karnam2020}. In this figure symbols represent the experiments, and solid lines correspond to the simulations for various NPRs. The initial spreading of the jets seems to slightly increase in response to the increment in NPR. From $x/h \approx 12.5$ the maximum screech case (NPR = 3) exhibits higher spreading rates in concert with strong twin-jet coupling and enhanced jet flapping motions. This trend is maintained further downstream in the experiments. In comparison to LES, the experiments show higher spreading rate and faster decay of the jet potential core across the three operating conditions. The discrepancies may be in part due to inaccurate nozzle exit boundary layer state. Note, however, that the overall trend towards slower spreading at NPR 2.5 and increased spreading for the maximum screech case (NPR = 3) is consistent between LES and  experiment.

\begin{figure}[h]
\centering
\begin{tabular}{ccc}
  \includegraphics[width=0.3\textwidth]{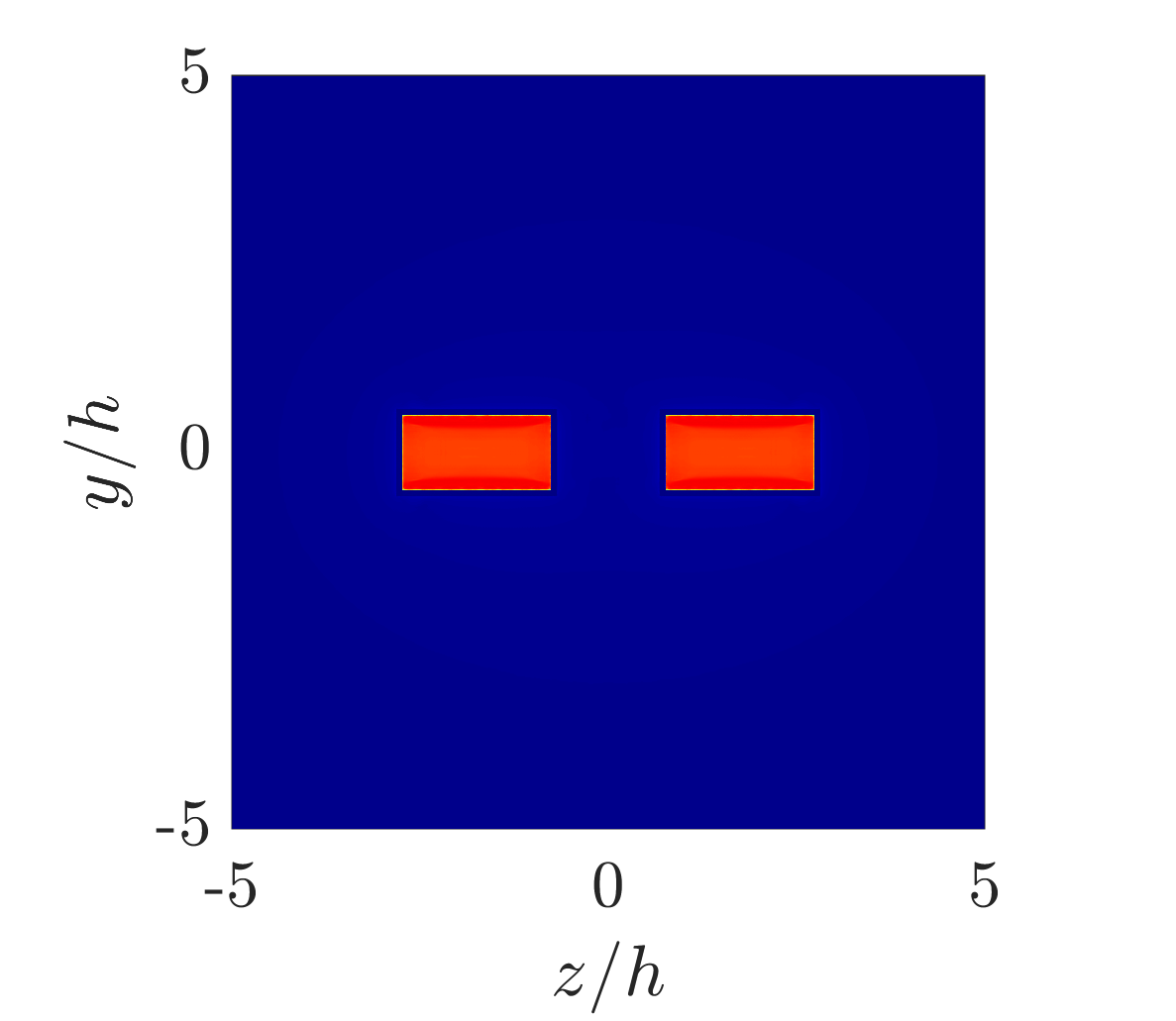} & \includegraphics[width=0.3\textwidth]{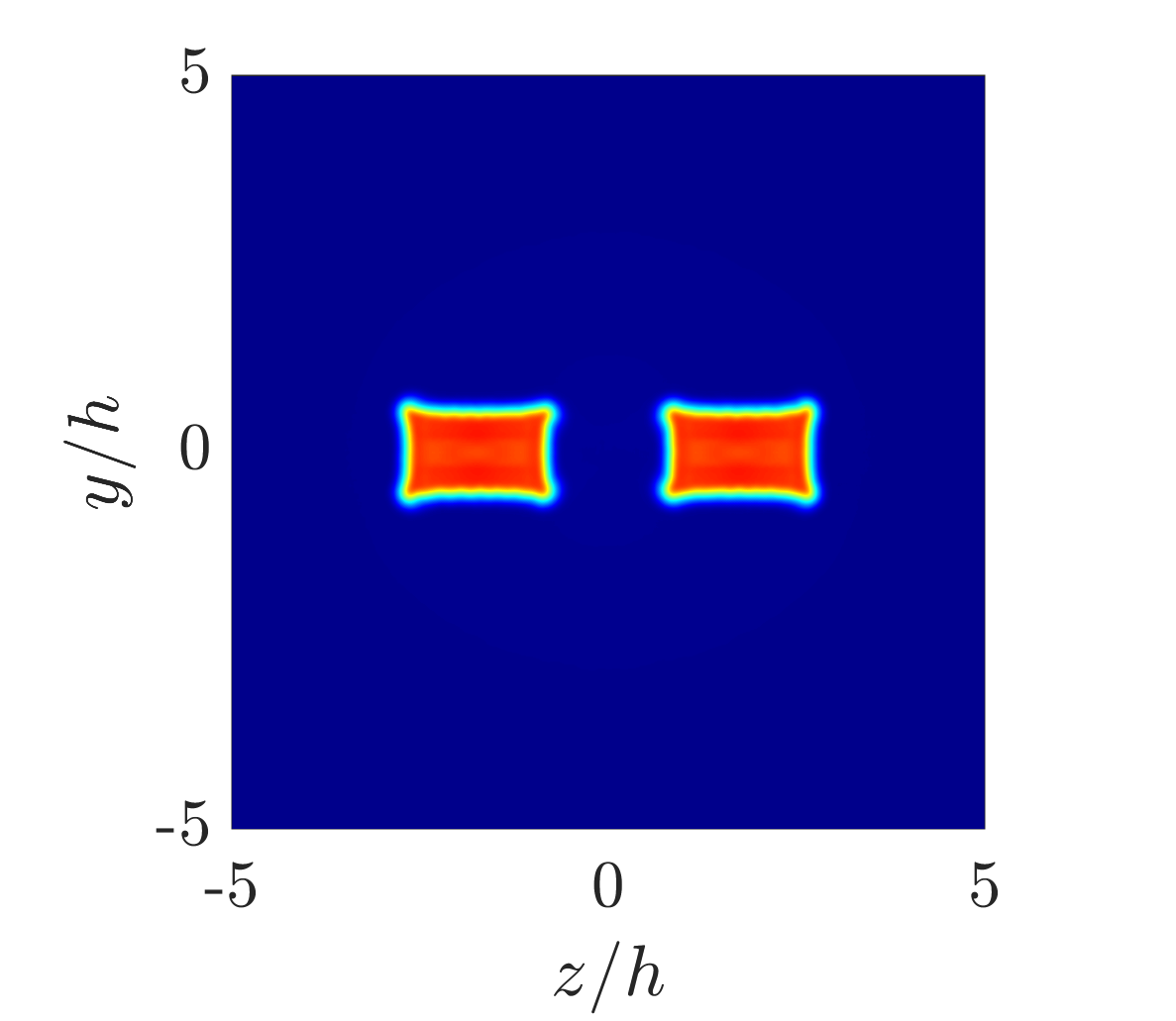} & \includegraphics[width=0.3\textwidth]{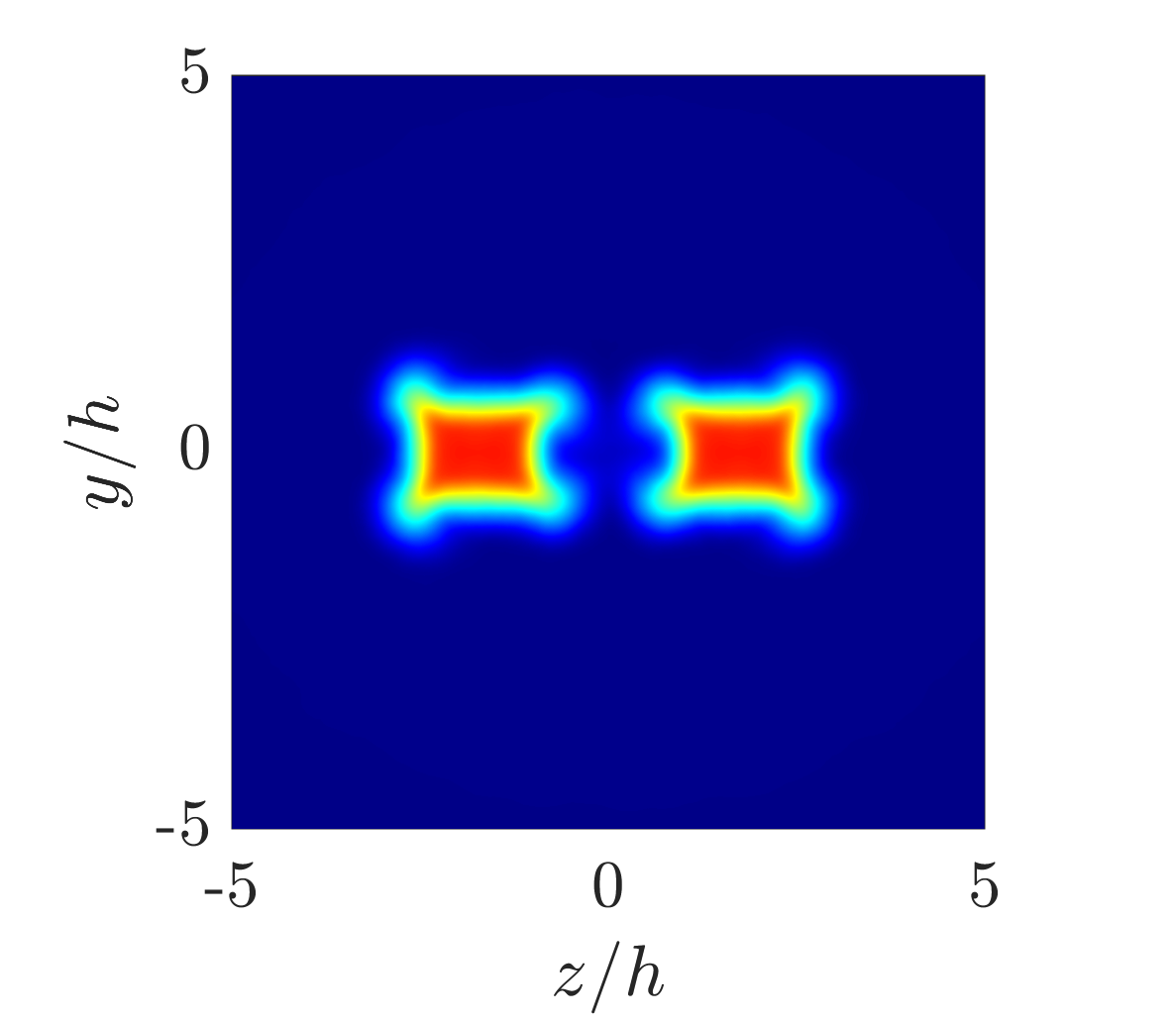}\\
  a) $x/h$ = 0 & b) $x/h$ = 2 & c) $x/h$ = 6 \vspace{-1mm} \\
  \includegraphics[width=0.3\textwidth]{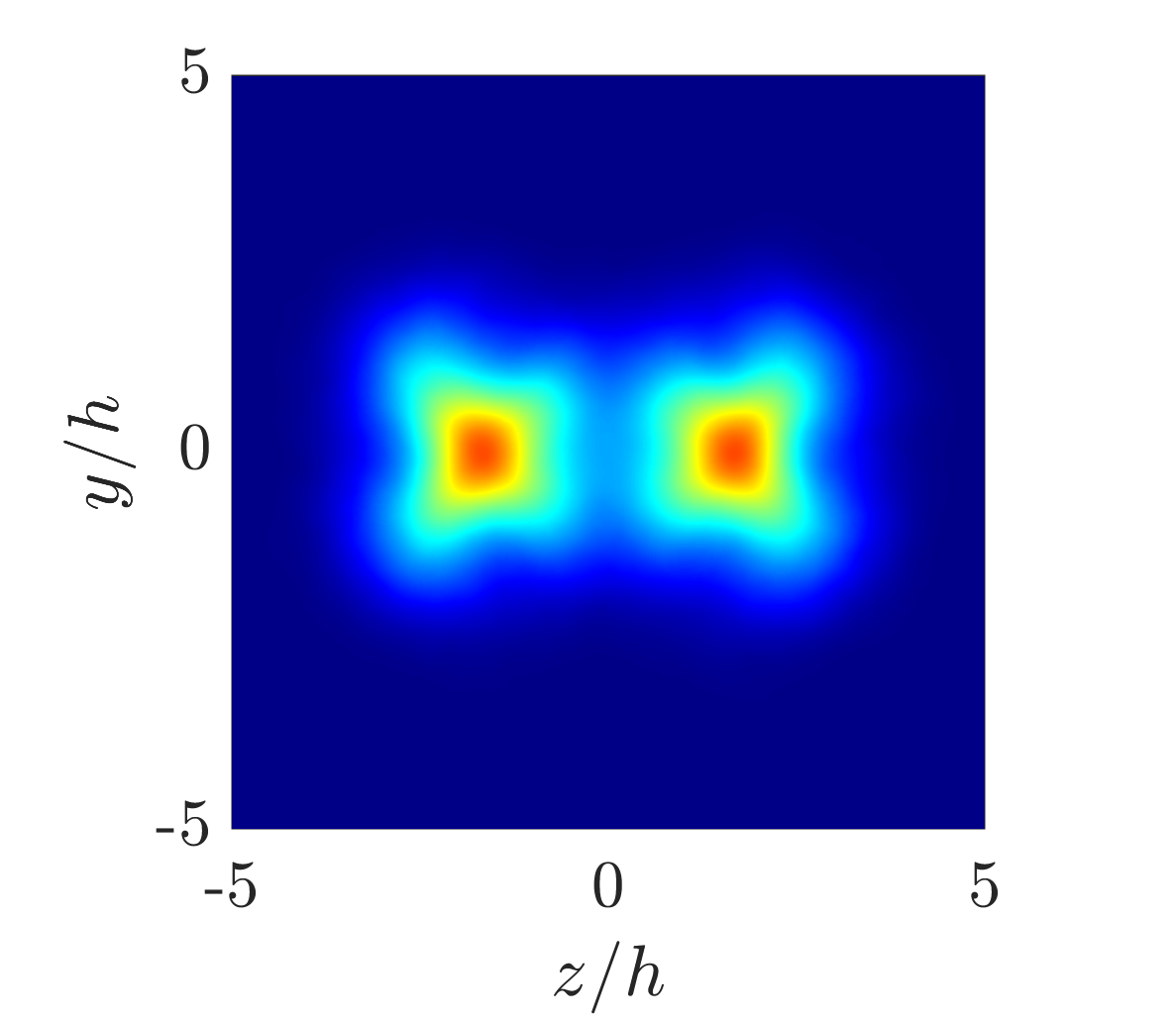} & \includegraphics[width=0.3\textwidth]{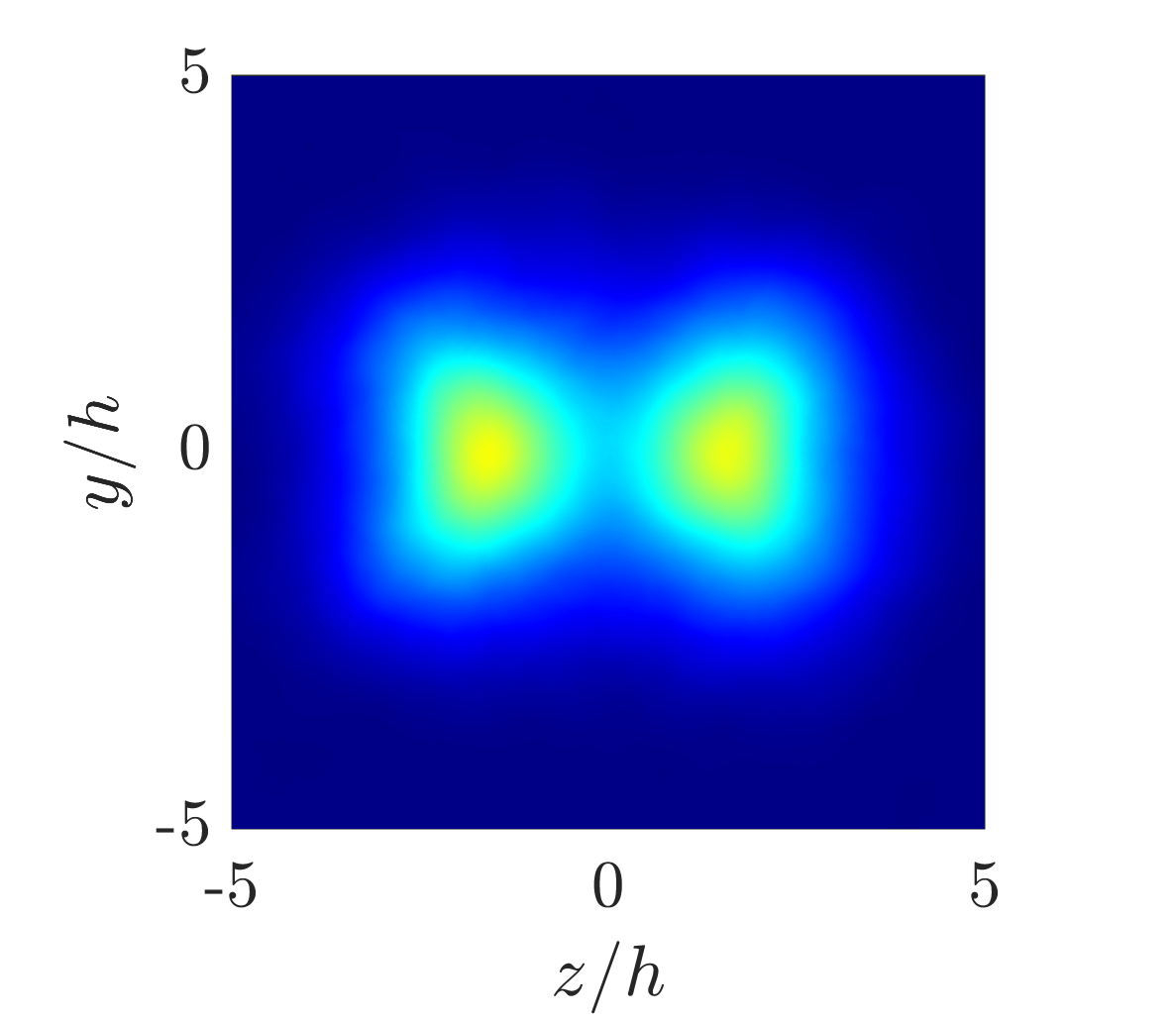} & \includegraphics[width=0.3\textwidth]{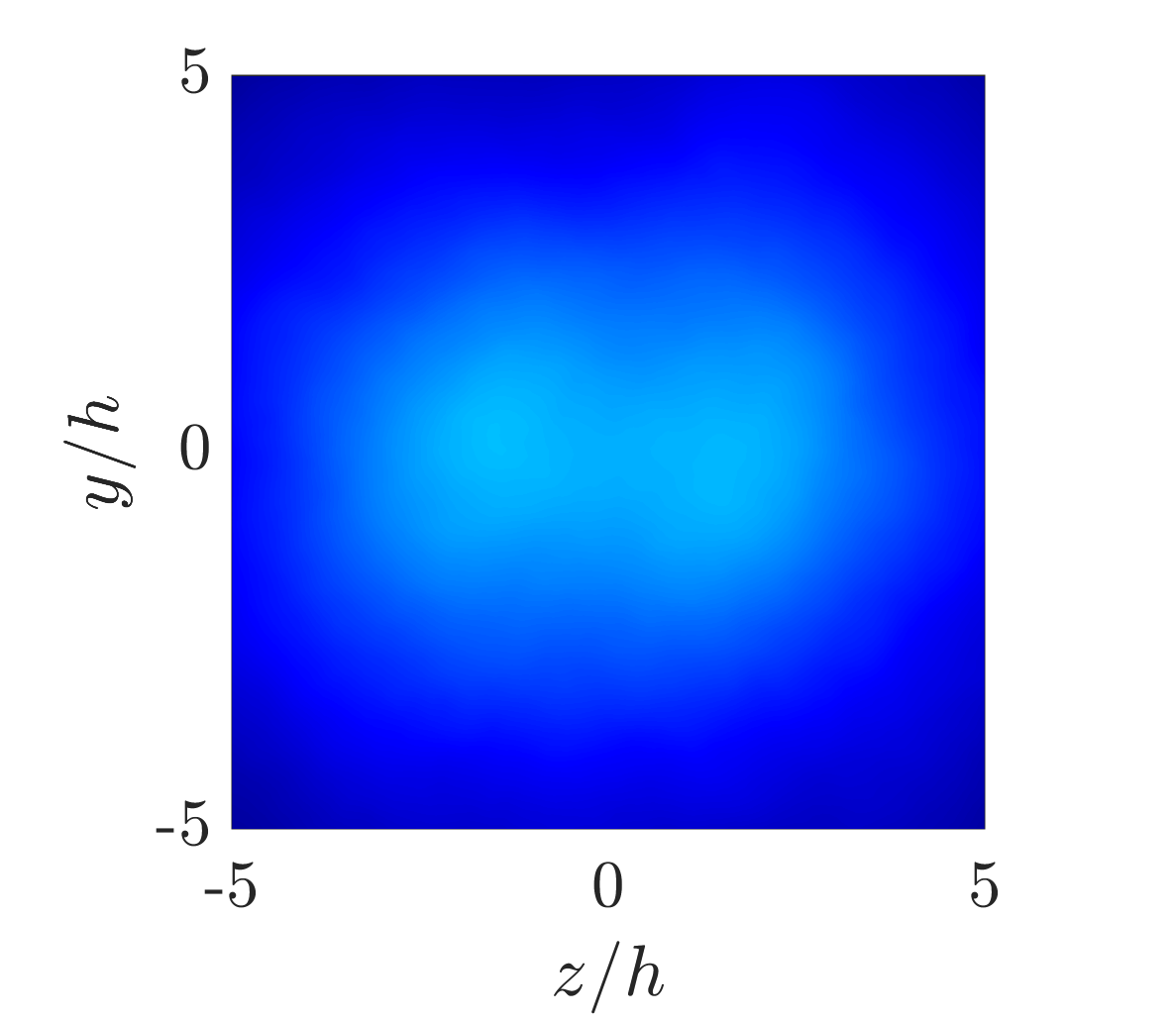}\\
  d) $x/h$ = 14 & e) $x/h$ = 20 & f) $x/h$ = 45 \\
\end{tabular}
\caption{Mean streamwise velocity contours normalized by the fully expanded jet velocity for NPR = 3.67, viewed in the streamwise cross-section at various locations. The colorscale ranges from 0 to 1.2.}
\label{fig:Uavg_contours_streamwise}
\end{figure}

The merging of the twin jets in the nozzle downstream region is more apparent when examining the velocity contours in $zy$-plane at multiple streamwise locations, as shown for NPR = 3.67 in Fig.~\ref{fig:Uavg_contours_streamwise}. As they spread in the downstream and start merging into each other in the inter-nozzle region, the two jets create a distinct butterfly pattern as shown in Fig.~\ref{fig:Uavg_contours_streamwise}e. Finally at locations even further downstream (for example, at $x/h$ = 45 as shown in Fig.~\ref{fig:Uavg_contours_streamwise}f), the merging is complete with the cross-section resembling that of a single jet.

\subsubsection{Shock-cell spacing}
Figure~\ref{fig:Uavg_profiles} indicates that the shock-cell spacing becomes elongated as NPR increases in agreement with other experimental, numerical, and analytical studies. As the shear layer grows and the potential core collapses, this length decreases in the downstream. However, while many proposed relations for variations of the shock-cell spacing assume a monotonic behavior~\cite{harper1973}, complex shock/expansion wave interactions in the very vicinity of the nozzle exits predict shorter initial shock-cell length (in some cases including the current nozzle configuration), after which the spacing reduces downstream as expected.

Table~\ref{tab:spacing} shows the cell spacing for the first shock-cell ($L_{s,1}$) and the average spacing for the first six cells ($L_{s,avg}$) computed from the experiment and LES. Additionally provided are the first shock-cell lengths predicted by Tam's formula~\cite{tam1988} for rectangular jets, which was proposed based on a vortex sheet model, as following:
\begin{equation}
    L_s = 2(M_j^2 - 1)^{1/2} h_j / [1+h_j^2/b_j^2]^{1/2}.
\end{equation}
Here, the jet width ($b_j$) and height ($h_j$) at fully expanded condition are computed as:
\begin{equation}
    b_j / b = [(A_j/A_d) - 1][h / (h+b)] + 1
\label{eq:b_j}
\end{equation}
and
\begin{equation}
    h_j / h = [(A_j/A_d) - 1][b / (h+b)] + 1,
\label{eq:h_j}
\end{equation}
by considering the effective size of a fully expanded jet such that:
\begin{equation}
    \left( \frac{A_j}{A_d} \right)^2 = \frac{M_d^2}{M_j^2}\left[\left(1+\frac{\gamma - 1}{2}M_j^2\right) \Bigg\slash \left(1+\frac{\gamma - 1}{2}M_d^2 \right)\right]^{(\gamma+1)/(\gamma-1)}.
\end{equation}
The numerically and experimentally predicted first shock-cell spacings are quite different from the first shock-cell length computed by Tam's model. It should be noted, however, that his model was based on infinitely thin vortex sheets and intended for a single rectangular jet. The shock-cell spacings predicted by LES are found to be slightly shorter than those measured in the experiment with a peak variation in average cell spacing of just 3.8\% among all test cases.

\begin{table}
\centering
\caption{Summary of the first and averaged shock-cell spacing measured by the LES, experiments~\cite{karnam2020}, and Tam's model~\cite{tam1988}, scaled by $h$.}
\begin{tabular}{cccccc}
\hline\hline
  \multirow{2}{*}{NPR} & \multicolumn{2}{c}{LES} & \multicolumn{2}{c}{Experiments} & Tam's model \\ \cline{2-6}
  & $L_{s,1}$ & $L_{s,avg}$ & $L_{s,1}$ & $L_{s,avg}$ & $L_{s}$ \\
  \hline
  2.5 & 1.01 & 0.975 & 1.03 & 1.01 & 1.17 \\
  3 & 1.22 & 1.37 & 1.42 & 1.42 & 1.56 \\
  3.67 & 2.06 & 1.76 & 2.01 & 1.81 & 2.00 \\
\hline\hline
\end{tabular}
\label{tab:spacing}
\end{table}

\subsubsection{Convection velocity}
\label{subsubsec:convection_vel}
\begin{figure}[t]
\centering
\begin{tabular}{cc}
  \includegraphics[width=0.45\textwidth]{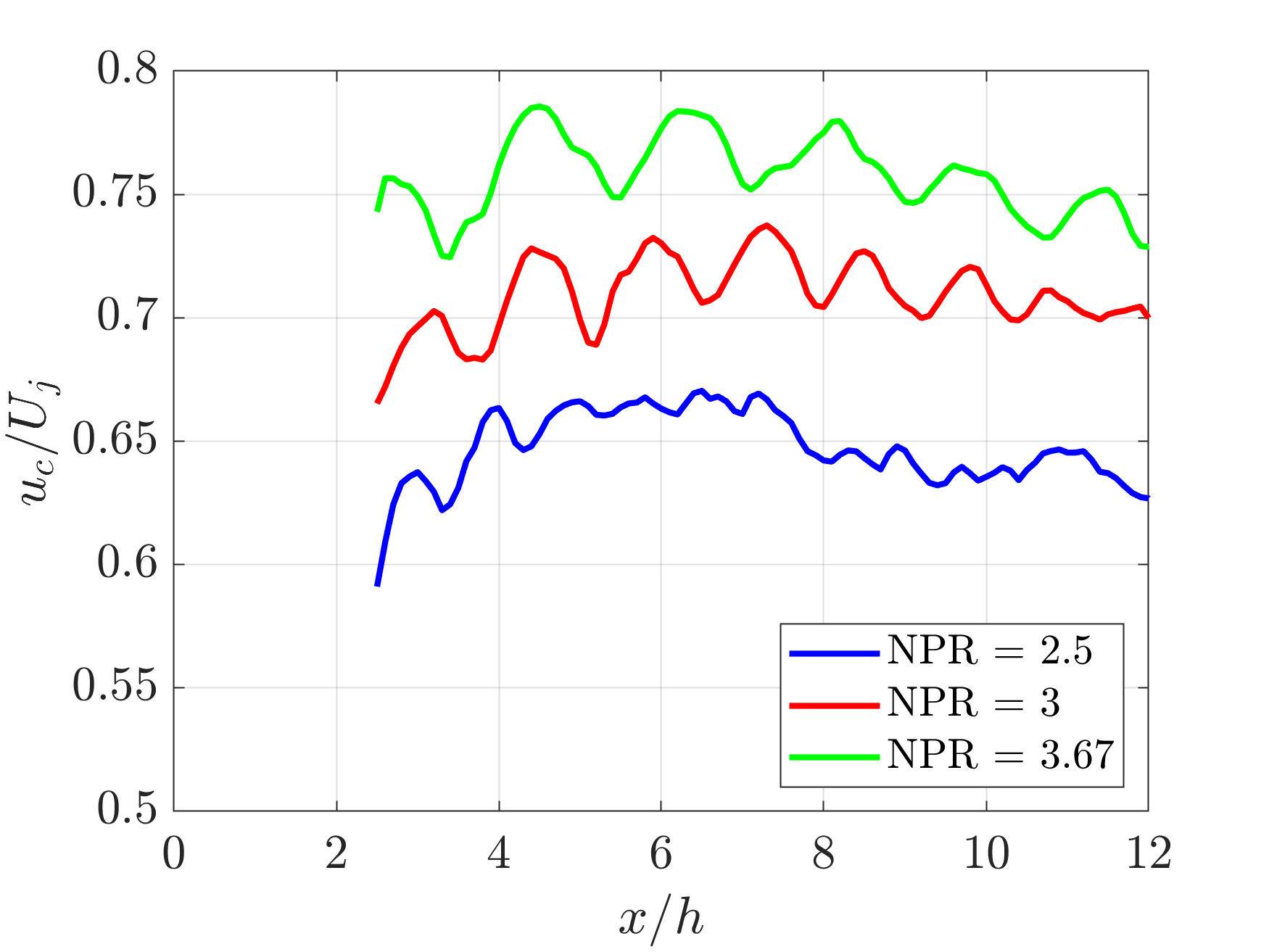} & \includegraphics[width=0.45\textwidth]{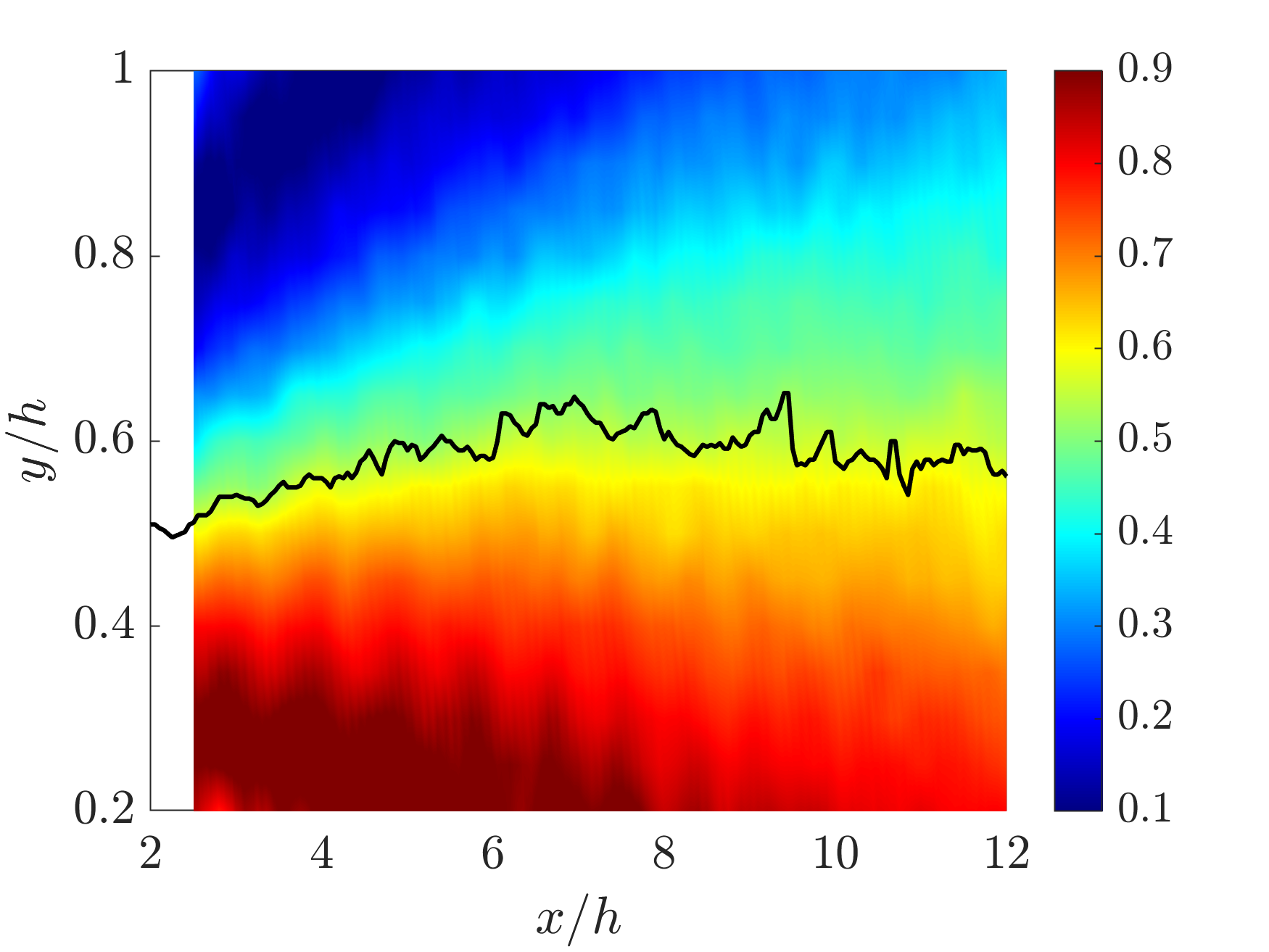} \\
  a) & b) NPR = 2.5 \\
  \includegraphics[width=0.45\textwidth]{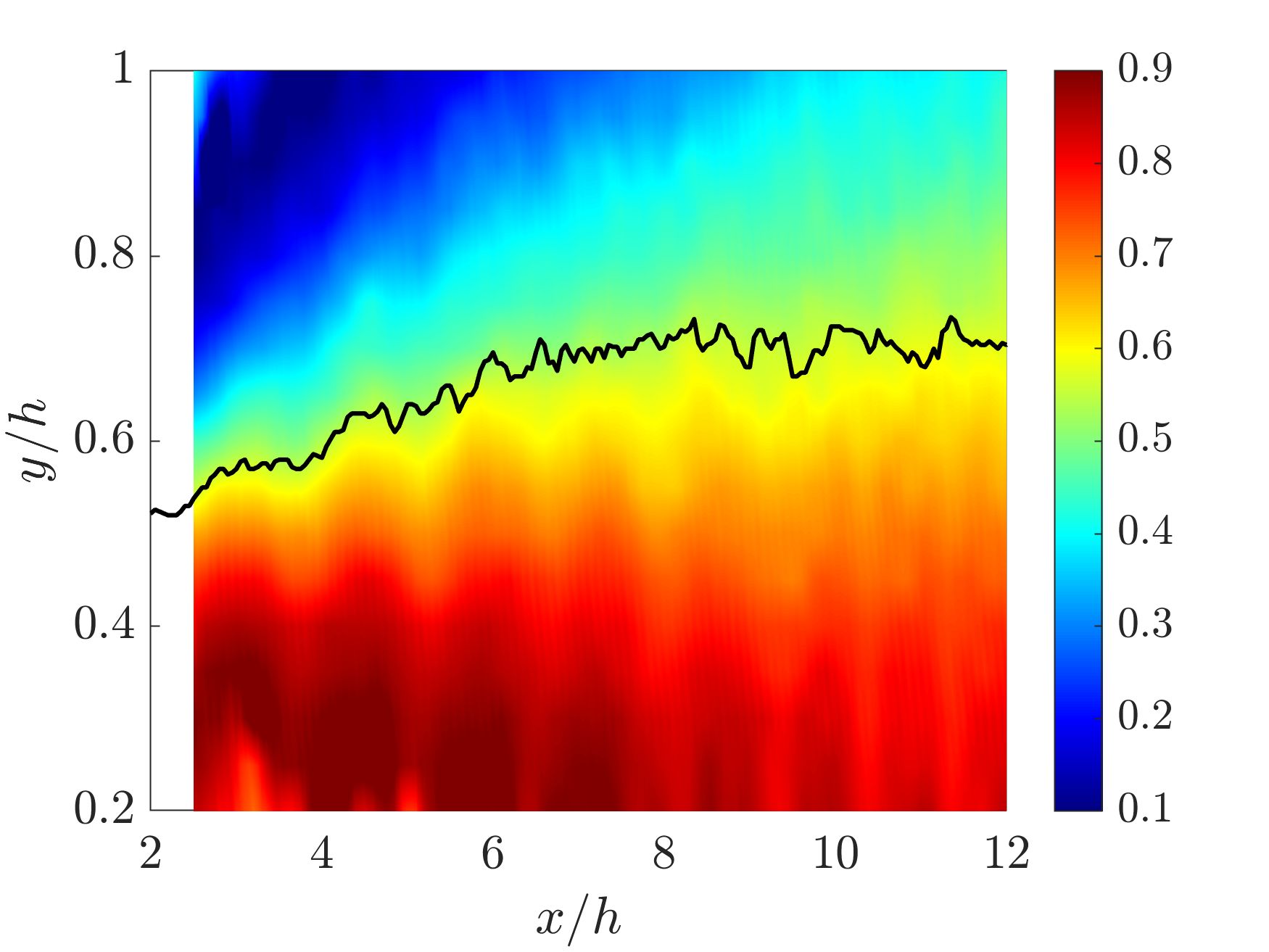} & \includegraphics[width=0.45\textwidth]{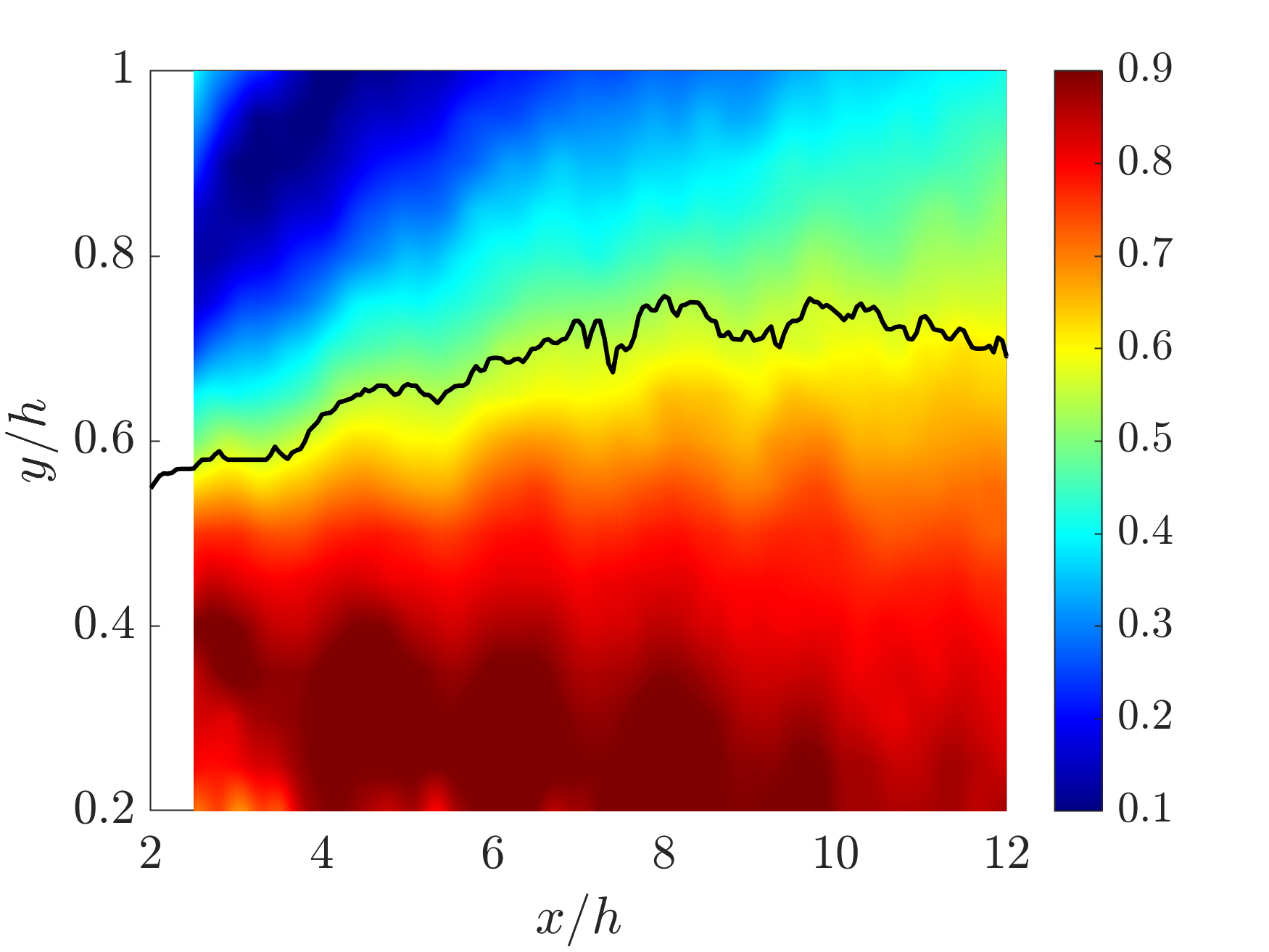} \\
  c) NPR = 3 & d) = NPR = 3.67 \\
\end{tabular}
\caption{a) Convection velocities for all three jet operating conditions, estimated by computing space-time cross-correlations of the mean streamwise velocity with respect to the reference points chosen along the lipline ($y/h$ = 0.5). b-d) Contours of the convection velocities computed by varying the reference locations. At the center of the jet shear layer (black solind line), the convection velocity is approximately 0.55$U_j$ for all three cases.}
\label{fig:convection_velocity}
\end{figure}

The local convection velocity ($u_c$) of turbulence is typically measured by calculating space-time cross-correlations of the streamwise velocity fluctuations. About the mean fields shown in Fig.~\ref{fig:Uavg_contours_minor}, the normalized cross-correlation coefficient is defined as following:
\begin{equation}
    R(\mathbf{x},\tau) = \frac{<u(\mathbf{x}_0,t_0)><u(\mathbf{x}_0 + \mathbf{x},t_0 + \tau)>}{\sqrt{<u^2(\mathbf{x}_0,t_0)>}\sqrt{<u^2(\mathbf{x}_0 + \mathbf{x},t_0 + \tau)>}},
\end{equation}
where $\textbf{x}$ and $\tau$ represent the spatial and time lags, respectively, with respect to a reference point ($\textbf{x}_0$ and $t_0$). The separation points are chosen between $x/h$ = 0 to 20 along the upper shear layer ($y/h$ = 0.5) in the same minor axis plane in which the reference points reside. After the normalization, the cross-correlation contours can be obtained with the maximum coefficient of 1 at the corresponding reference point itself. The slope of the peak correlation in space and time then determines the convection velocity. As quantified in Fig.~\ref{fig:convection_velocity}a, the local convection velocity varies periodically following the shape of the shock-cell structure. As the mean velocity increases, the average convection velocity reaches approximately 0.66$U_j$, 0.71$U_j$, and 0.76$U_j$ for NPR = 2.5, 3, and 3.67, respectively, in $4 \le x/h \le 10$. These values are well within the typically assumed convection velocity for turbulent jets (0.55-0.75$U_j$). Here, the convection velocity changes as a function of NPRs since we use the same transverse position as reference for all three cases, despite the variations in the jet spreading rates for different NPRs. By taking the reference points along the shear layer centerline (approximated by peak velocity fluctuations), the average convection velocity remains unchanged as 0.55$U_j$ for all three cases as shown in Figs.~\ref{fig:convection_velocity}b-d.

\subsubsection{Velocity fluctuations and turbulent kinetic energy}
\begin{figure}
\centering
\begin{tabular}{c}
  \includegraphics[width=0.5\textwidth]{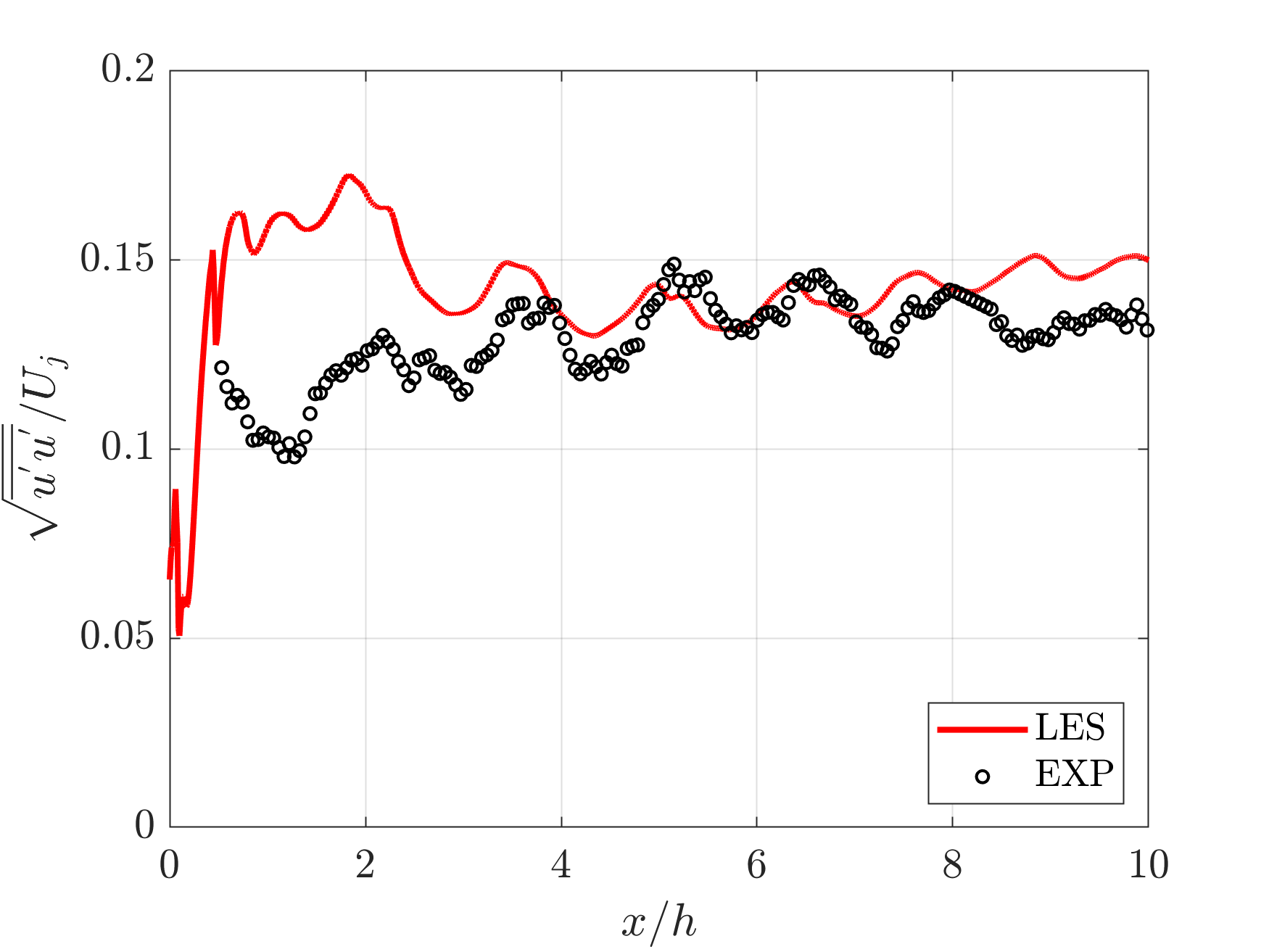} \\
\end{tabular}
\caption{Comparison between the LES (red solid line) and experimental (black circles) RMS streamwise velocities along the nozzle lipline in the minor axis for NPR = 3.}
\label{fig:urms_profile}
\end{figure}

Figure~\ref{fig:urms_profile} represents the RMS streamwise velocity profiles along the lipline at NPR = 3. The initial shear layer is somewhat disturbed by numerical artifacts attributed to the grid transitions, but it is not fully turbulent yet. While the experiments (circles) register high fluctuation amplitudes that periodically varies depending on the shock/expansion wave formation, the LES (solid lines) shows an overshoot around $x/h$ = 0.5, which is associated with transition to turbulence from an initially laminar shear layer state~\cite{bres2018a}. 

\begin{figure}
\centering
\begin{tabular}{c}
  \includegraphics[width=0.8\textwidth]{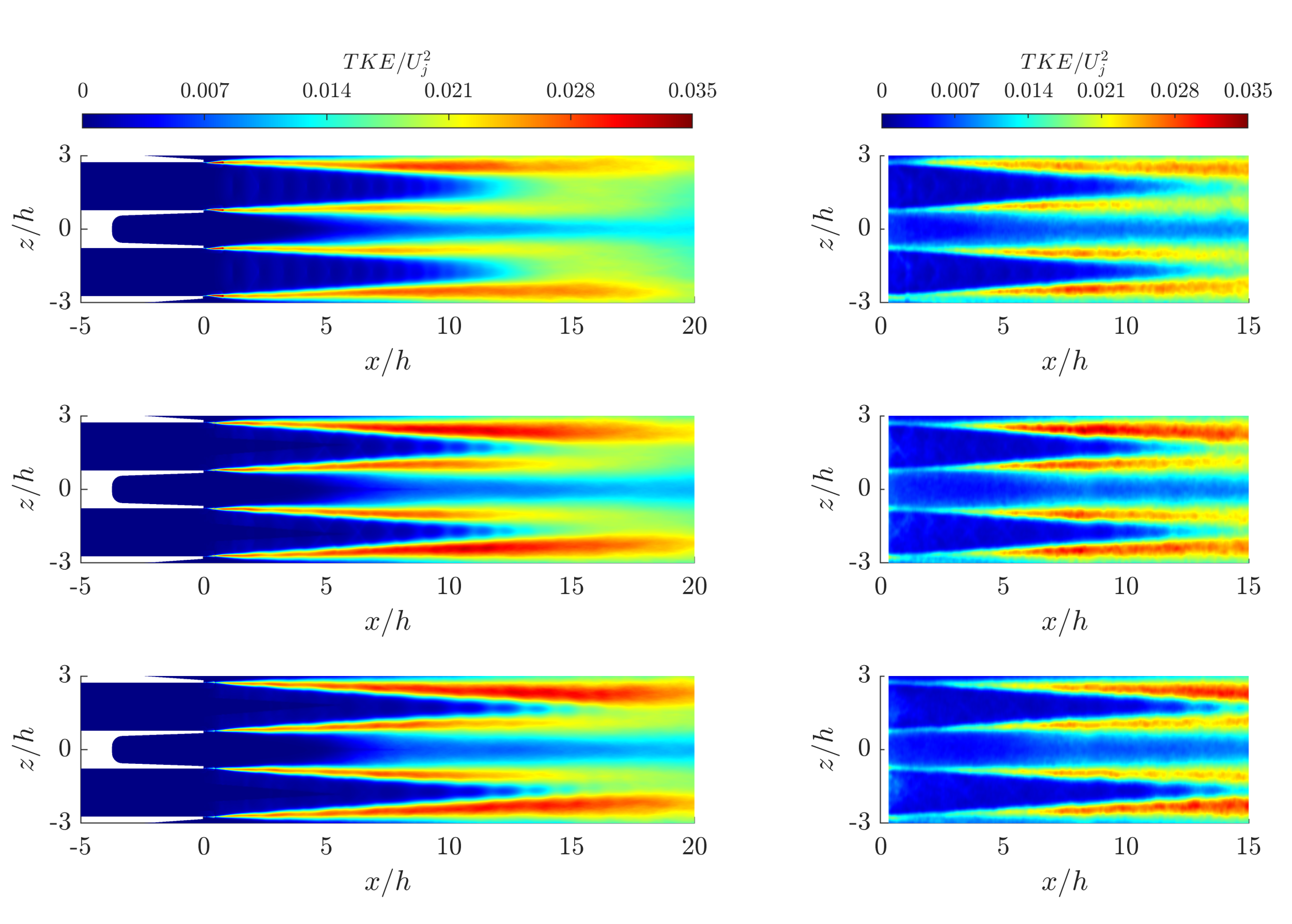} \\
\end{tabular}
\caption{Comparison of the TKE contours normalized by the squares of the fully expanded jet velocity obtained by the (left) LES and (right) experiments: (top) NPR = 2.5, (middle) NPR = 3, and (bottom) NPR = 3.67. \vspace{5mm}}
\label{fig:tke_contours_major}
\end{figure}

TKE contours obtained by the LES and experiments are compared in Fig.~\ref{fig:tke_contours_major}. TKE levels are computed using Eq.~\eqref{eq:tke}. Due to the twin configuration, the TKE contours exhibit outer and inner shear layers in this axis view. Contours in the other two axes are omitted in this paper since the differences in characteristics between the LES and experiments are kept similar to those observed in the major axis view. In all three conditions, LES produce significantly higher TKE levels in the vicinity of the nozzle exits. Such over-prediction near the nozzle lip is probably due to the sharp transition to turbulent from initially laminar shear layers. Except this, the two data sets share common flow features. The outer shear layers have higher TKE levels than the inner shear layers do. It is also clearly visible that the two jets pull towards each other as they spread in the downstream. The mismatch in the overall TKE levels may be alleviated once more accurate nozzle-exit boundary layer statistics are achieved in LES.

\begin{figure}
\centering
\begin{tabular}{ccc}
  \includegraphics[width=0.3\textwidth]{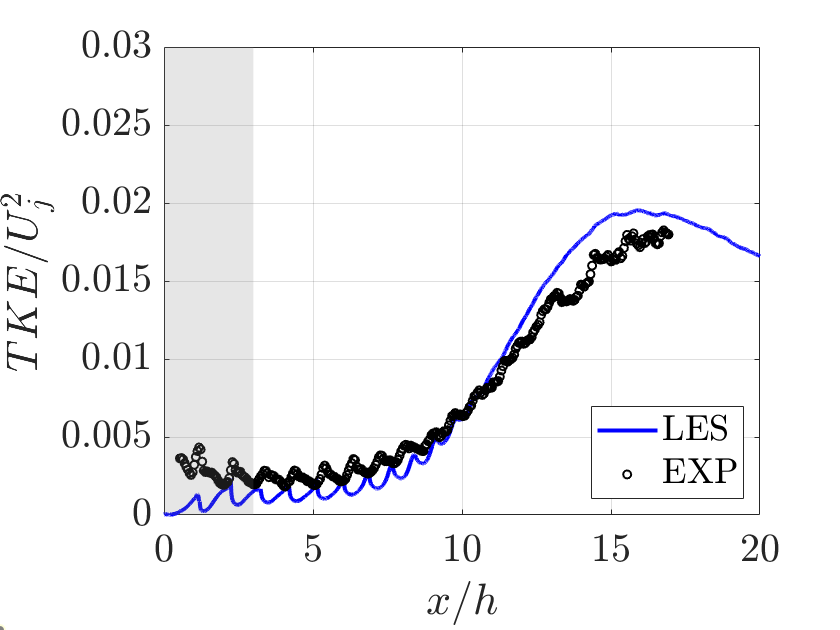} &
  \includegraphics[width=0.3\textwidth]{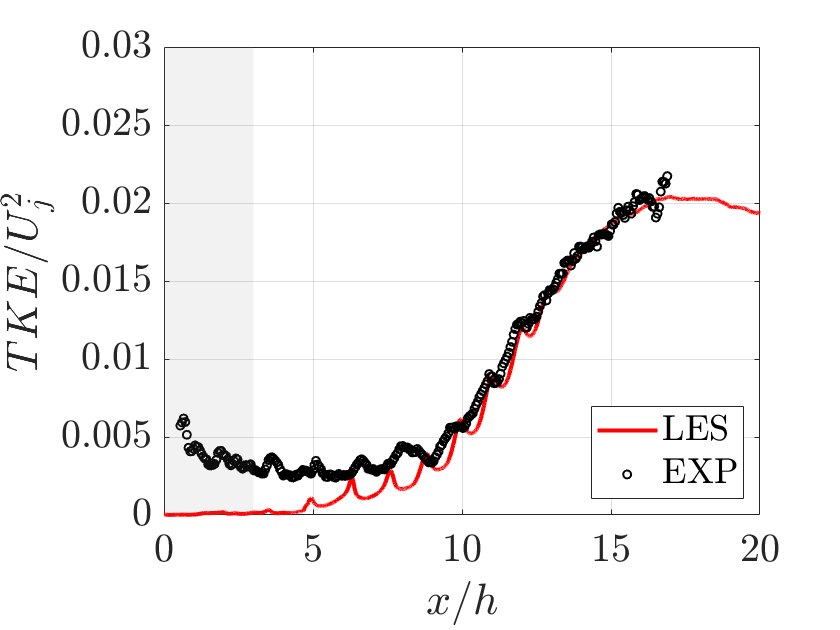} &
  \includegraphics[width=0.3\textwidth]{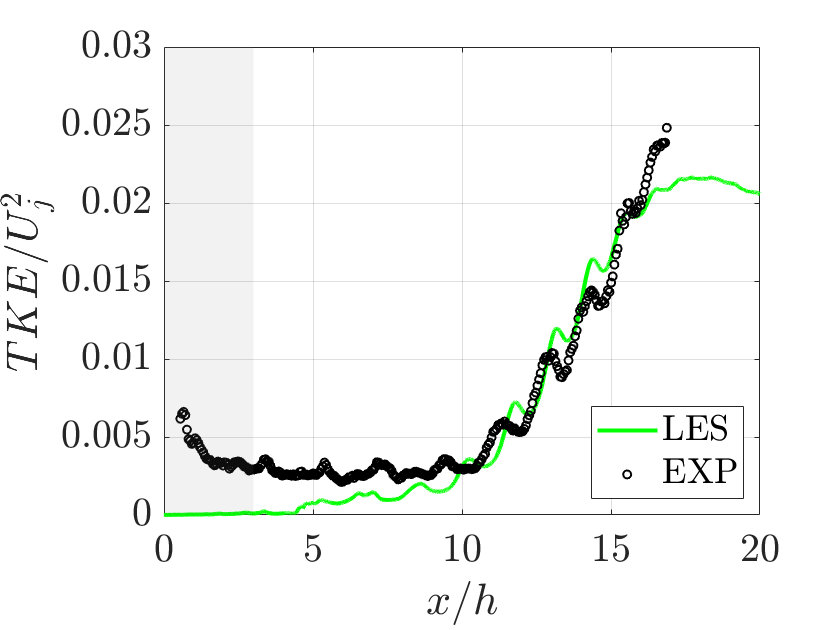} \\
  a) NPR = 2.5 & b) NPR = 3 & c) NPR = 3.67 \\
\end{tabular}
\caption{Comparison between the LES and experimental TKE profiles along the major axis jet centerlines.}
\label{fig:tke_profiles}
\end{figure}

Figure~\ref{fig:tke_profiles} further compares the TKE profiles between the LES and the experiments along the major axis, in the same way that the mean velocity profiles are compared in Fig.~\ref{fig:Uavg_profiles}. The TKE levels are computed using Eq.~\eqref{eq:tke}. To achieve more smooth profiles for better comparisons between the LES and experiments, profiles extracted along each (geometric) centerline of the two jets are averaged. Compared to the mean statistics, TKE is generally slow to converge, but the numerical and experimental TKE levels match each other quite well in this direction except the region close to the nozzle exit. LES also captures variations in the TKE levels induced by the shock-cell systems in a fashion similar to that observed in the experiments. The TKE profiles spanning up to $x/h$ = 10-14 correspond to the region inside the jet potential core. Overshoots registered in the experiment are, in part, artifacts due to residual seed agglomeration. On the other hand, even though a knife edge was also used to limit the spread of the laser sheet to prevent strong reflections from the nozzle lip bleeding into the camera frame, the FOV of the current PIV setup still seems to capture more of the diverging laser sheet that impinges on the inner wall of the second jet in the major orientation, which leads to the TKE spikes (shown by the gray regions in Fig.~\ref{fig:tke_profiles}).

Compared to the two high NPR cases, flow statistics are slow to converge for NPR = 2.5. It seems that the formation of Mach stems in the jet plume and slower convection velocities associated with low NPR require much longer time horizon to resolve turbulent statistics. At NPR = 3.67, LES shows slight under-prediction of the TKE values around the end of the potential core. By recalling that this case showed the longest potential core length, further mesh refinement may help to achieve improved agreement with the measurements. Inflow forcing to achieve turbulent boundary layer state at nozzle exit may also be useful further development.

\subsection{Far-field acoustics}
\label{subsec:far_field}
\subsubsection{Sound pressure levels}
\begin{figure}
\centering
  \includegraphics[width=0.8\textwidth]{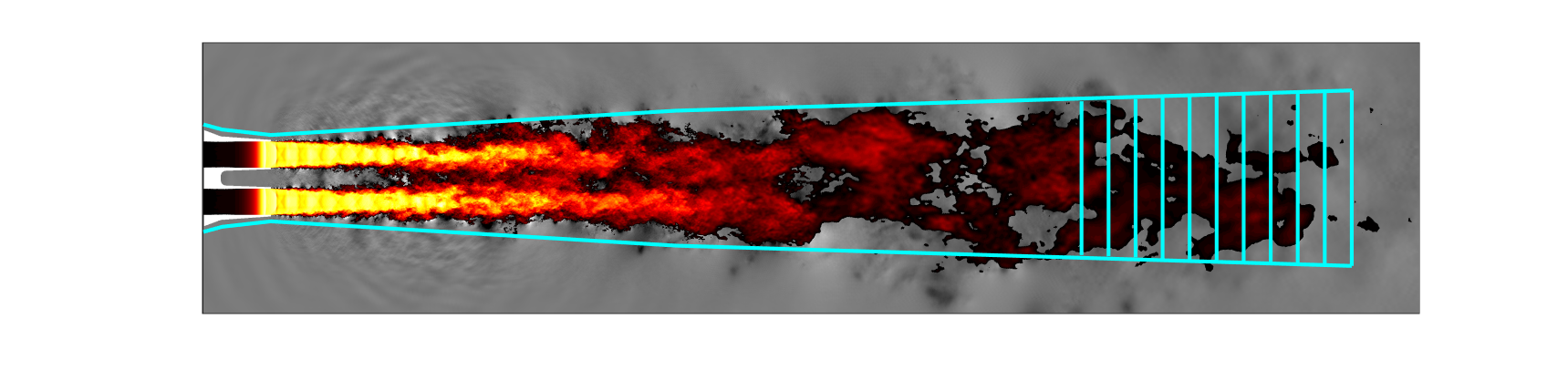} \\
  \includegraphics[width=0.8\textwidth]{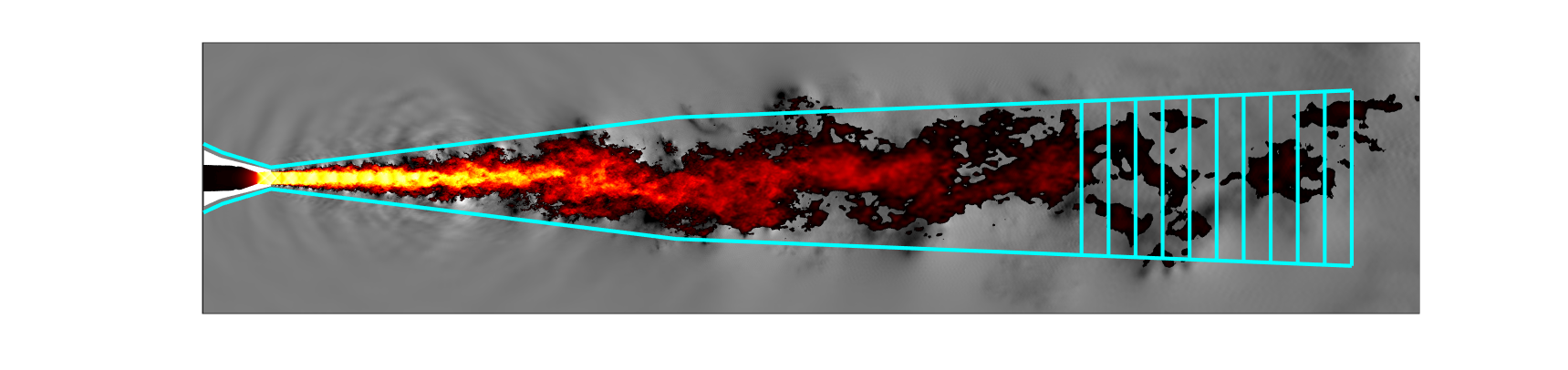} \\
\caption{Outline of a FW-H surface closed with 11 end-caps, overlaid on instantaneous velocity and pressure contours in the (top) major and (bottom) minor axes. Figures drawn to scale.}
\label{fig:fwh_surfaces}
\end{figure}

For a direct comparison with the experiments, far-field sound is computed by projecting the near-field flow data onto a circular arc microphone array that is 61.5$D_e$ = 93.2$h$ away from the system origin in both the major and minor directions as described in Fig.~\ref{fig:mic}. The near-field data to be projected are collected from the FW-H surface shown in Fig.~\ref{fig:fwh_surfaces} at an interval of $\Delta t c_{\infty} / h = 0.05$. Sound pressure levels (SPL) are estimated using Welch's method by splitting the raw data into several blocks and applying a Hann windowing with 75$\%$ overlap between blocks to match a minimum bin size of 50 Hz in the experiments. By defining the Strouhal number based on the fully expanded jet velocity and equivalent diameter such that $St = f U_j / D_e$, the minimum bin sizes are $\Delta St$ = 0.0025, 0.0023, and 0.0021 for NPR = 2.5, 3, and 3.67, respectively. 

\begin{figure}
\centering
  \begin{tabular}{ccc}   
    \includegraphics[width=0.3\textwidth]{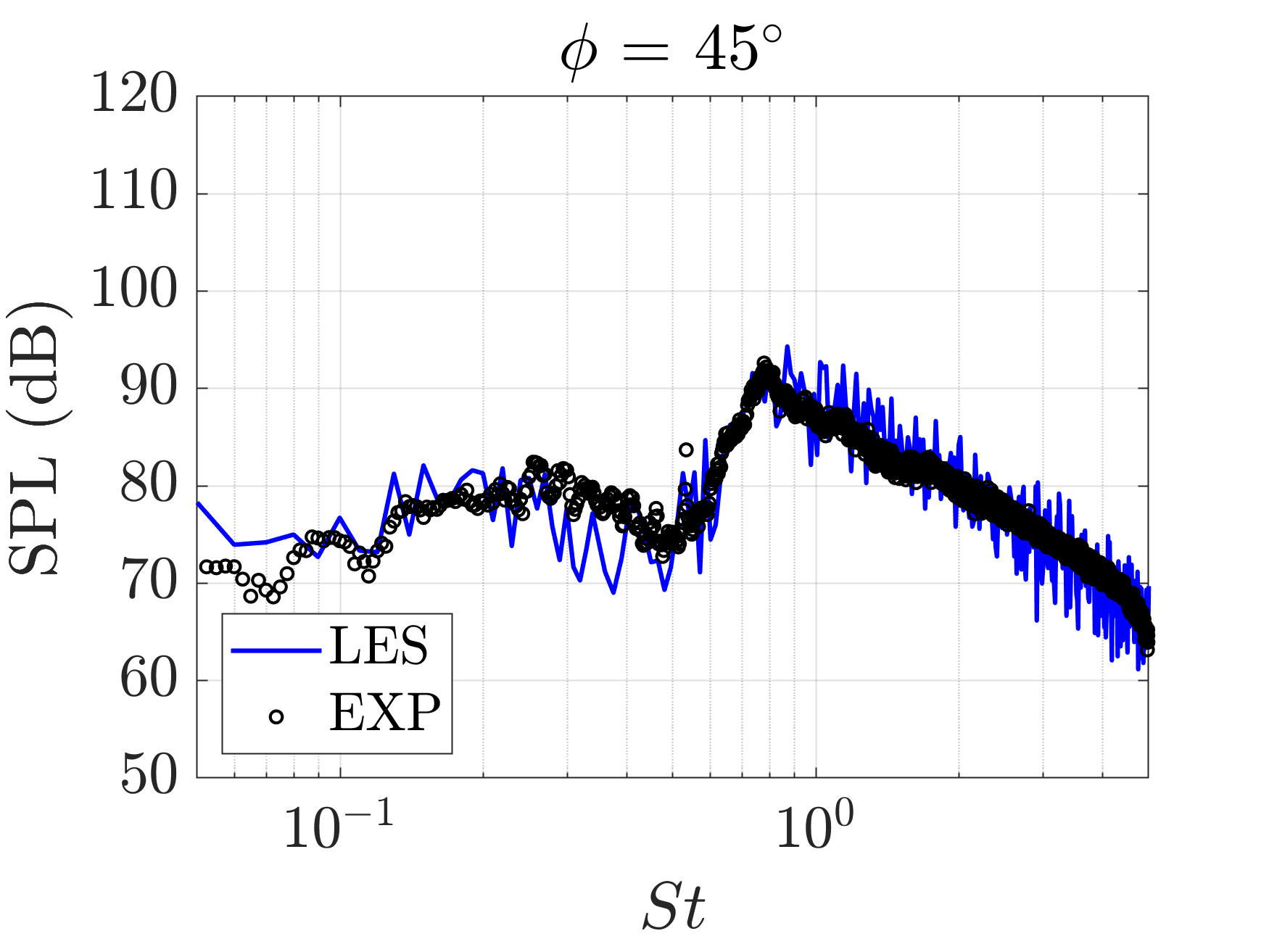} & \includegraphics[width=0.3\textwidth]{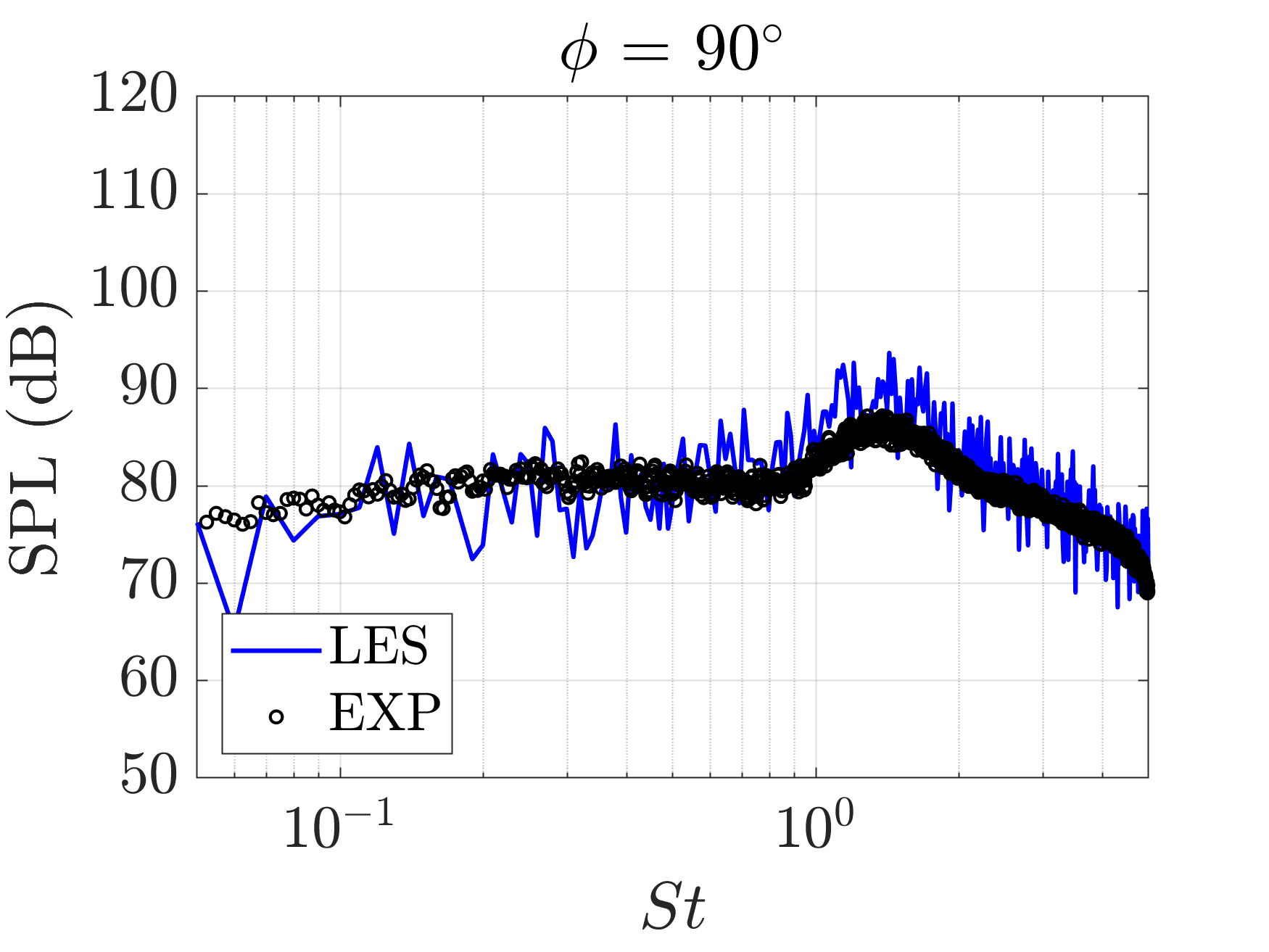} & \includegraphics[width=0.3\textwidth]{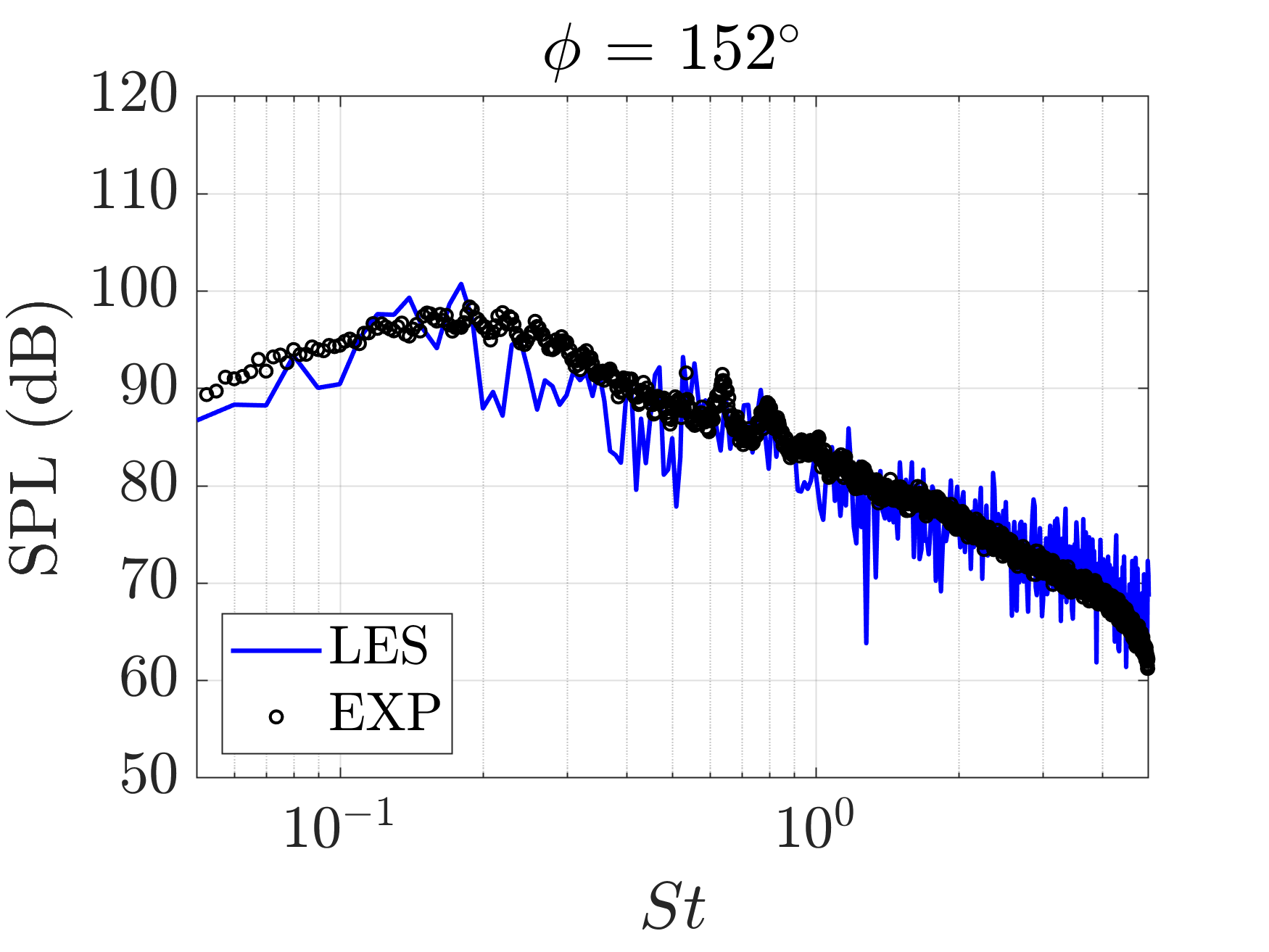} \\
    \includegraphics[width=0.3\textwidth]{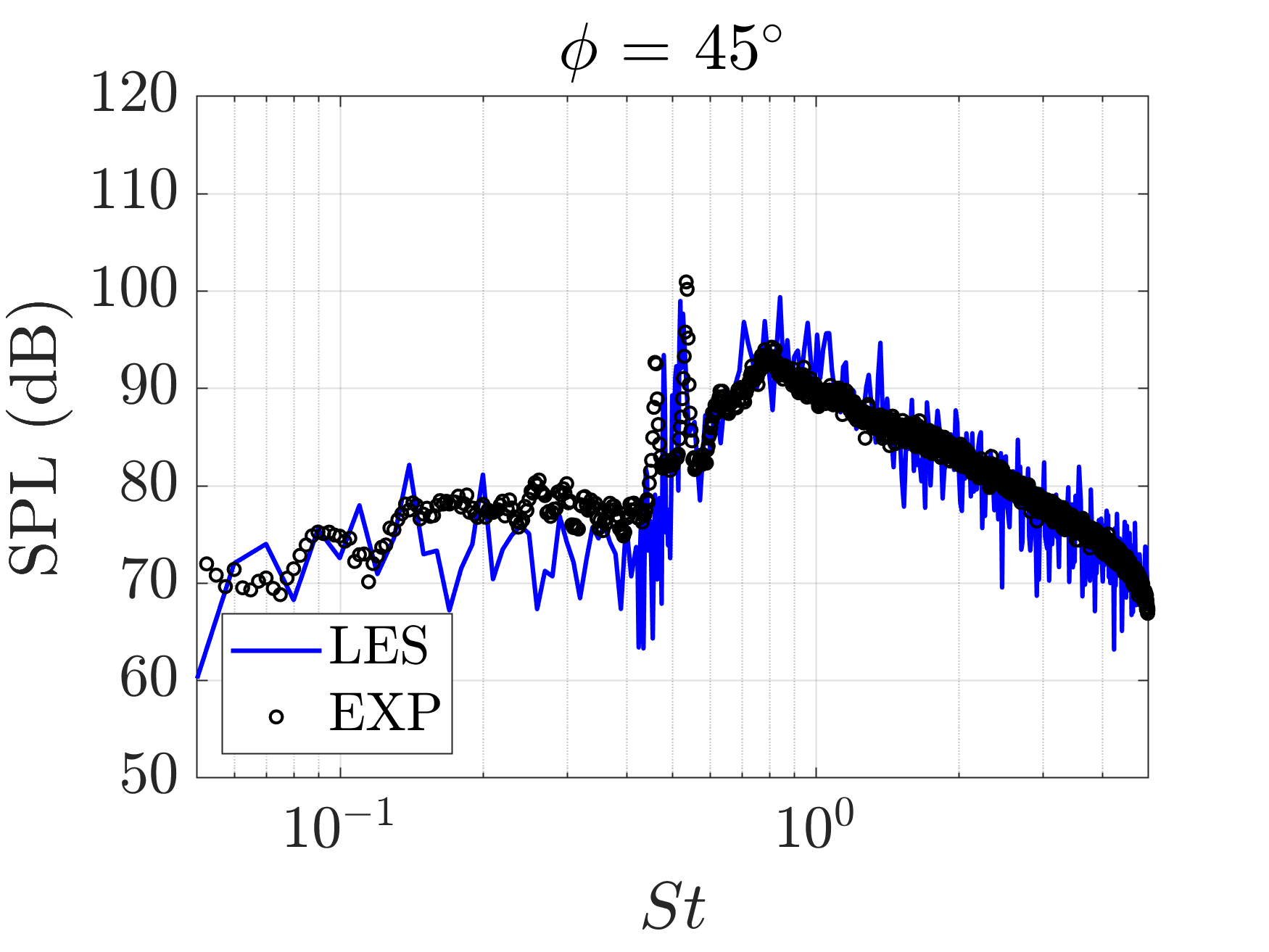} & \includegraphics[width=0.3\textwidth]{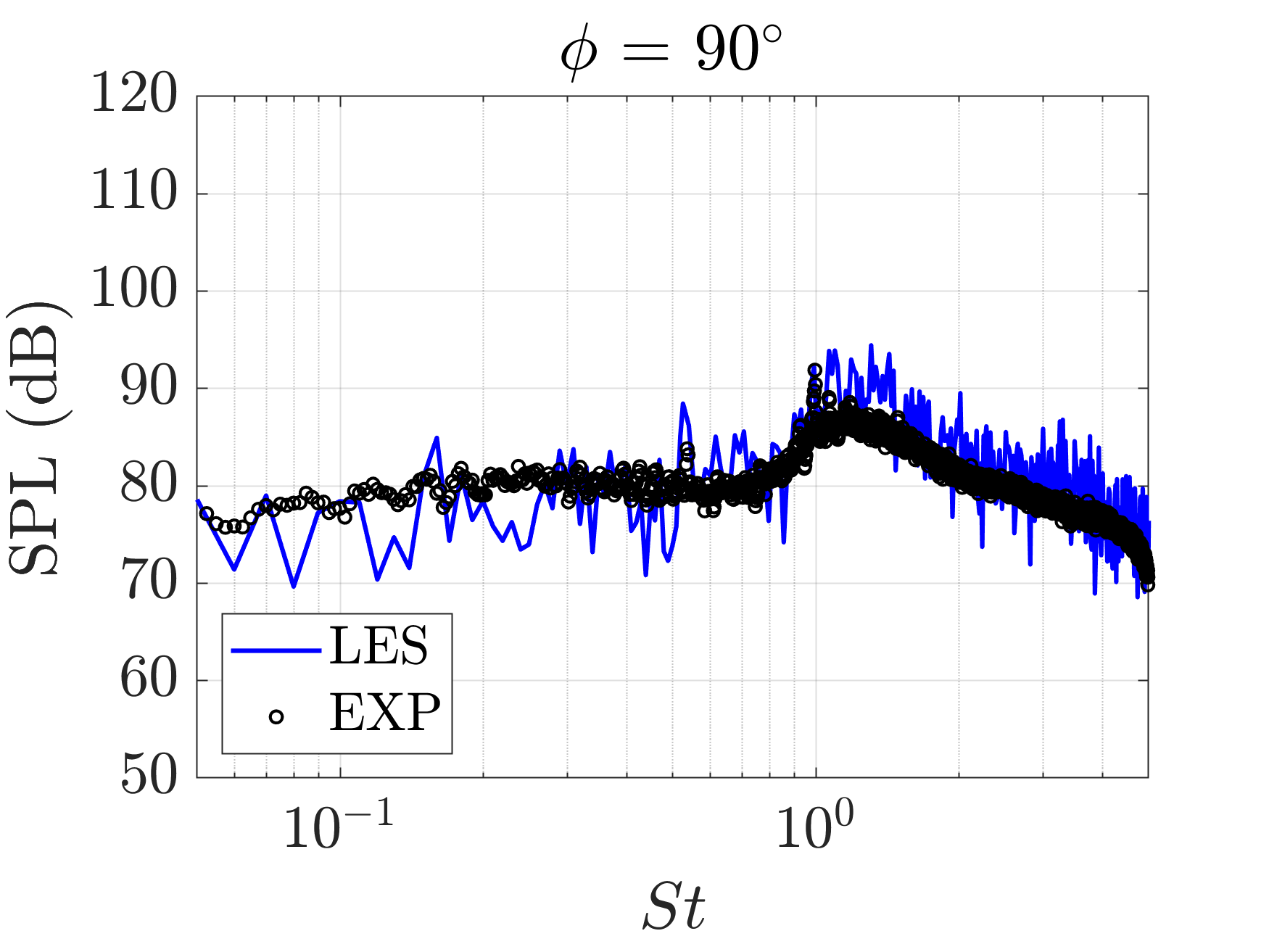} & \includegraphics[width=0.3\textwidth]{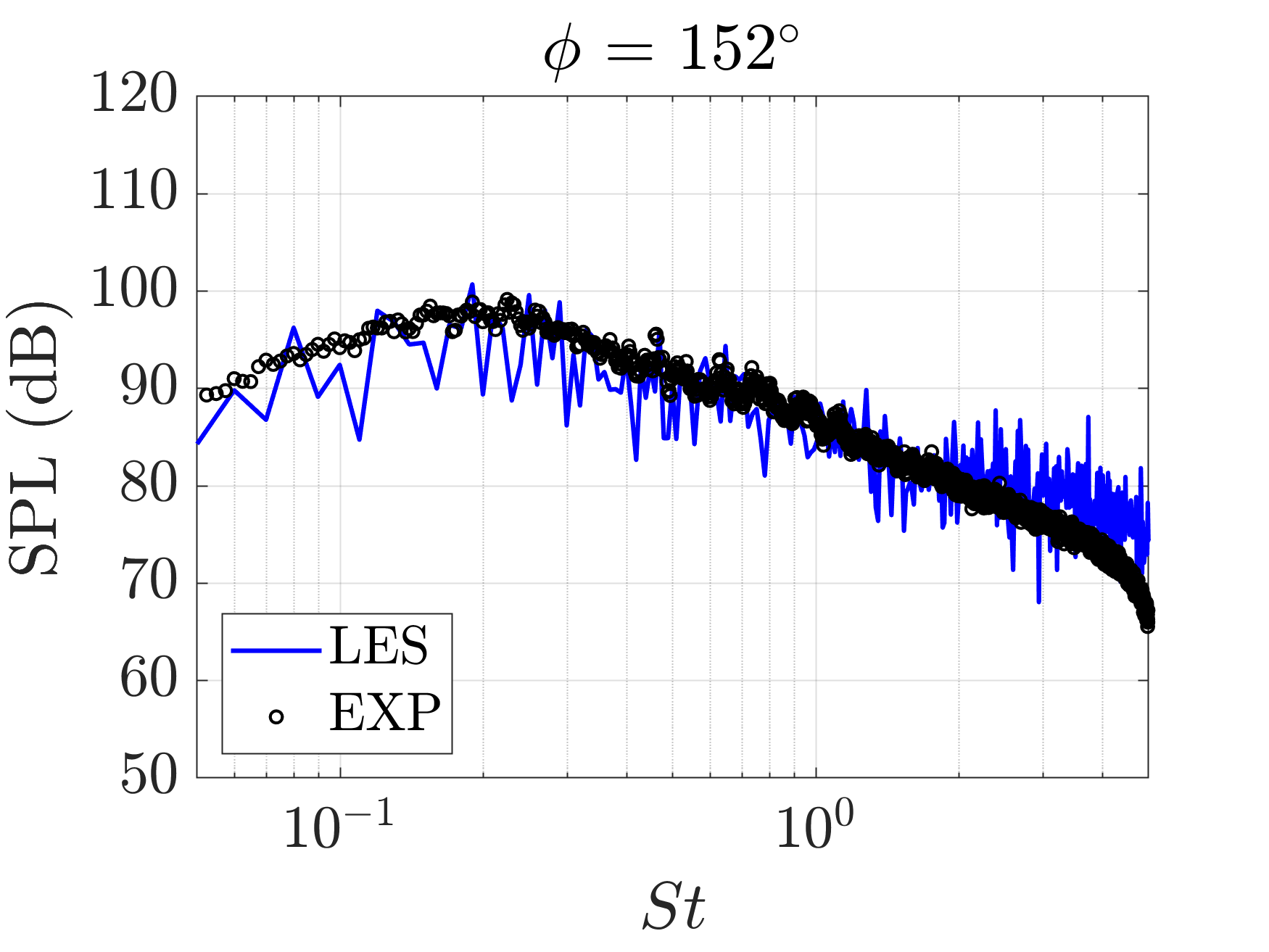} \\
  \end{tabular}
\caption{Comparison of the sound pressure levels between the LES (colored solid lines) and experiments (black circles) at $61.5D_e$ away from the nozzle exit: (top) in the major axis and (bottom) in the minor axis for NPR = 2.5.}
\label{fig:spl_npr2p5}
\end{figure}

\begin{figure}
\centering
  \begin{tabular}{ccc}   
    \includegraphics[width=0.3\textwidth]{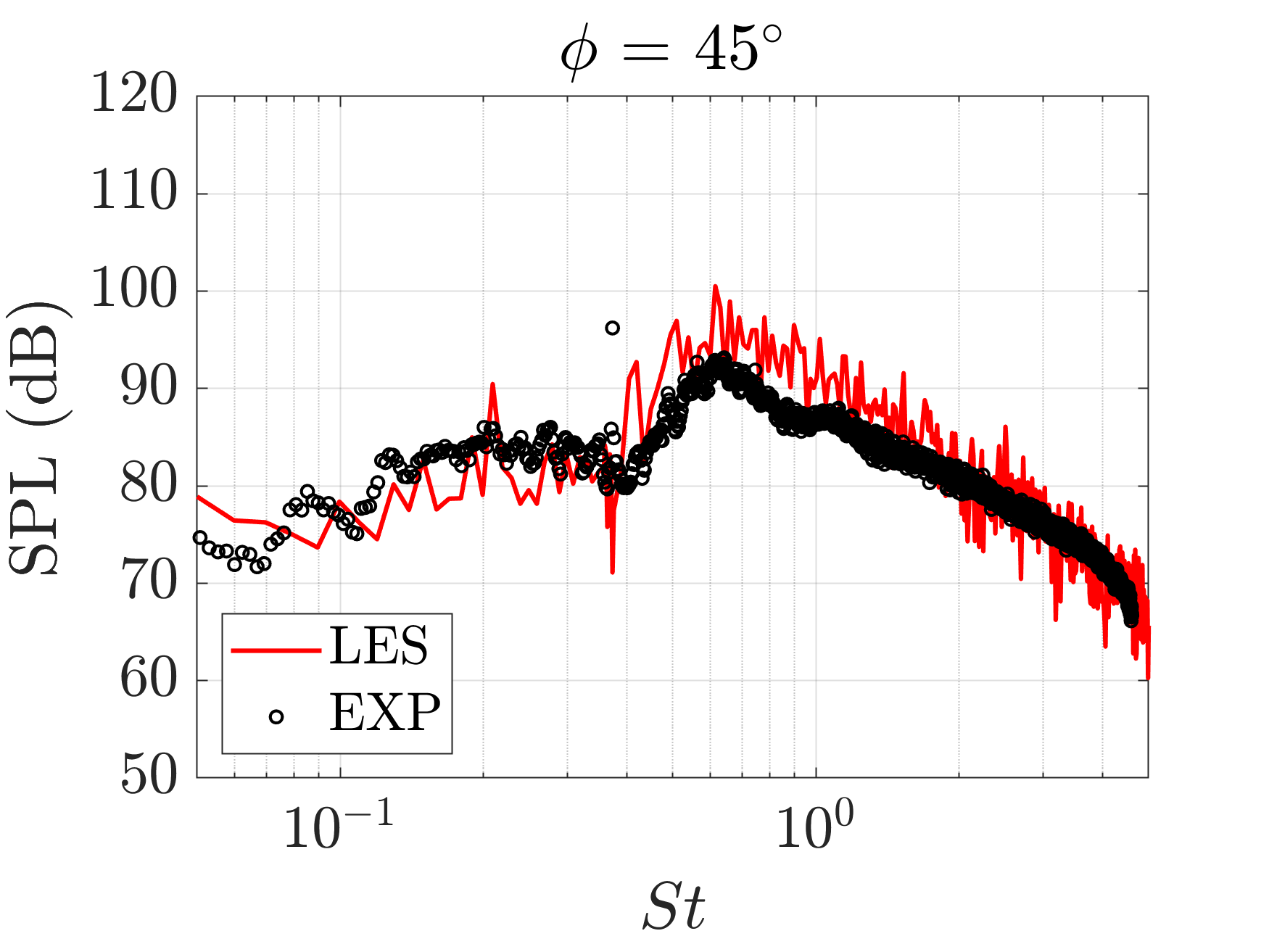} & \includegraphics[width=0.3\textwidth]{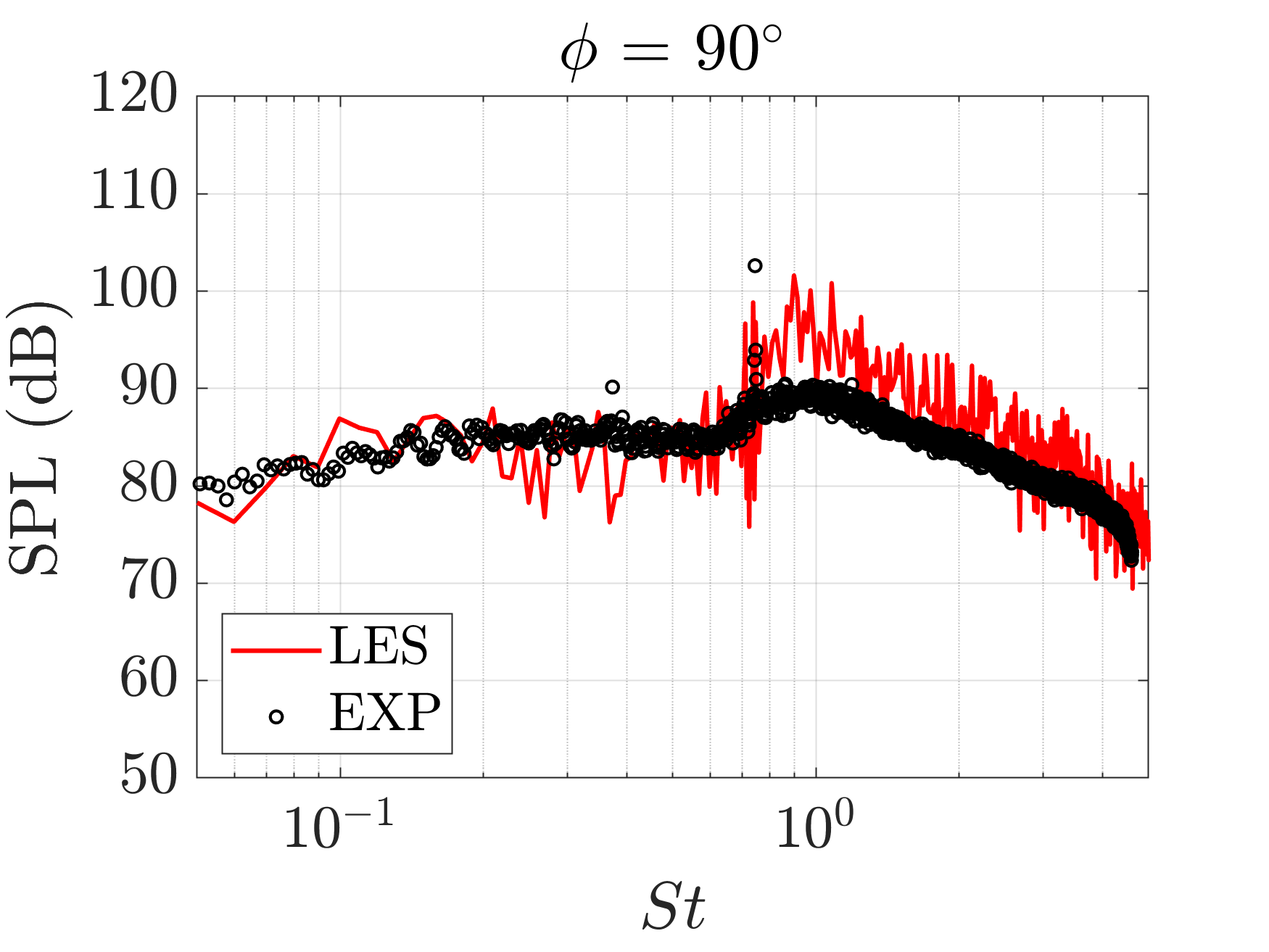} & \includegraphics[width=0.3\textwidth]{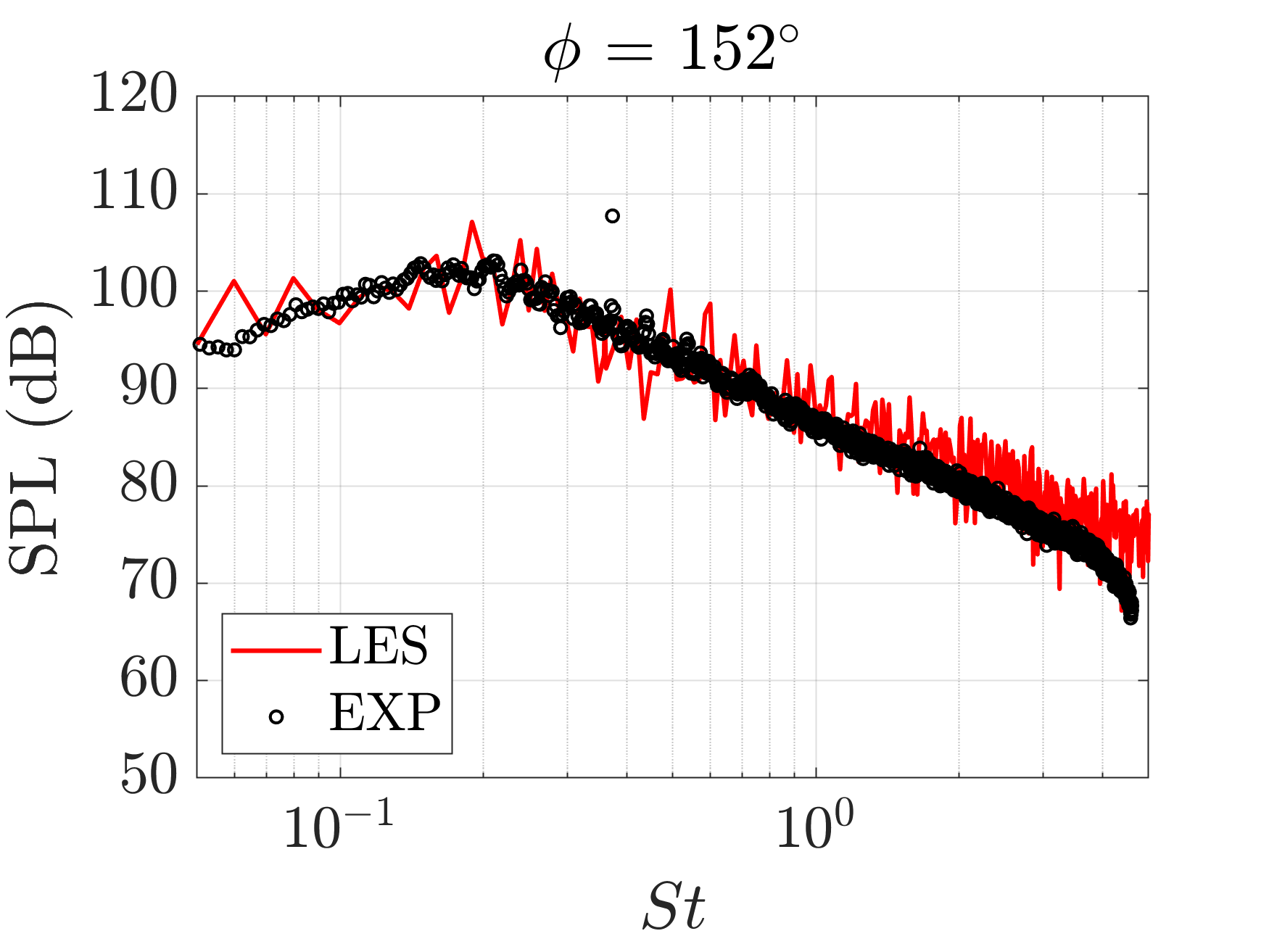} \\
    \includegraphics[width=0.3\textwidth]{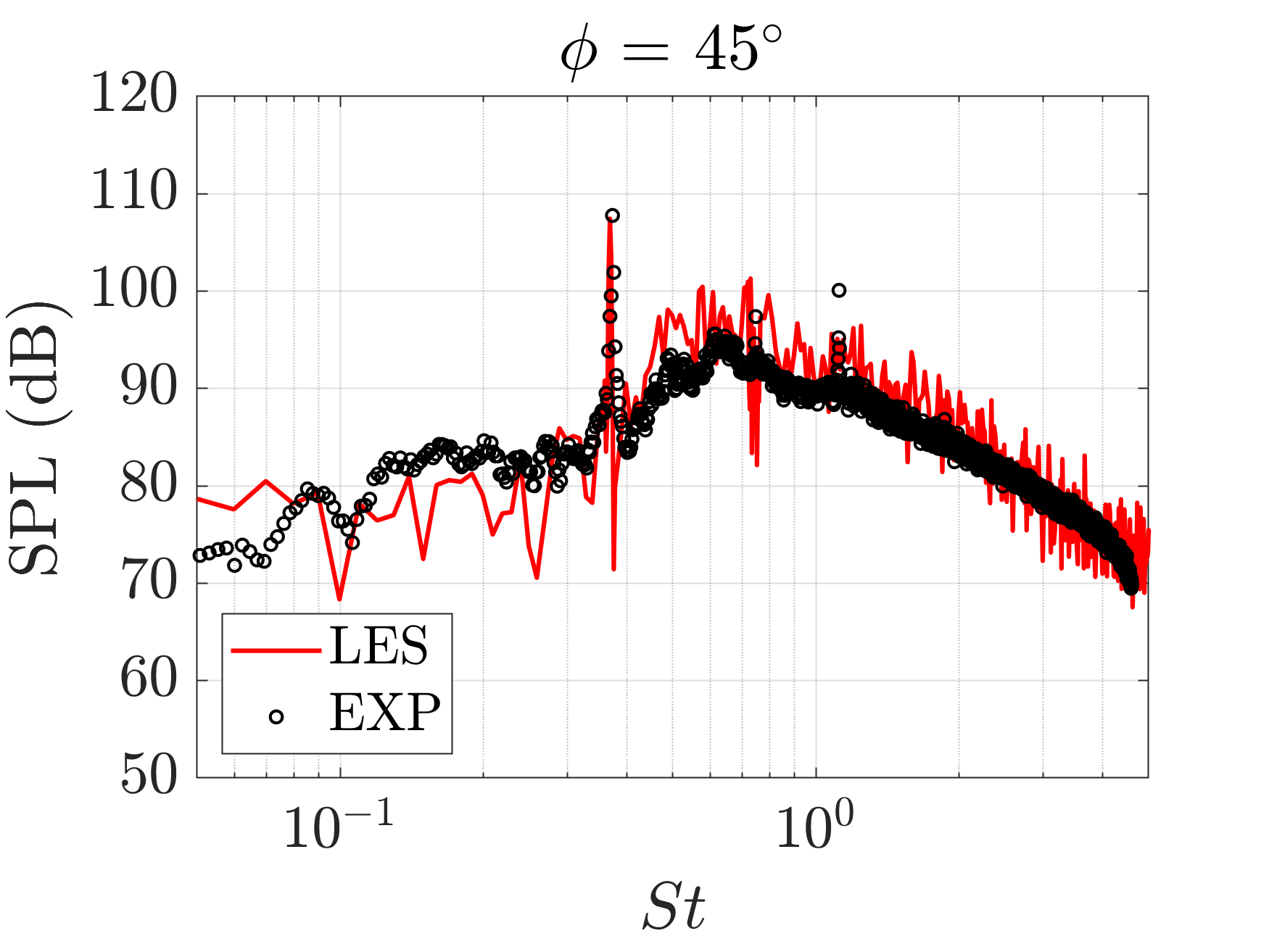} & \includegraphics[width=0.3\textwidth]{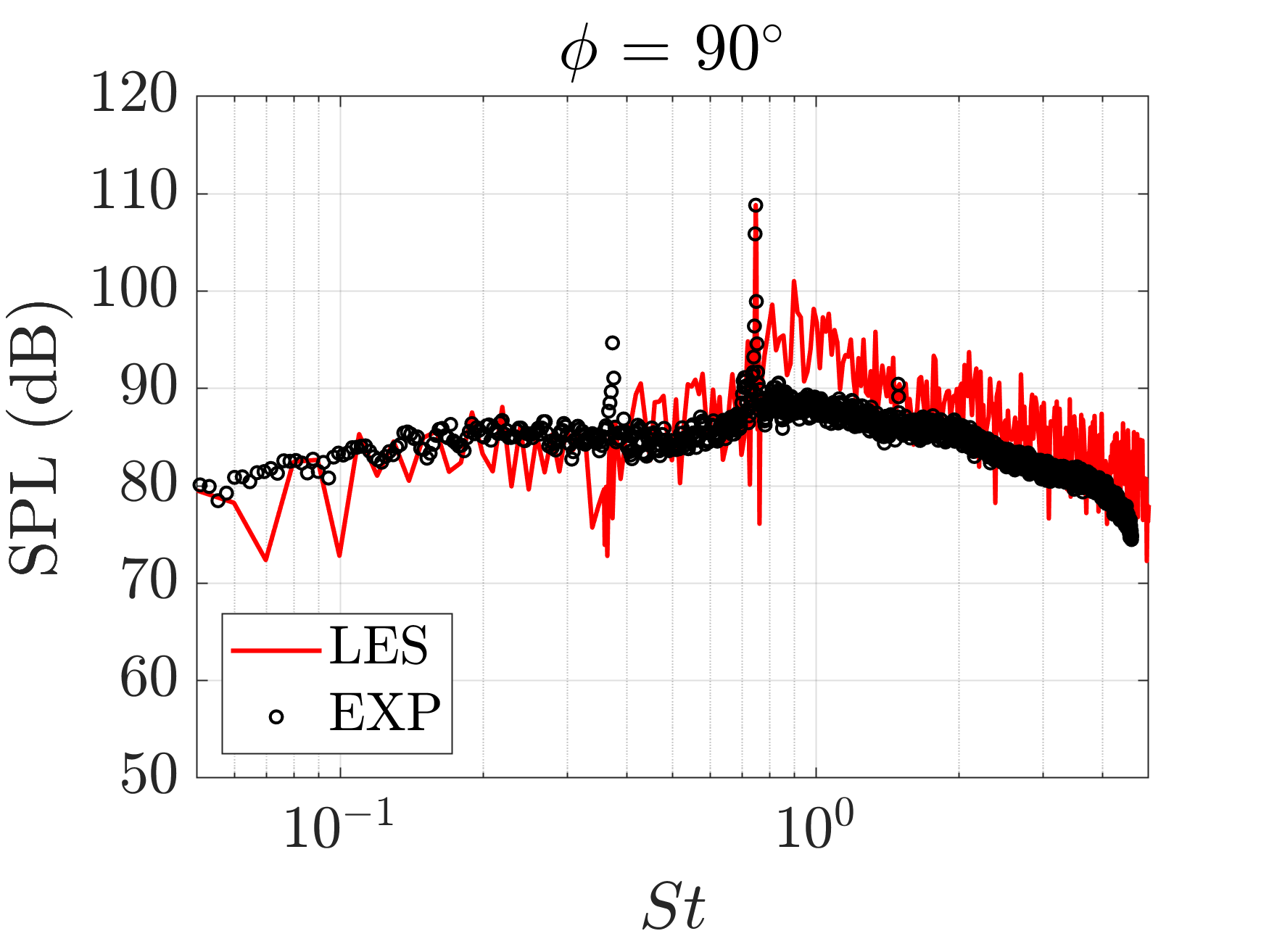} & \includegraphics[width=0.3\textwidth]{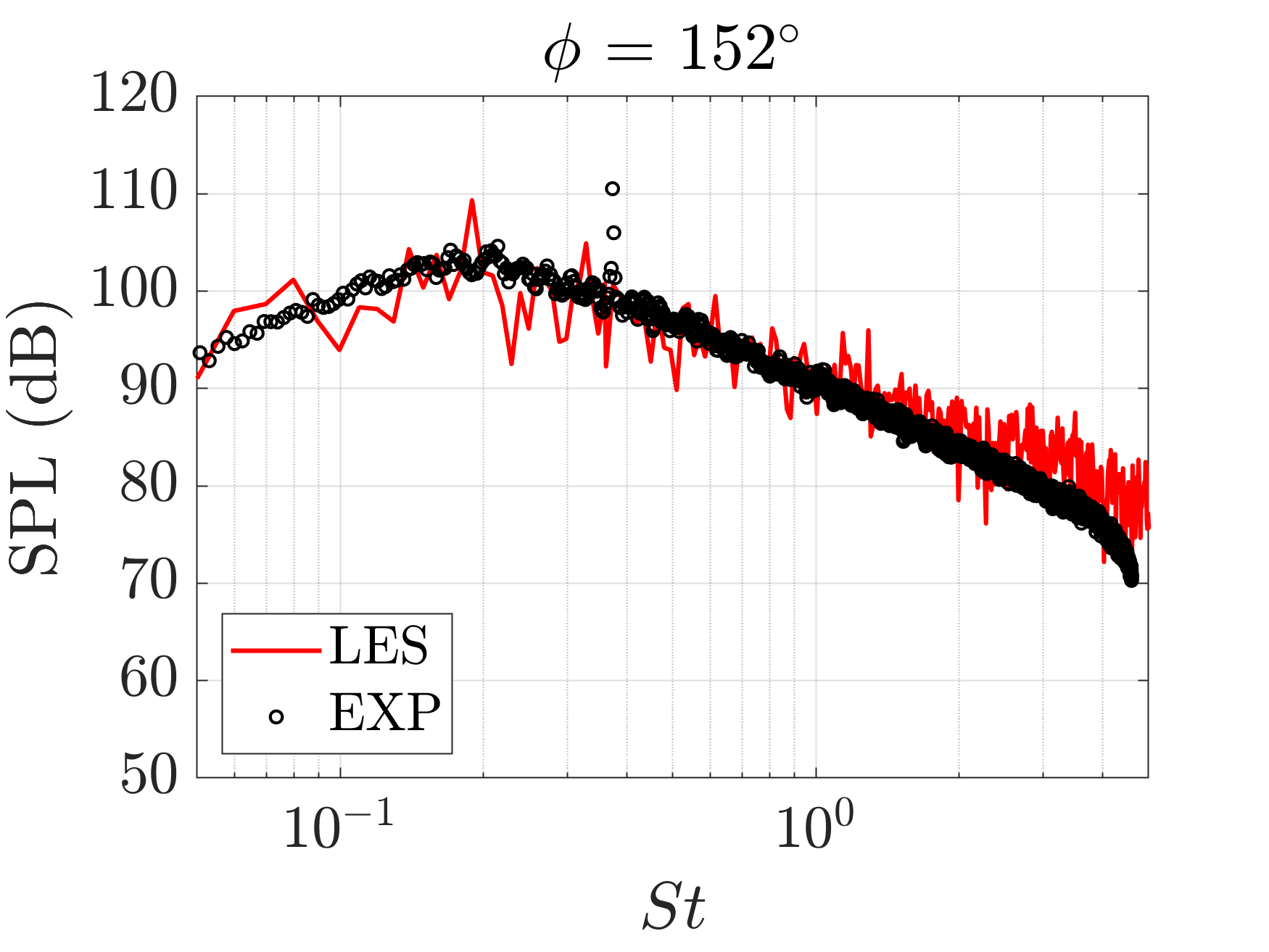} \\
  \end{tabular}
\caption{Comparison of the sound pressure levels between the LES (colored solid lines) and experiments (black circles) at $61.5D_e$ away from the nozzle exit: (top) in the major axis and (bottom) in the minor axis for NPR = 3.}
\label{fig:spl_npr3}
\end{figure}

\begin{figure}
\centering
  \begin{tabular}{ccc}   
    \includegraphics[width=0.3\textwidth]{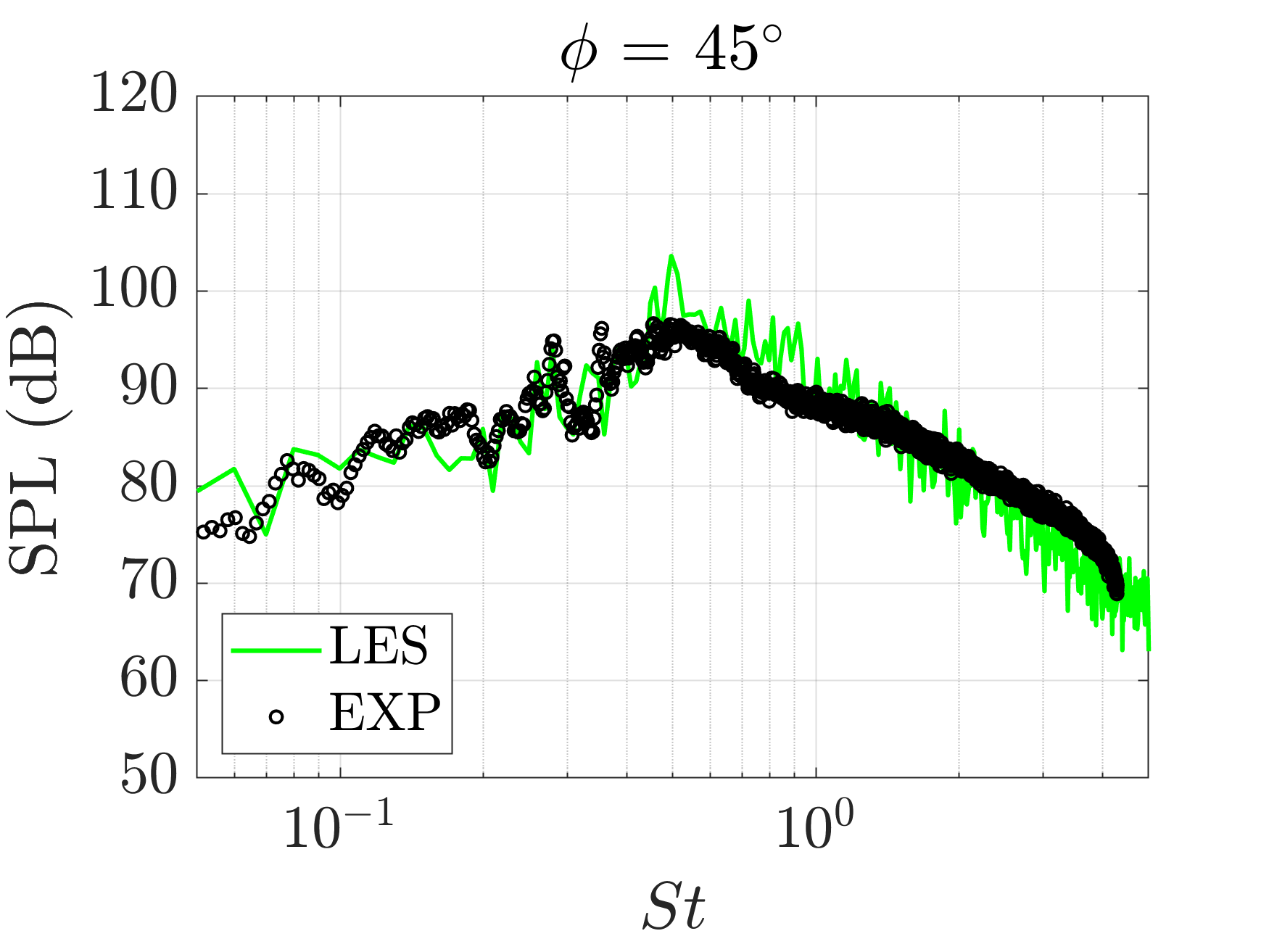} & \includegraphics[width=0.3\textwidth]{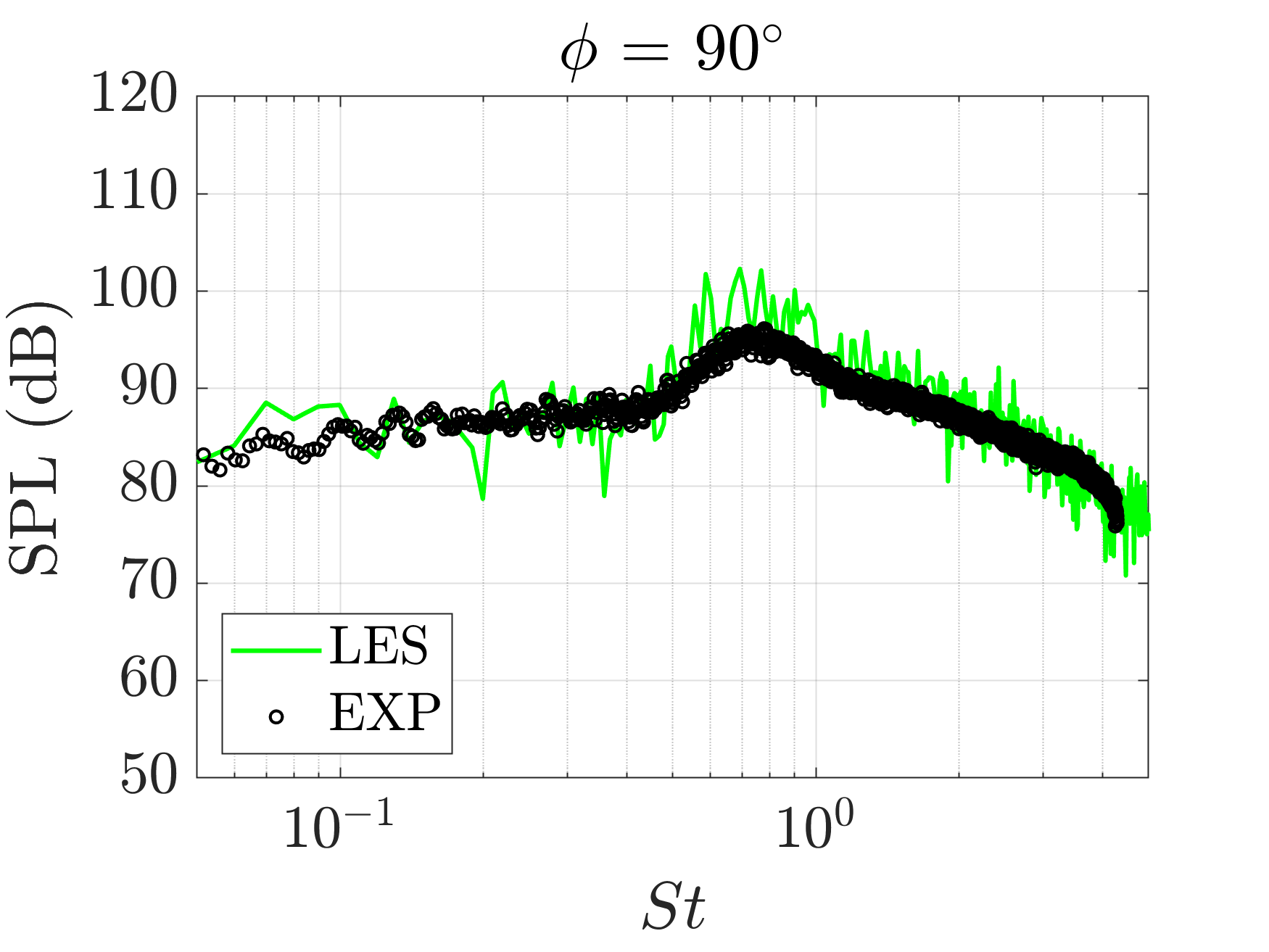} & \includegraphics[width=0.3\textwidth]{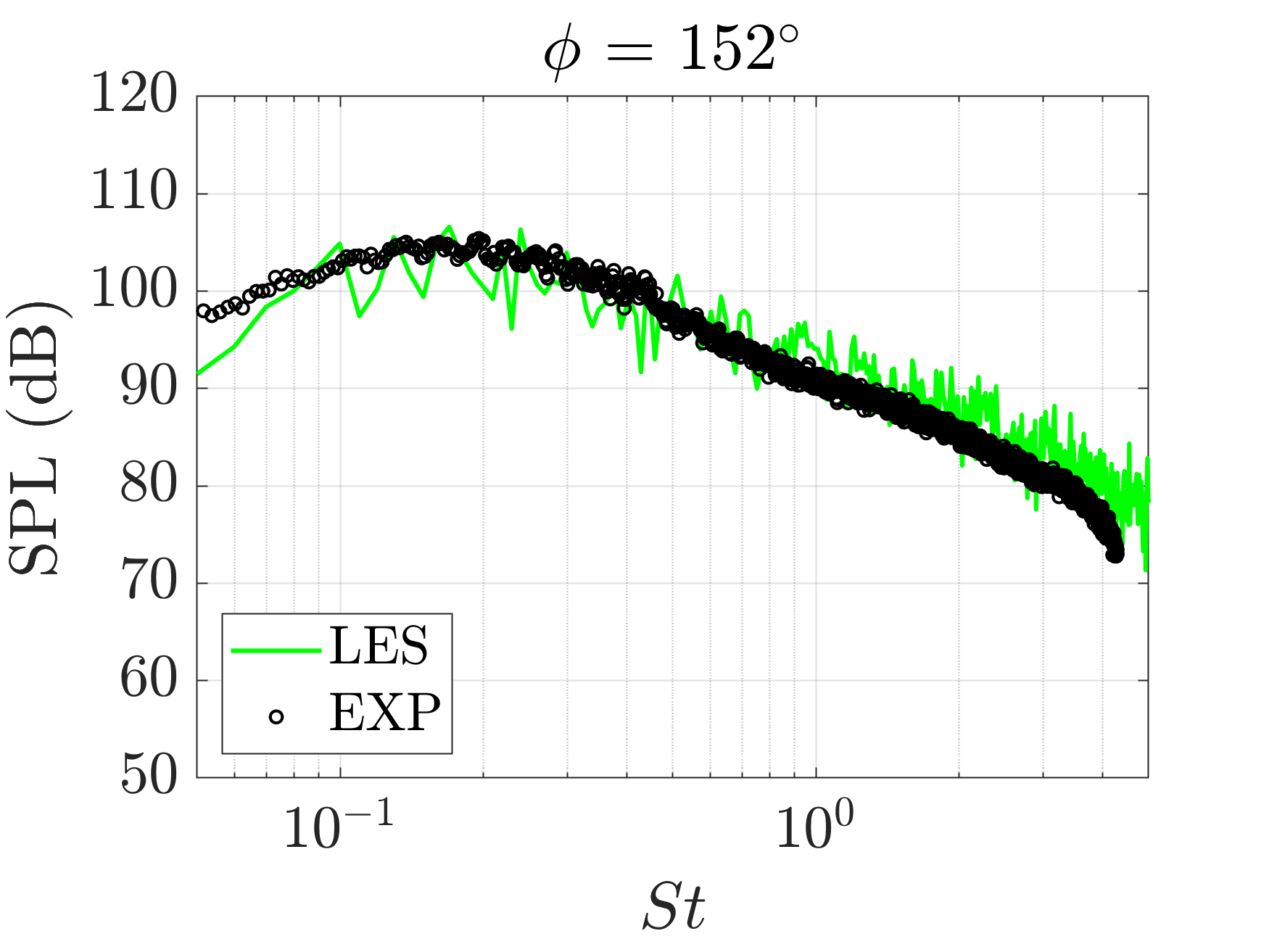} \\
    \includegraphics[width=0.3\textwidth]{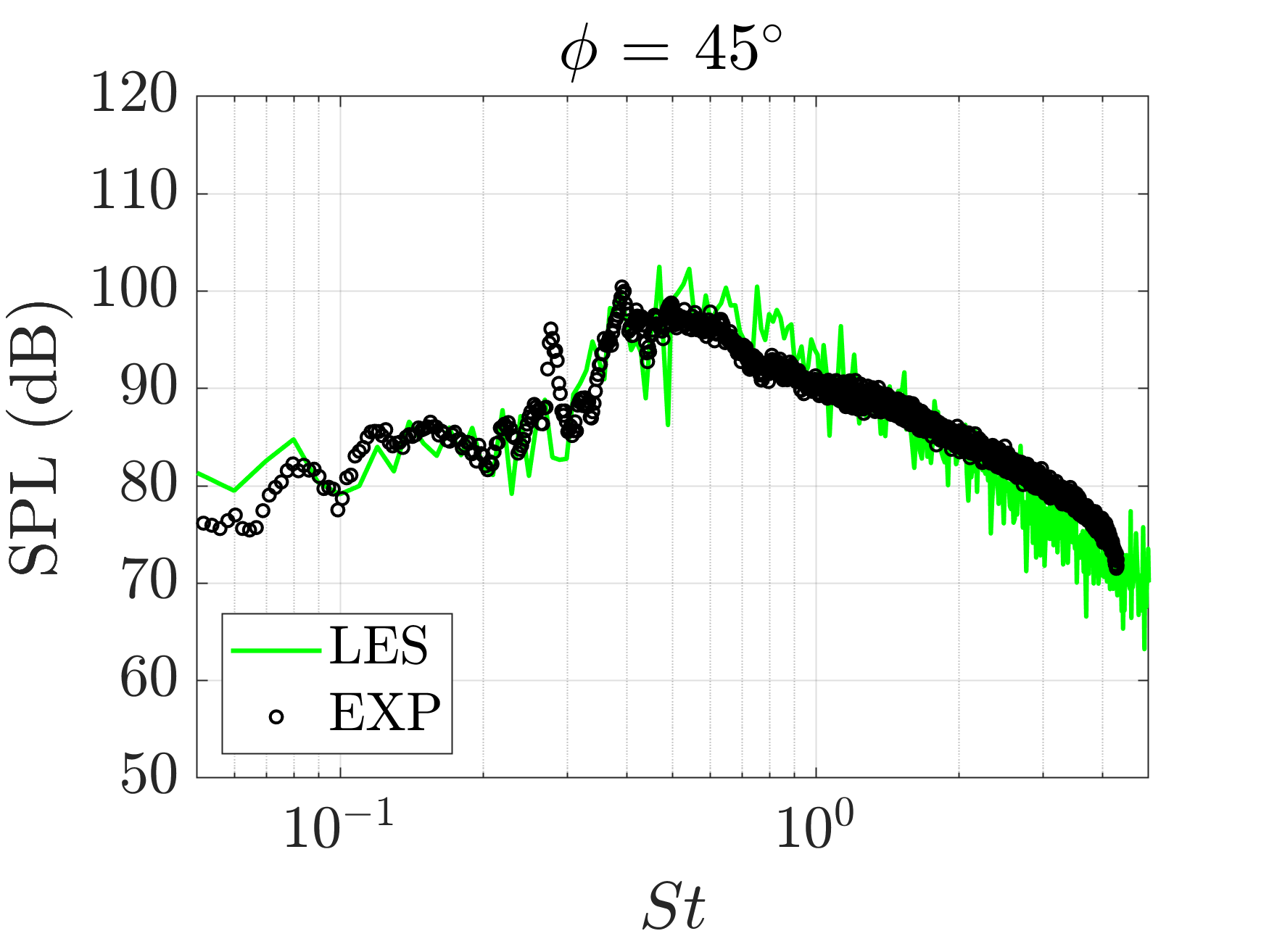} & \includegraphics[width=0.3\textwidth]{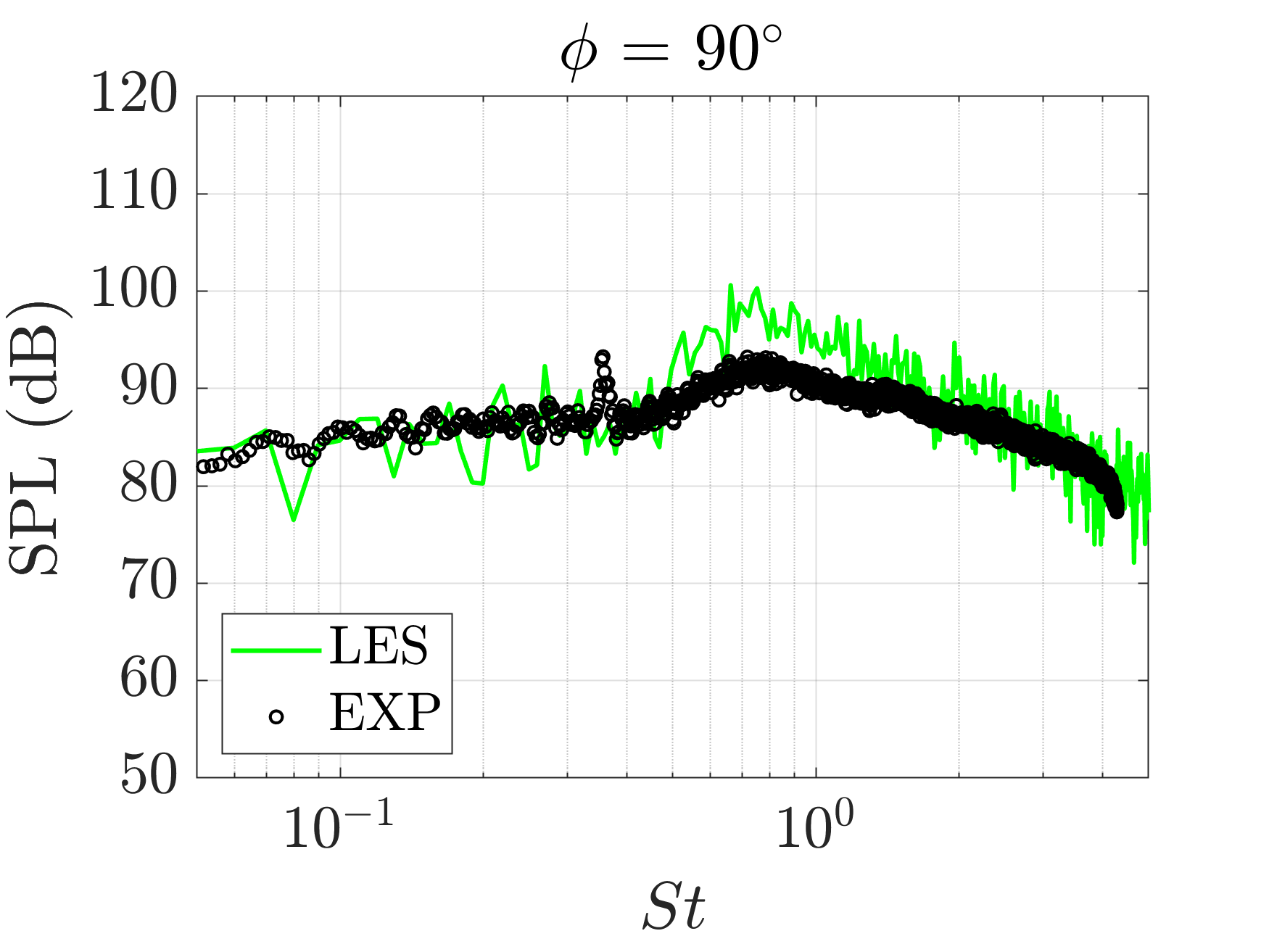} & \includegraphics[width=0.3\textwidth]{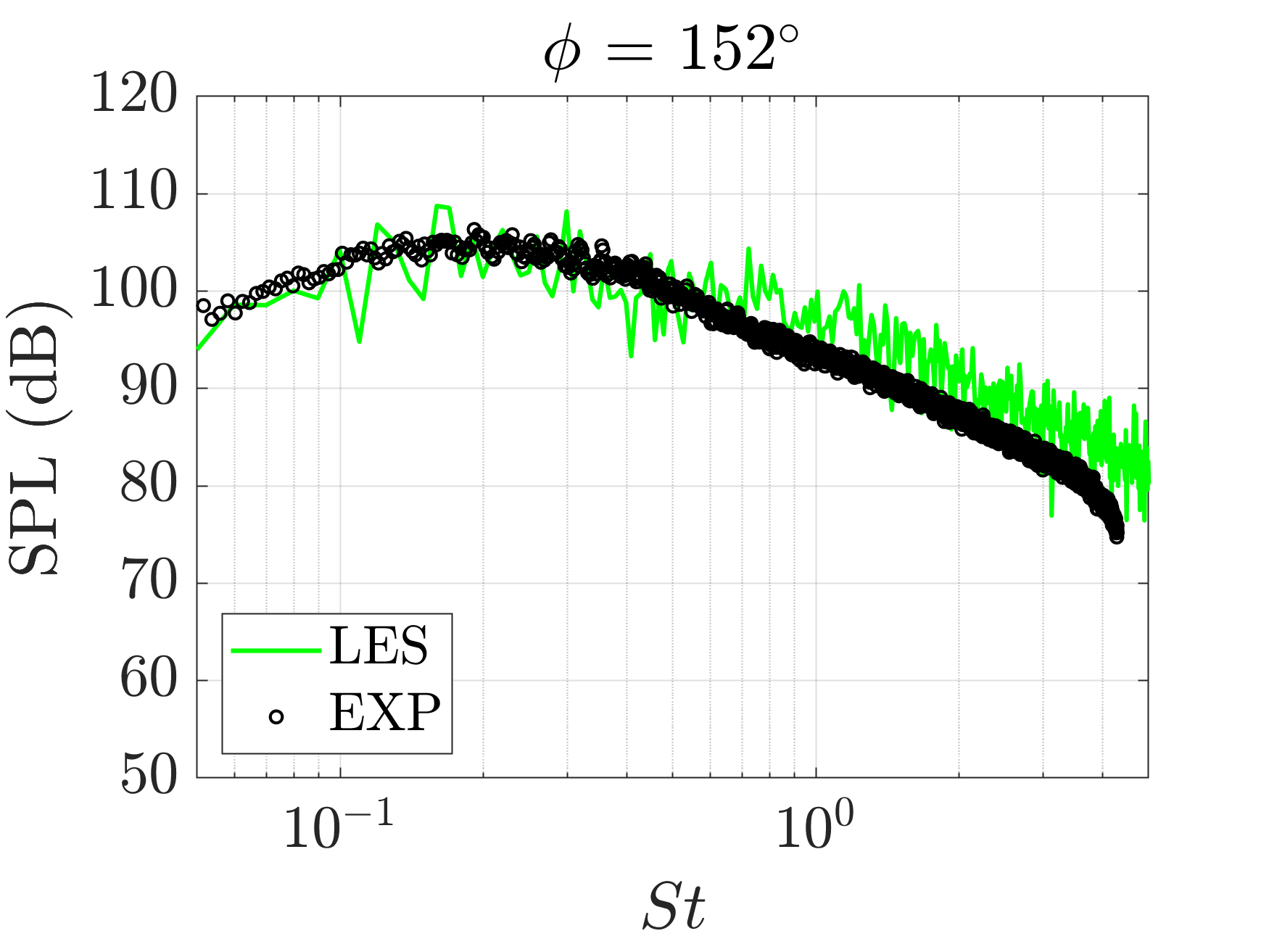} \\
  \end{tabular}
\caption{Comparison of the sound pressure levels between the LES (colored solid lines) and experiments (black circles) at $61.5D_e$ away from the nozzle exit: (top) in the major axis and (bottom) in the minor axis for NPR = 3.67.}
\label{fig:spl_npr3p67}
\end{figure}

Figures~\ref{fig:spl_npr2p5}-\ref{fig:spl_npr3p67} compare the sound spectra predicted from the LES (colored solid lines) with the experimental measurements (black circles). In each figure the narrow-band spectra predicted by LES are bin-averaged over a bandwidth $\Delta St = 0.01$ for low frequencies ($St < 0.5$) and $\Delta St = 0.015$ for high frequencies ($St > 1$) except the screech frequencies, for better comparison with smoother experimental measurements while avoiding damping of screech tones. Additionally, the lack of azimuthal homogeneity of the rectangular nozzles allows to use two-way symmetry only along each direction. 

Despite some deviations in the shock-cell strengths and locations discussed in the previous section, the broadband shock-associated noise and turbulent mixing noise predicted by the simulations agree very well with the experimental measurements for all three conditions. For NPR = 2.5, Fig.~\ref{fig:spl_npr2p5} shows that multiple screech tones are captured in the upstream angles by LES, but two adjacent screech frequencies in the minor axis are slightly off from the experiments ($St$ = 0.46 and 0.53). Between the two, the higher frequency tone is deemed to be the fundamental with the screech amplitude higher than the other by almost 10 dB in the minor axis. This frequency is also closely aligned with the screech frequency prediction based on Eqs.~\eqref{eq:fs} and~\eqref{eq:fs2}, as will be discussed in the following sub-section. The reason for the existence of two neighboring screech tones at this operating condition is currently unknown, but they do not seem to originate from the twin-jet coupling as similar observations were noticed in a single rectangular jet with the same aspect ratio~\cite{karnam2021}. At $\phi = 90^\circ$, the second harmonic tone starts appearing in the minor axis, registering much lower amplitude than those of the upstream tones. At $\phi = 152^\circ$, no significant screech tones are observed in LES, but several moderate tones are detected in the experimental data in both axes. Interestingly, some of them correspond to the two screech tones found in the upstream direction, and the others are shown to arise completely randomly.

The twin jets at NPR = 3 produce the loudest screech tones among the three conditions simulated, as seen in Fig.~\ref{fig:spl_npr3}. In the experiments (black circles) many fundamental and second harmonic tones are measured at all observer locations both in the major and minor axis directions. Fundamental screech tones are found to be very strong even at the downstream angles ($\phi$ = 152$^\circ$). In the minor axis direction a strongly damped fourth harmonic tone at $St = 1.11$ is also observed. The LES data (red solid lines) contain multiple fundamental and harmonic tones captured in the experiments, although in some cases their ampltidues are weaker than the measured values. At $\phi = 45^\circ$ in the minor direction, LES perfectly matches the fundamental screech tone in terms of both the frequency ($St$ = 0.37) and amplitude (almost 110 dB), but agreements are less favorable for higher harmonics. Tones in the major direction are under-predicted as well. At the sideline angle ($\phi = 90^\circ$) LES predicts the second harmonic tones only. The amplitude of the second harmonic tone is weaker than the measured value in the major axis direction, but it is predicted very accurately in the other direction. At much downstream ($\phi$ = 152$^\circ$) tones are completely missed in the LES data. The absence of screech tones in the LES results at these angles may be due to insufficient downstream grid resolutions pointed out in Sec.~\ref{subsubsec:mesh}, while the slight under-prediction of some screech tones may be attributed to insufficient simulation duration. The errors may also result from the mismatch in turbulent statistics in the near-field shock cell systems, although they do not affect the screech frequencies, in general.

As shown in Fig.~\ref{fig:spl_npr3p67}, the experimental data captures several moderate tones at upstream angles in both axes as well as in the sideline direction of the minor axis even at the design condition, whereas no distinctive tones are shown in the LES results. Although the screech tones detected in the experiment are not as appreciably strong as those observed in previous two conditions, the comparatively weak shock system is responsible for broadband noise recorded around the mid-frequencies.

\begin{figure}
\centering
  \begin{tabular}{ccc}   
    \includegraphics[width=0.3\textwidth]{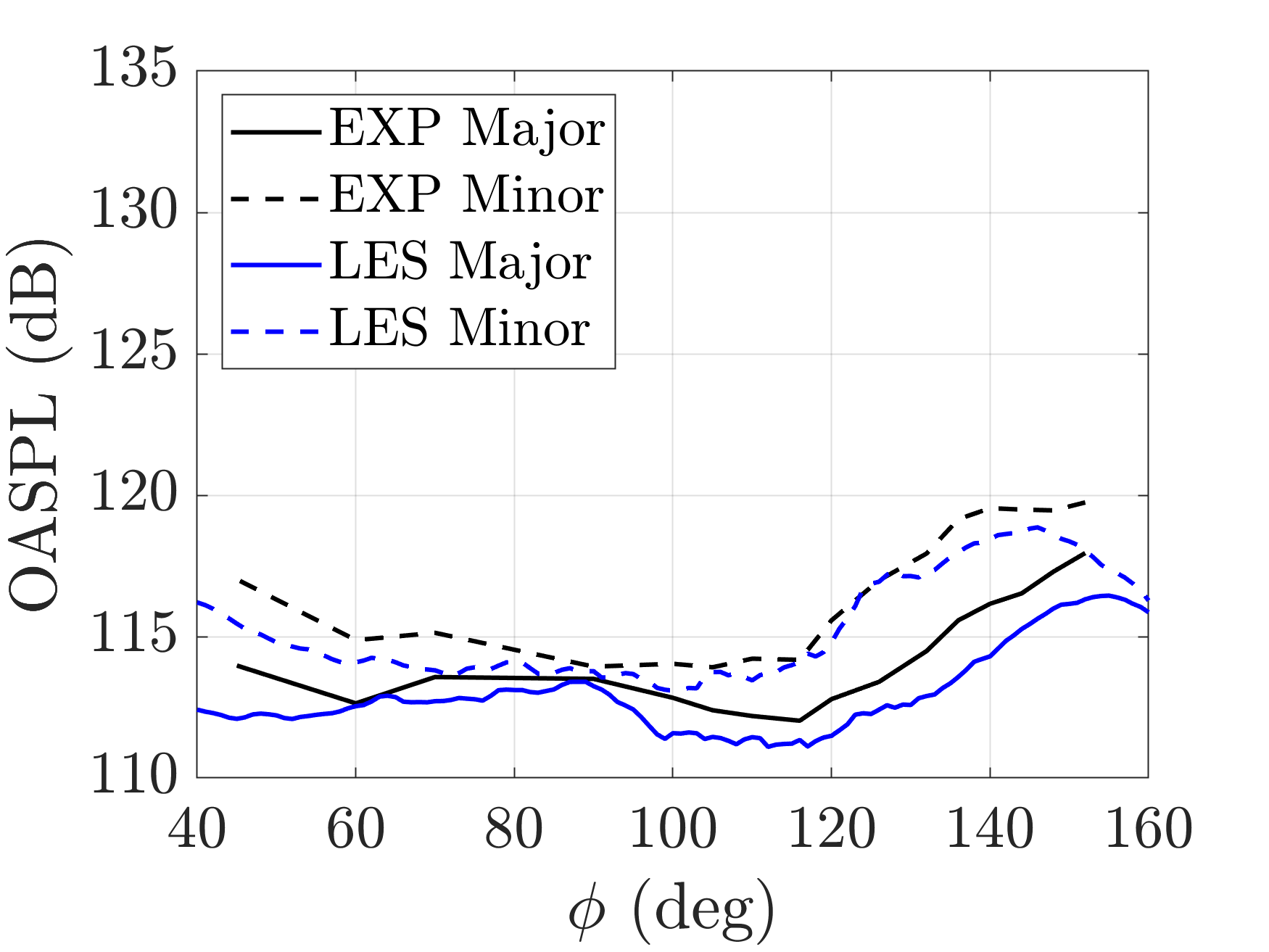} & \includegraphics[width=0.3\textwidth]{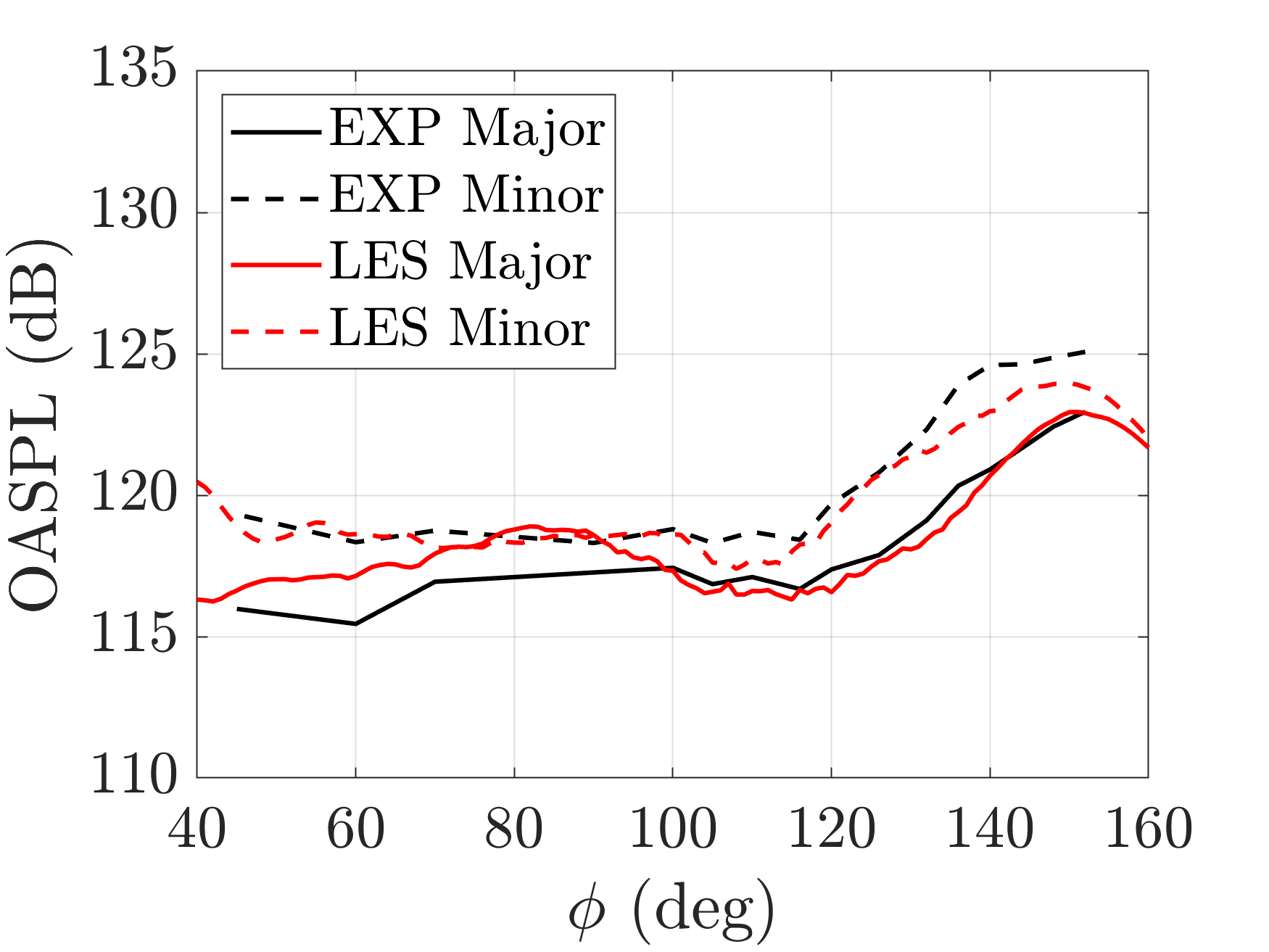} & \includegraphics[width=0.3\textwidth]{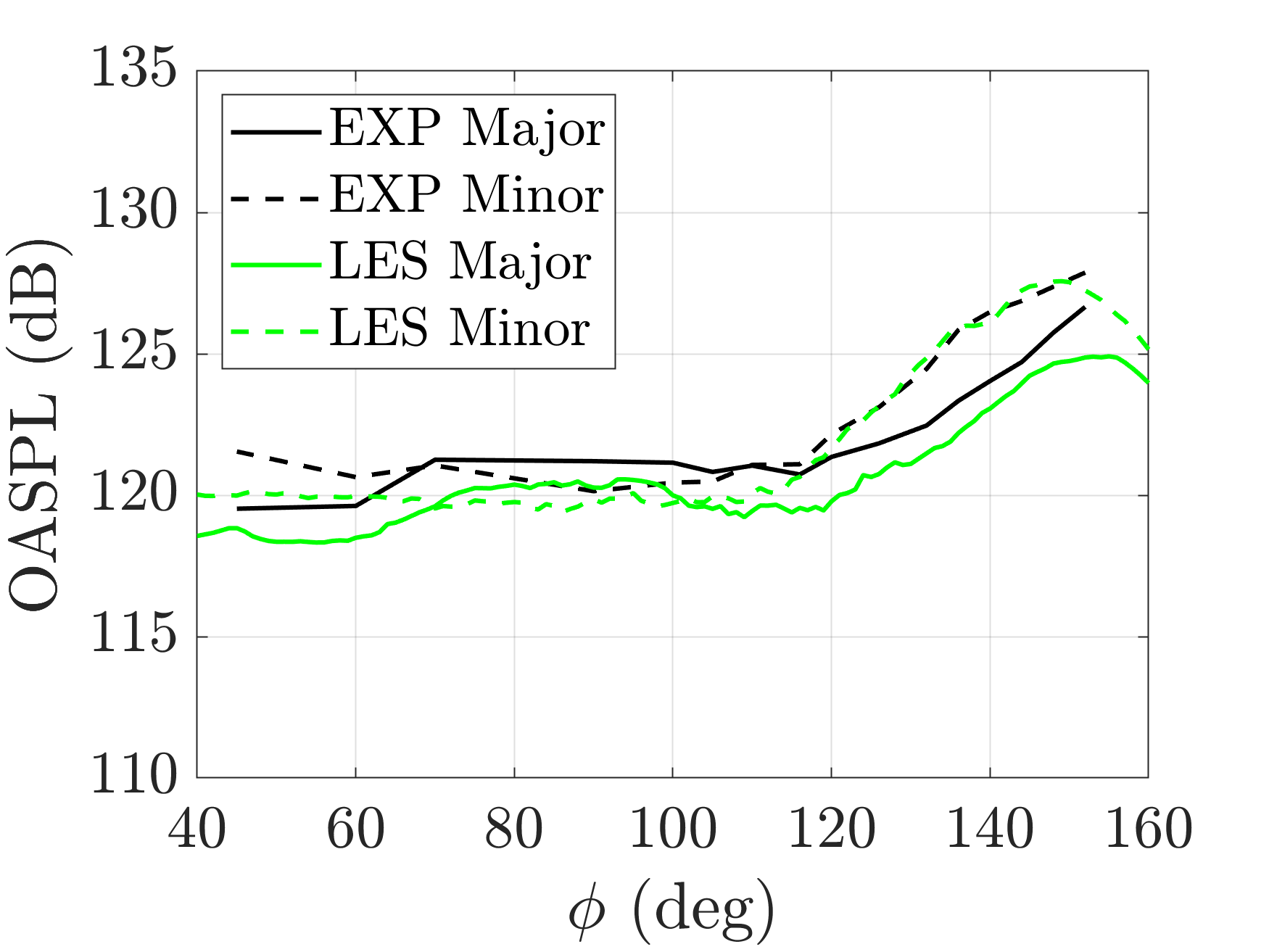} \\
    a) NPR = 2.5 & b) NPR = 3 & c) NPR = 3.67 \\
  \end{tabular}
\caption{Comparison of the overall sound pressure levels (OASPL) between the LES (colored solid lines) and experiments (black dashed lines).}
\label{fig:oaspl}
\end{figure}

The OASPLs are given in Figure~\ref{fig:oaspl} for all three conditions. While the true Nyquist limit of the simulations is approximately $St$ = 13, the frequency range used for computation of the OASPL is limited to, for example, $St$ = [0.023, 4.3] for NPR = 3, in line with the anechoic chamber threshold. In spite of several screech tones missing or slightly under-predicted in the individual spectra, the LES results agree very well with the experimental data with only 1-2 dB differences in both major and minor axis planes for all of the observer angles considered. Also note that the OASPL increases with increasing NPR, as expected and measured in the experiments.

\subsubsection{Screech frequency predictions}
It is known that screech frequencies of axisymmetric jets can be predicted pretty well based on the convective Mach number and appropriate length scales. For non-axisymmetric jets, Tam~\cite{tam1988} proposed a prediction formula based on the shock-cell spacing using a vortex sheet model as following:
\begin{equation}
    \frac{f h}{u_j} = \frac{u_c/u_j}{2(1+u_c/c_{\infty}) (M_j^2-1)^{1/2}} \left[ \left( \frac{h_j}{b_j} \right)^2 + 1 \right]^{1/2} \quad \Bigg\slash \quad \Bigg\{ \Bigg[ \left( \frac{1+\frac{\gamma-1}{2}M_j^2}{1+\frac{\gamma-1}{2}M_d^2} \right)^{\frac{(\gamma+1)}{2(\gamma-1)}} \frac{M_d}{M_j}-1 \Bigg] \frac{b}{b+h}+1 \Bigg\}.
\label{eq:fs}
\end{equation}
Here, the dimensions of jets at fully expanded conditions ($b_j$ and $h_j$) can be computed by Eqs.~\eqref{eq:b_j} and~\eqref{eq:h_j}. By assuming the convection velocity of the jets to be $0.7 U_j$ as in \citet{tam1988}, Table.~\ref{tab:screech_frequency} compares the screech frequencies estimated by LES, experiments, and Tam's model. It should be pointed out that Tam's formula was suggested for a single rectangular jet. Nevertheless, Table.~\ref{tab:screech_frequency} shows fairly good agreement between the estimation based on his model and the LES/experiments at the two over-expanded conditions, suggesting that the similar linear mechanism in a single jet can be applied to twin jets. 

On the other hand, by employing values computed by LES, the screech frequencies can be estimated by some characteristic length scale ($L$) and the convection velocity ($u_c$) such as: 
\begin{equation}
    f = \frac{u_c}{L(1+M_c)},
    \label{eq:fs2}
\end{equation}
where $M_c$ = $u_c/c_\infty$ is the convective Mach number ~\cite{powell1953,tam1986}. By using the average shock-cell spacing in Table~\ref{tab:spacing} as the relevant length scale and setting $u_c$ = $0.55U_j$ as computed in Sec.~\ref{subsubsec:convection_vel}, the screech frequency can be predicted as shown in the rightmost column of Table~\ref{tab:screech_frequency}. Based on the convection velocity estimated appropriately along the centerline of the jet shear layer, the predicted screech frequencies match quite well with those captured in the far-field acoustic spectra from the LES (Figs.~\ref{fig:spl_npr2p5} and~\ref{fig:spl_npr3}) at both off-design conditions.

\begin{table}
\centering
\caption{Screech frequency prediction using Eqs.~\eqref{eq:fs} and \eqref{eq:fs2}}
\begin{tabular}{ccccc}
\hline\hline
  NPR & LES & Experiments~\cite{karnam2020} & Prediction by Eq.~\eqref{eq:fs}~\cite{tam1988} & Prediction by Eq.~\eqref{eq:fs2}~\cite{powell1953,tam1986} \\ 
\hline
  2.5 & 0.518 & 0.533 & 0.521 & 0.538\\
  3 & 0.373 & 0.373 & 0.373 & 0.371\\
\hline\hline
\end{tabular}
\label{tab:screech_frequency}
\end{table}

\subsubsection{Intermittent nature of over-expanded twin-jet screech}
\label{subsubsec:intermittency}
In the authors' previous publication~\cite{jeun2021,karnam2021} we identified the coupling modes of the two over-expanded jets by calculating two-point spatial cross-correlations. In the minor axis  direction, both the LES and experiments captured anti-symmetric coupling between the twin jets. Intense jet flapping motions in the minor axis of rectangular jets aided clear illustration of the anti-symmetric coupling in this case. In the major axis direction, the cross-correlation modes looked somewhat complicated. The choice of the reference point influenced the resulting modes, and they were much less ordered compared to those in the minor axis view. In some cases LES extracted anti-symmetric modes also in the major axis direction, but the experiments did not provide clear evidence of out-of-phase coupling. These observations have drawn a question about the nature of screech tone, leading to the time-frequency analysis of the pressure measurements to characterize its intermittency. 

\begin{figure}[t]
\centering
  \begin{tabular}{cc}   
    \includegraphics[width=0.4\textwidth]{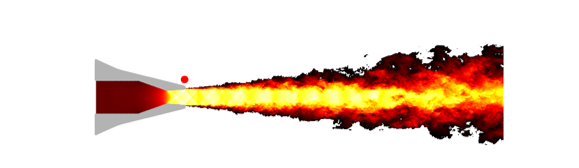} & \includegraphics[width=0.4\textwidth]{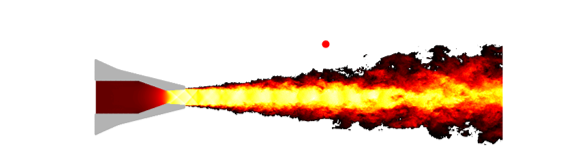} \\
    \includegraphics[width=0.4\textwidth]{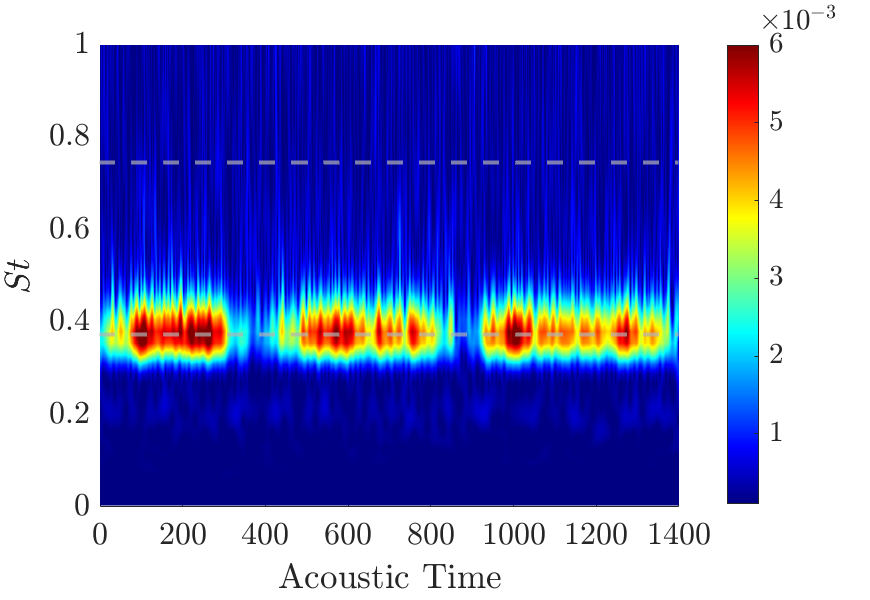} & \includegraphics[width=0.4\textwidth]{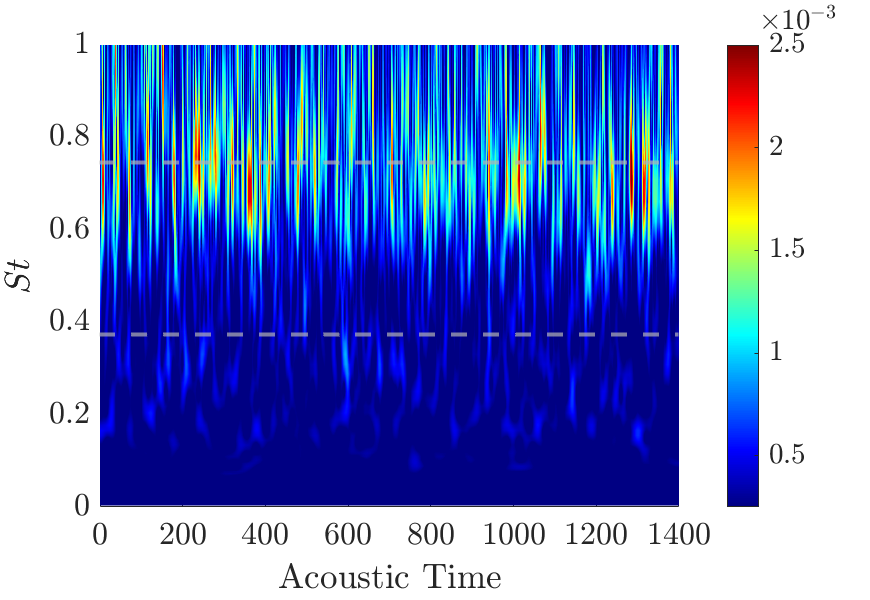} \\
    \includegraphics[width=0.4\textwidth]{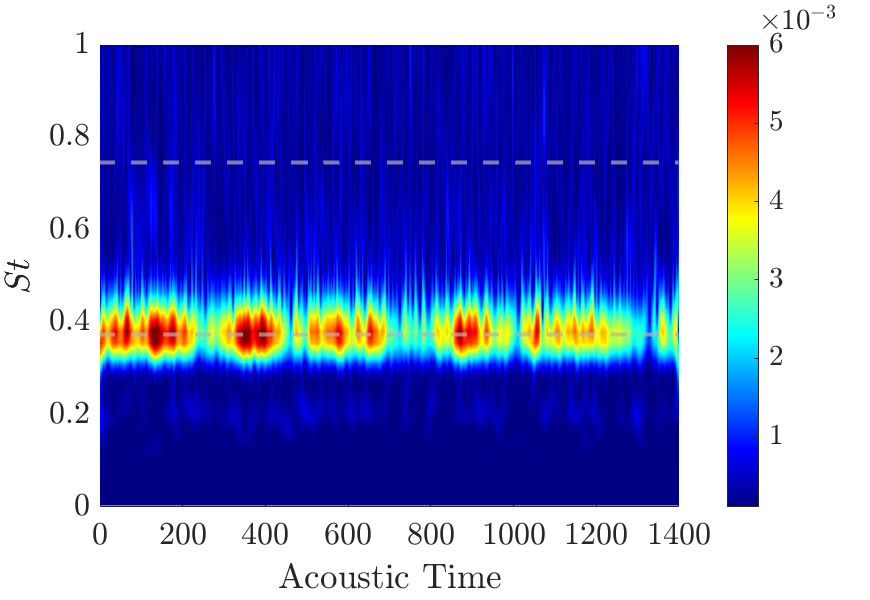} & \includegraphics[width=0.4\textwidth]{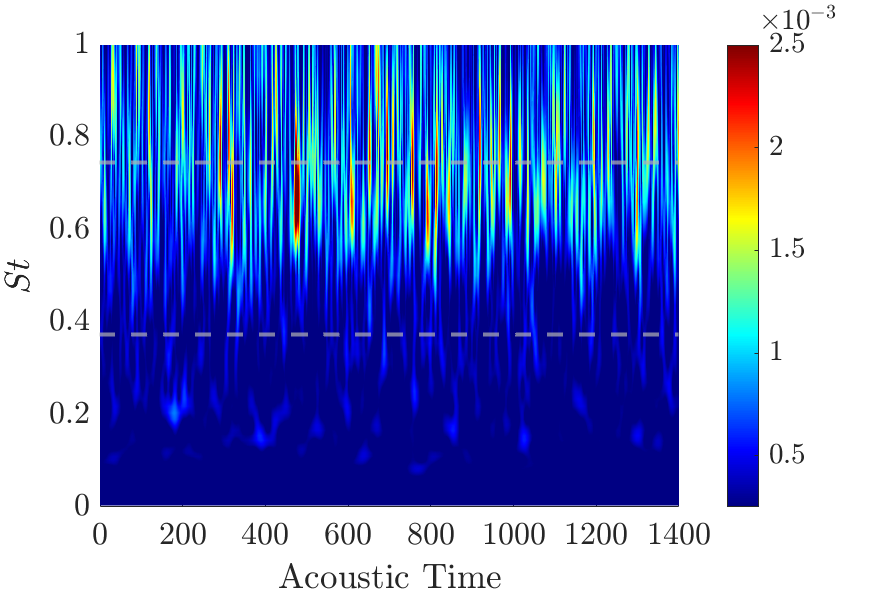} \\
    a) Probe at $(x,y,z)$ = $(0h,1h,-1.75h)$ & b) Probe at $(x,y,z)$ = $(8h,5h,-1.75h)$ \\
  \end{tabular}
\caption{Scalograms of the acoustic signals for NPR = 3: (top) Jet 1 centered at $z/h$ = 1.75 and (bottom) Jet 2 centered at $z/h$ = -1.75. White dashed lines indicate the fundamental and second harmonic screech frequencies. Red dots above the jets represent the probe locations. Adapted from~\cite{jeun2021b}.}
\label{fig:scalogram}
\end{figure}

We apply the continuous wavelet transform of the pressure data using the Morse wavelet. As a family of complex wavelets, it can provide both amplitude and phase information. To examine the fundamental tone ($f_s$), the probe is set to a point just above one jet exit, i.e., $(x,y,z)$ = $(0h,1h,\pm1.75h)$. At this position, the signal is assumed to be almost purely acoustic. The harmonic tone ($2f_s$) is known to radiate in the sideline so another probe is placed at $(x,y,z)$ = $(8h,5h,\pm1.75h)$ outside of the jet turbulence. The data are collected with the same sampling frequency used in the far-field acoustic extrapolation. The resulting scalograms, which visualize the absolute value of the continuous wavelet transform coefficient, are shown in Fig.~\ref{fig:scalogram}. White dashed lines correspond to the fundamental and second harmonic screech frequencies in order. The scalograms support that both the fundamental and harmonic tones are indeed intermittent, but the second harmonic tone appears to be severely more irregular than the fundamental screech. It is also shown that the intermittency patterns for the two jets are not the same. Such intermittent screech tones seem to be associated with the switching of coupling modes between the two jets~\cite{jeun2021b,bell2021}. 

\subsection{SPOD analysis and twin-jet coupling modes}
\label{subsec:spod}
Spatial cross-correlation analysis is simple yet powerful to determine the coupling modes of the twin jets~\cite{knast2018,jeun2021}. Nevertheless, this approach cannot resolve multiple modes that possibly exist at many different frequencies and is not suitable to investigate mode-switching of jet screech. In this regard we perform SPOD analysis using various flow variables data obtained from the high-fidelity simulations to examine dominant coherent flow structures at frequencies of interest, i.e., the fundamental screech frequency and its harmonics. SPOD analysis is therefore limited only to the two over-expanded cases which include (relatively) strong screech. However, it is found that SPOD analysis of the NPR = 2.5 jets produces energy spectra and mode shapes that greatly resemble those yielded for NPR = 3. Hence, we discuss SPOD analysis for the NPR = 3 twin jets only, as this case generates much louder screech tones.

The probe points are distributed in either the minor or major axis planes with a uniform spacing of $\Delta x$ = $\Delta y$ = $\Delta z$ = 0.05$h$. The probe planes range from $x/h$ = 0 to 30 in the streamwise direction and from -5 to 5 in the cross-streamwise directions, respectively. The sampling frequency in the experiments can be recovered by applying Hann windows over several blocks of the LES data with 75\% overlapping, in a similar way we perform post-processing of the FW-H data. The flow data are collected from these planes over 1,400 acoustic time units, totaling six SPOD modes at each frequency.

To identify 3D coupling modes of the twin jets, SPOD analysis is performed in both the major and minor axis planes. First, in the minor axis plane taken at each jet center ($z/h$ = $\pm$1.75), SPOD analysis is applied to the fluctuating pressure and transverse ($y$-) velocity data. In these cases SPOD analysis is expected to produce dominant leading modes, which represent flow structures related to the strong screech tones and jet flapping motions influenced by the twin-jet coupling in the minor direction. In contrast, SPOD analysis based on the out-of-plane ($z$-) fluctuating velocities may not be very useful to investigate the coupling modes in this direction.

\begin{figure}
\centering
  \begin{tabular}{cc}
  \includegraphics[width=0.4\textwidth]{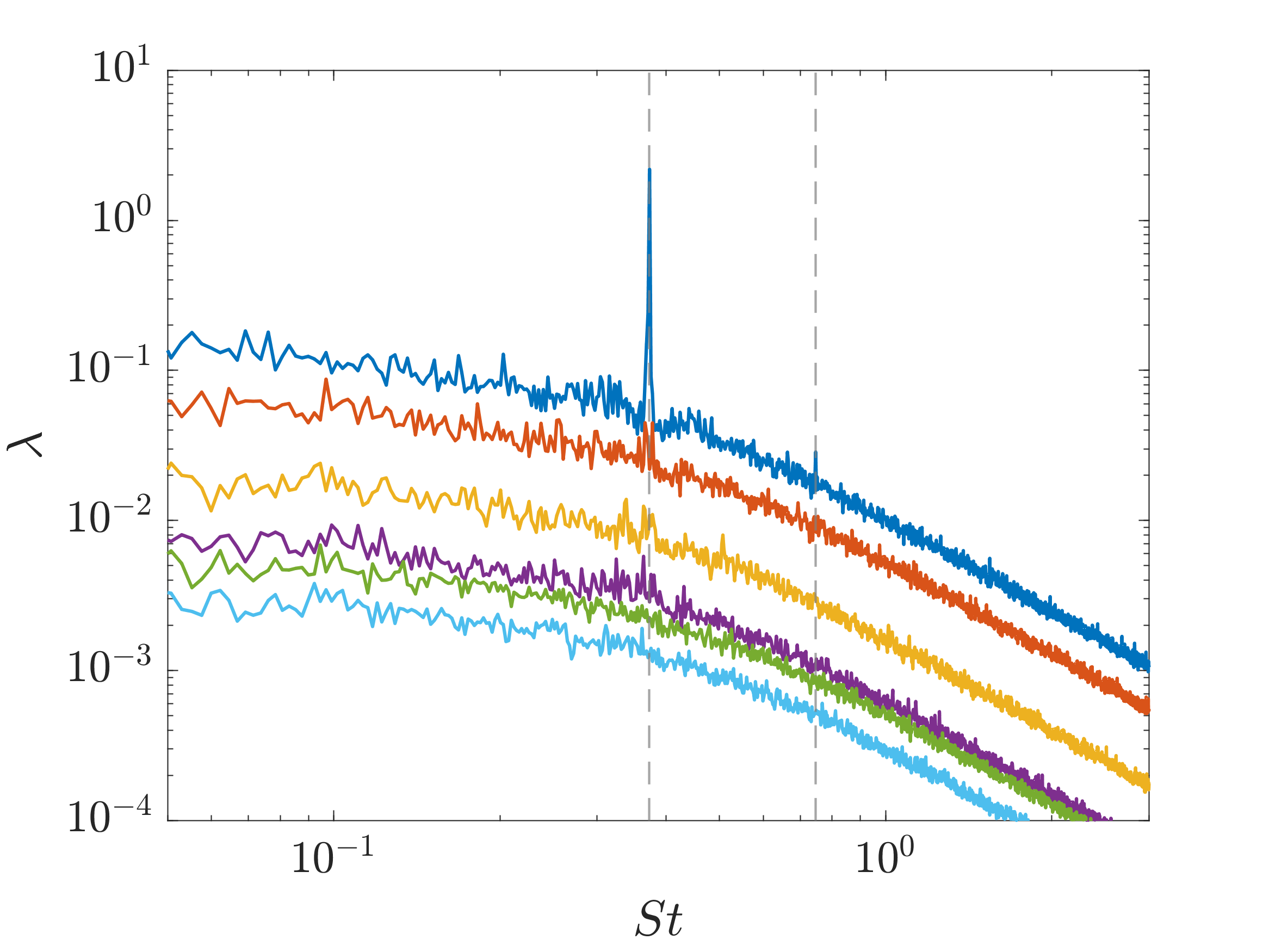} & \includegraphics[width=0.4\textwidth]{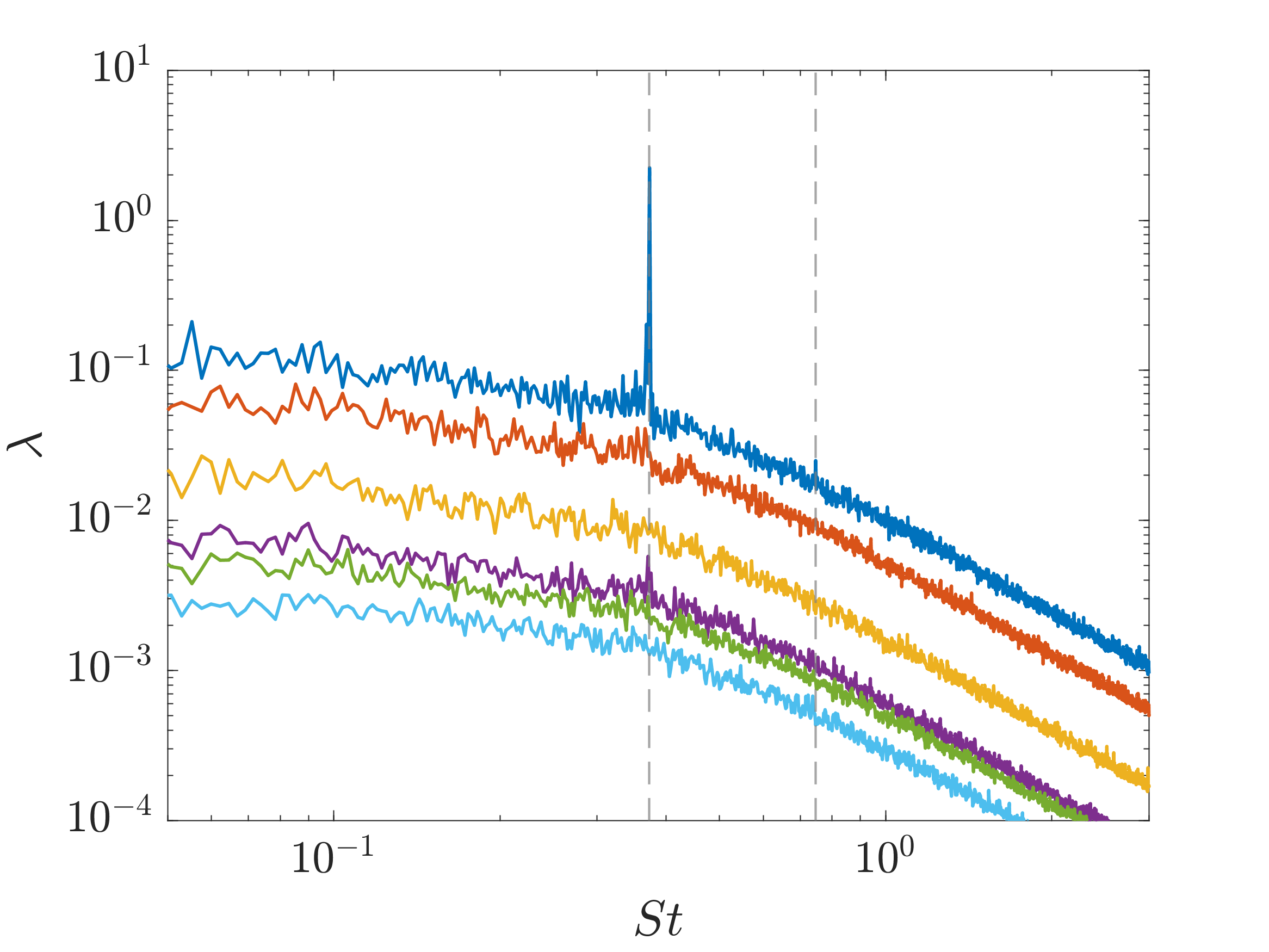} \\
  a) Jet 1 (centered at $z/h = 1.75$) & b) Jet 2 (centered at $z/h = -1.75$) \\
  \end{tabular}
  \caption{SPOD energy spectra obtained using pressure fluctuations for NPR = 3. Gray dashed lines indicate the fundamental and second harmonic screech frequencies.}
\label{fig:p_spod_spectra_xy}
\end{figure}

Figure~\ref{fig:p_spod_spectra_xy} visualizes SPOD eigenvalue ($\lambda$) spectra based on the pressure fluctuations for NPR = 3. Here, the fundamental and second harmonic screech frequencies are marked by gray dashed lines. For both jets, intense peaks are registered at the fundamental screech frequency ($St = 0.37$) in the leading SPOD modes. Additional peaks are found at the second harmonic ($St = 0.74$), but their amplitudes are almost two orders of magnitudes smaller than those at the fundamental frequency. As shown in Fig.~\ref{fig:scalogram} the wavelet analysis reveals that the screech tones corresponding to the second harmonic arise more irregularly in time, which may result in small SPOD eigenvalues at this frequency. Compared to the leading mode, eigenspectra of the higher modes look smooth, monotonically decreasing as the frequency increases. SPOD analysis using the fluctuating minor axis velocity ($v'$-SPOD) yields similar energy spectra with particularly amplified leading SPOD modes at $St$ = 0.37 (not shown for simplicity), although the ratio of the leading to second mode eigenvalues at this frequency is a bit smaller than the value computed in the SPOD analysis of the fluctuating pressure ($p'$-SPOD). 

\begin{figure}
\centering
\begin{tabular}{cc}
  \includegraphics[width=.45\textwidth]{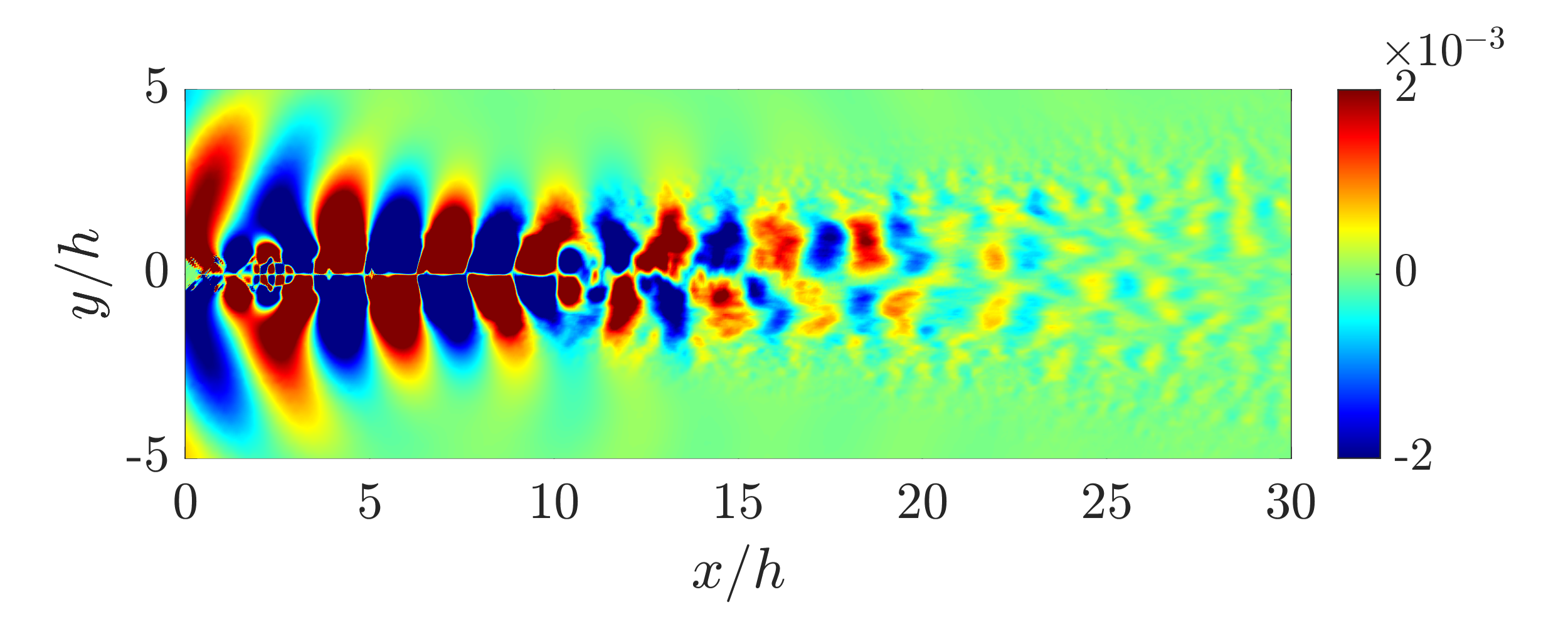} & \includegraphics[width=.45\textwidth]{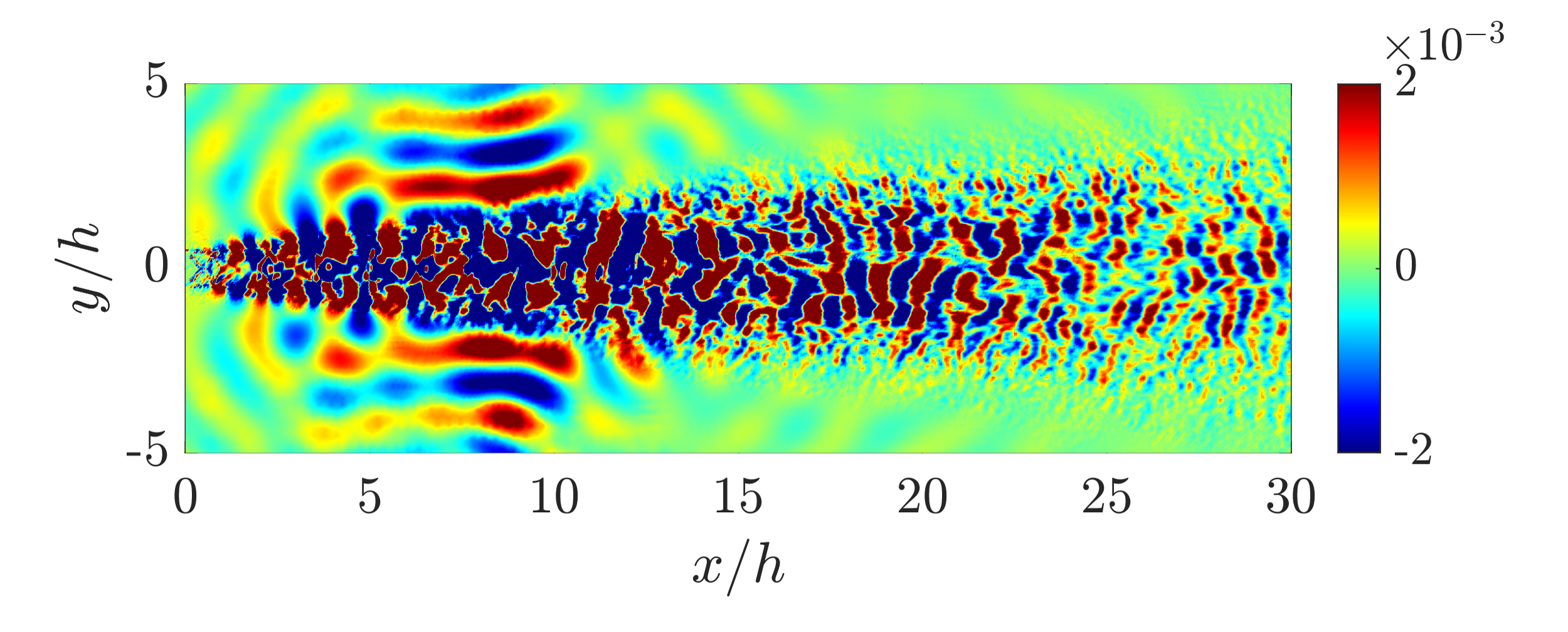} \\
  \includegraphics[width=.45\textwidth]{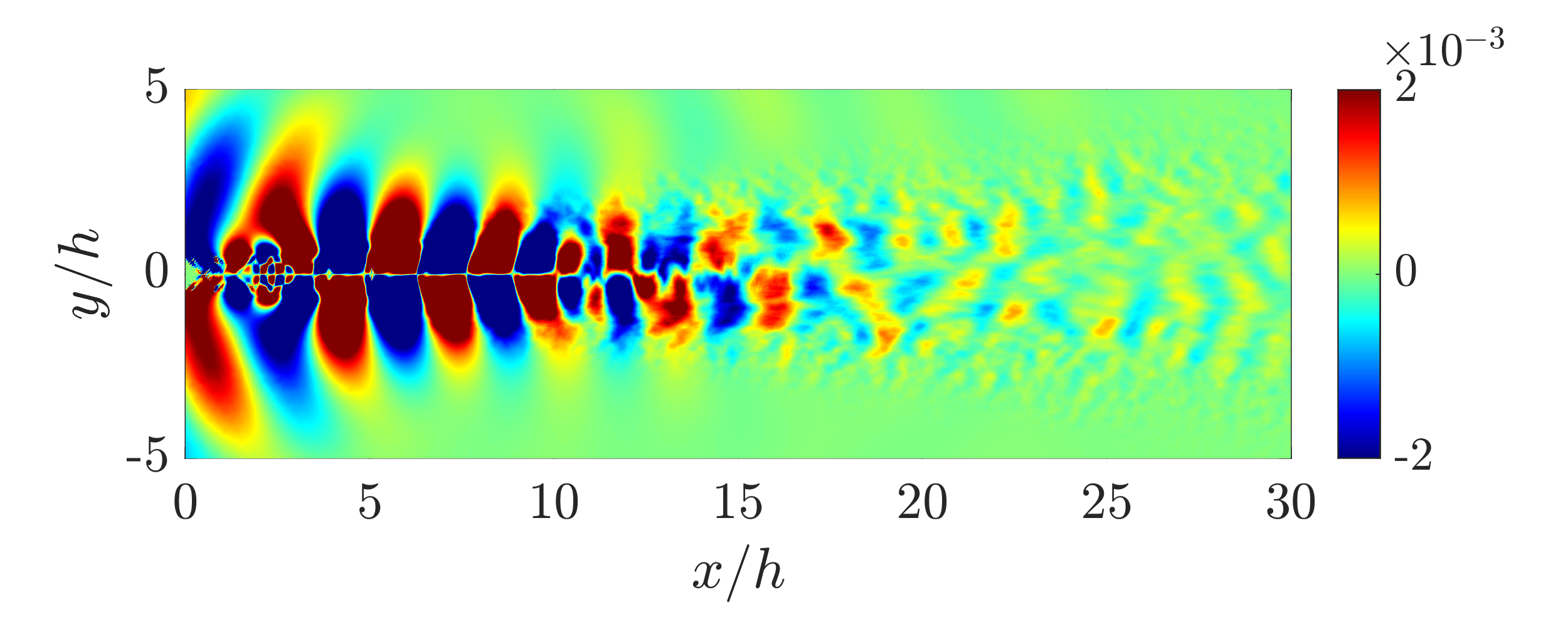} & \includegraphics[width=.45\textwidth]{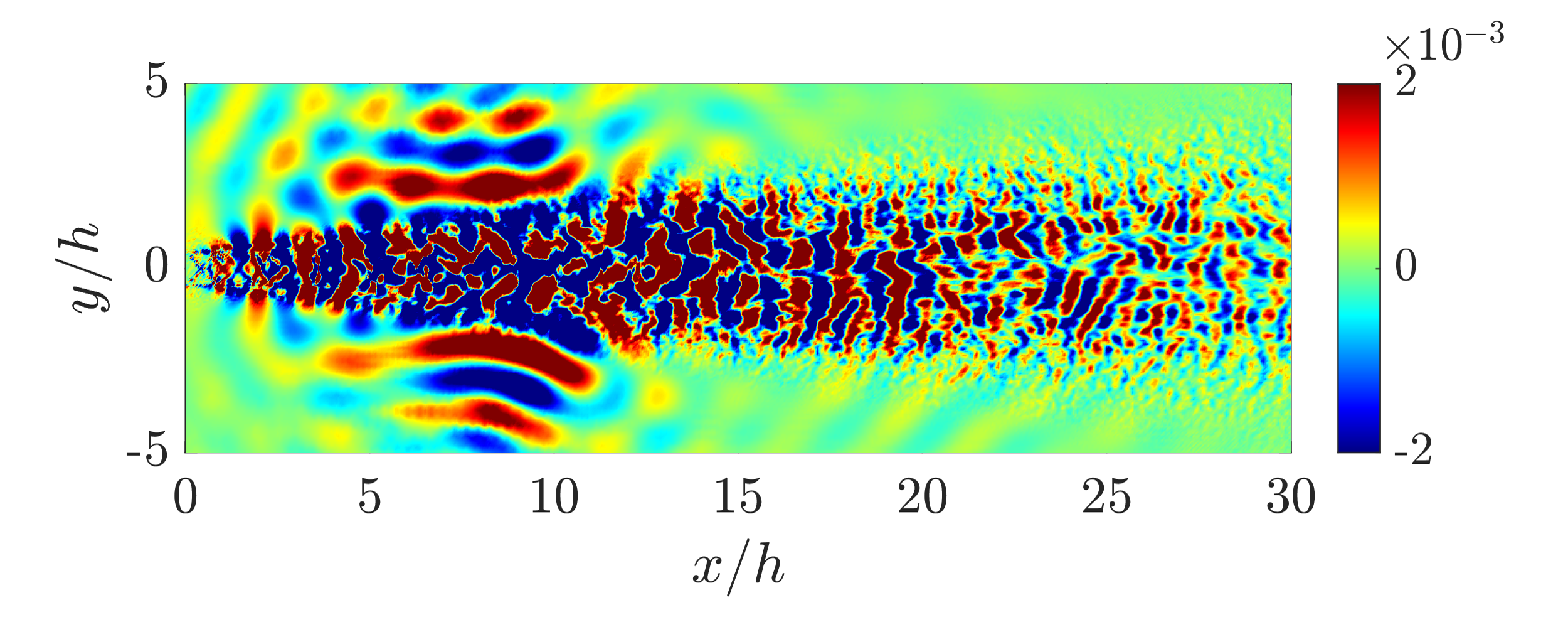} \\
  a) $St = 0.37$ & b) $St = 0.74$ \\
\end{tabular}
\caption{The leading SPOD modes visualized by the real part of the corresponding SPOD eigenfunctions computed by pressure fluctuations for NPR = 3: (top) Jet 1 centered at $z/h$ = 1.75 and (bottom) Jet 2 centered at $z/h$ = -1.75.}
\label{fig:p_spod_modes_xy}
\end{figure}

\begin{figure}
\centering
\begin{tabular}{cc}
  \includegraphics[width=.45\textwidth]{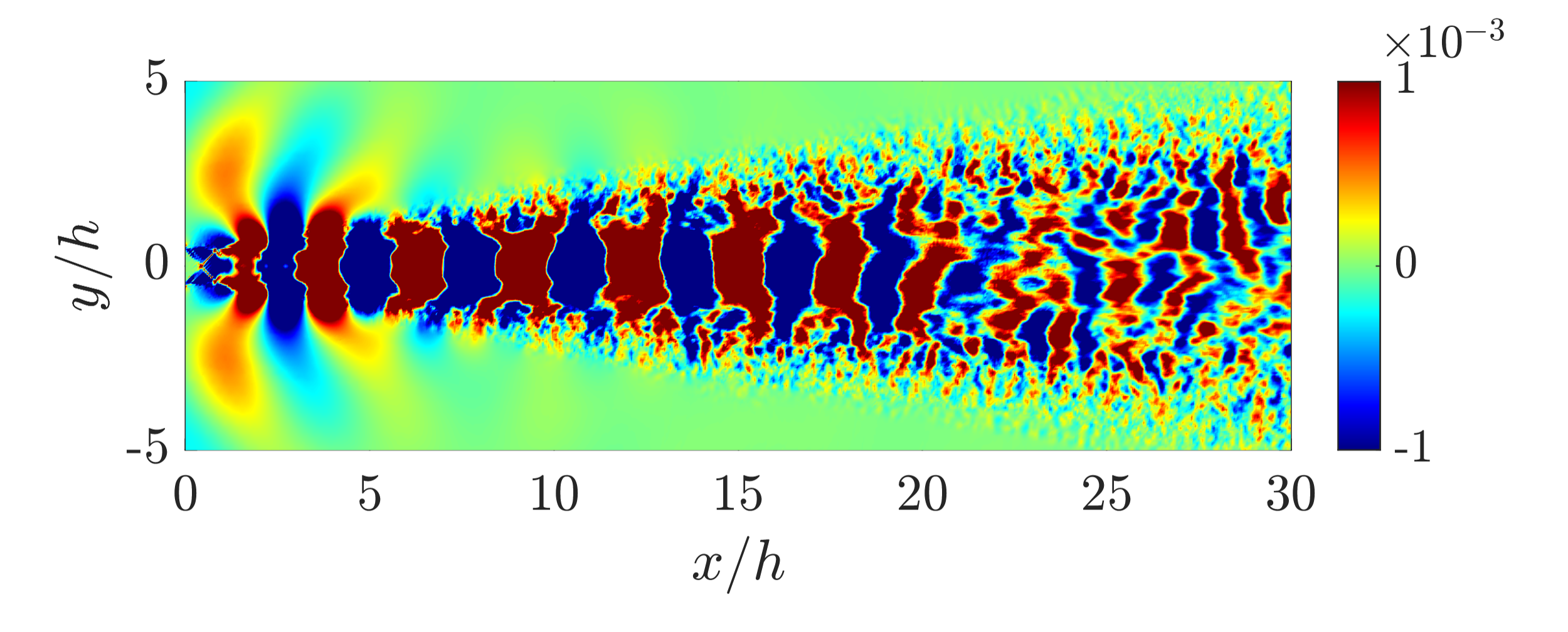} & \includegraphics[width=.45\textwidth]{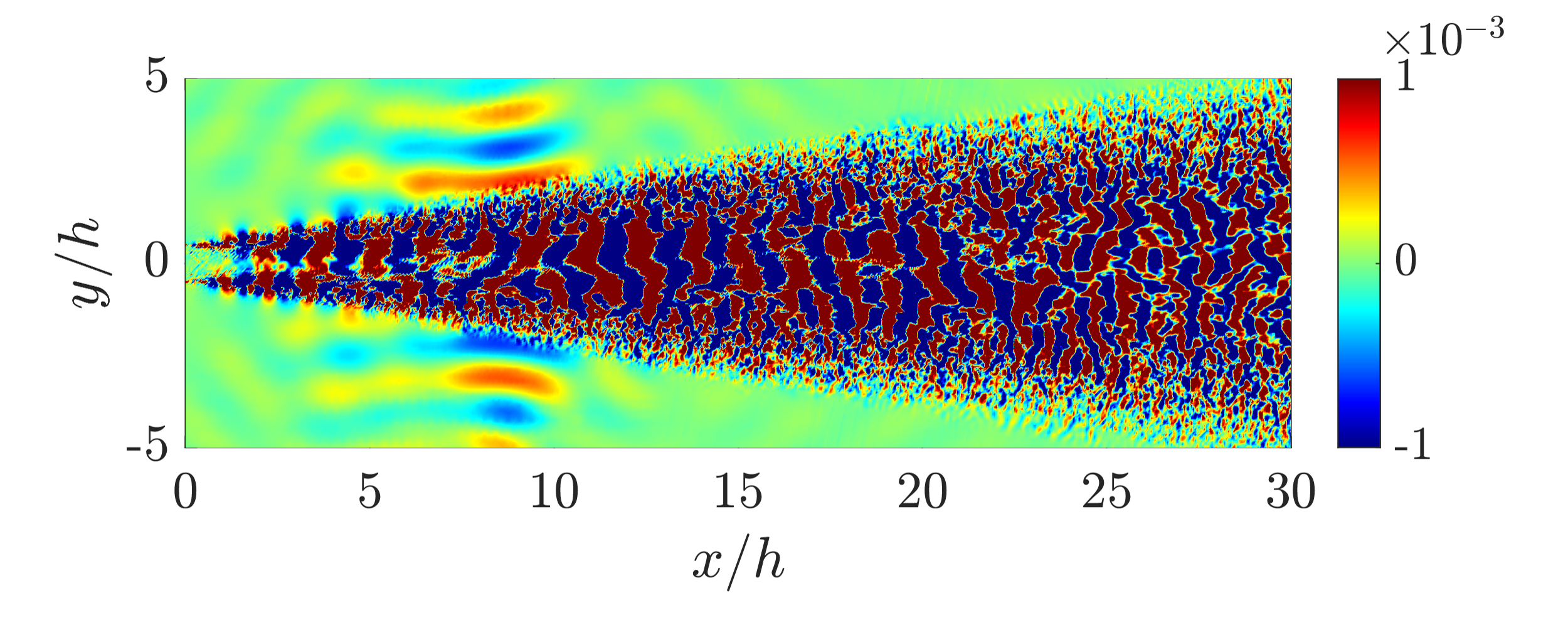} \\
  \includegraphics[width=.45\textwidth]{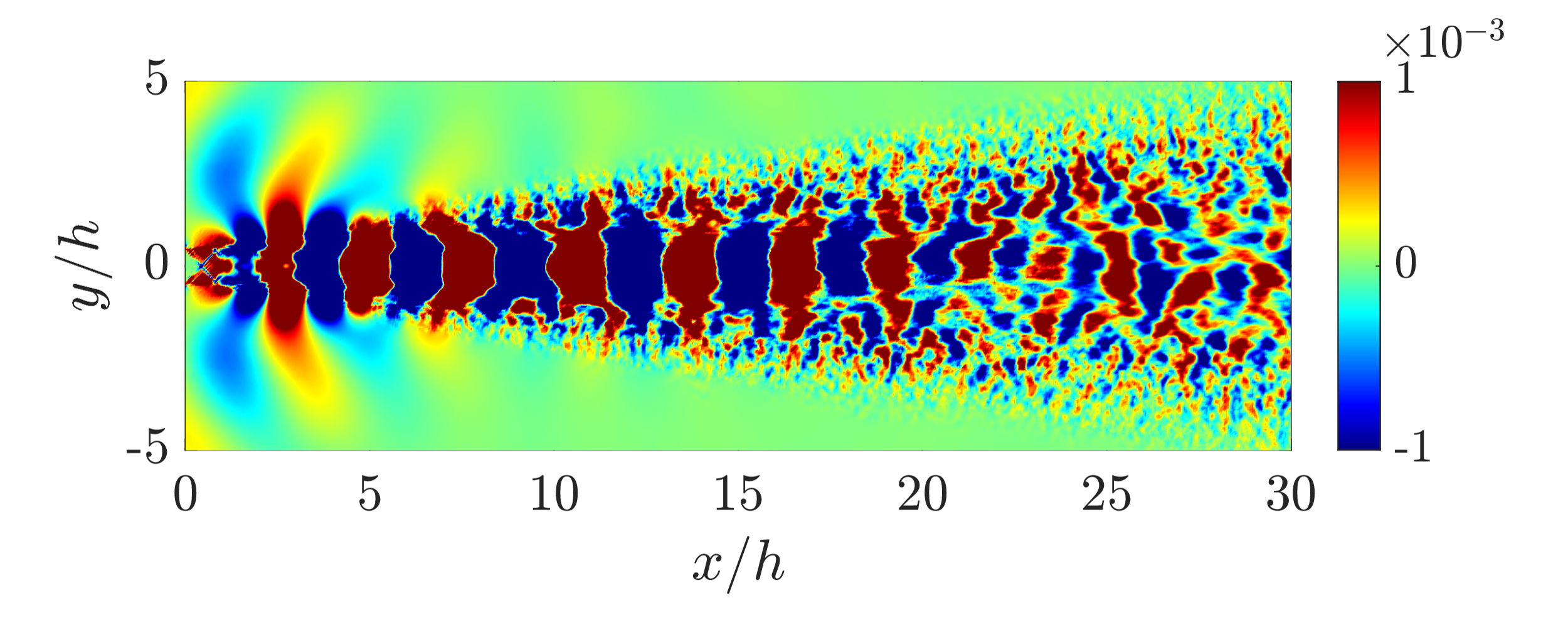} & \includegraphics[width=.45\textwidth]{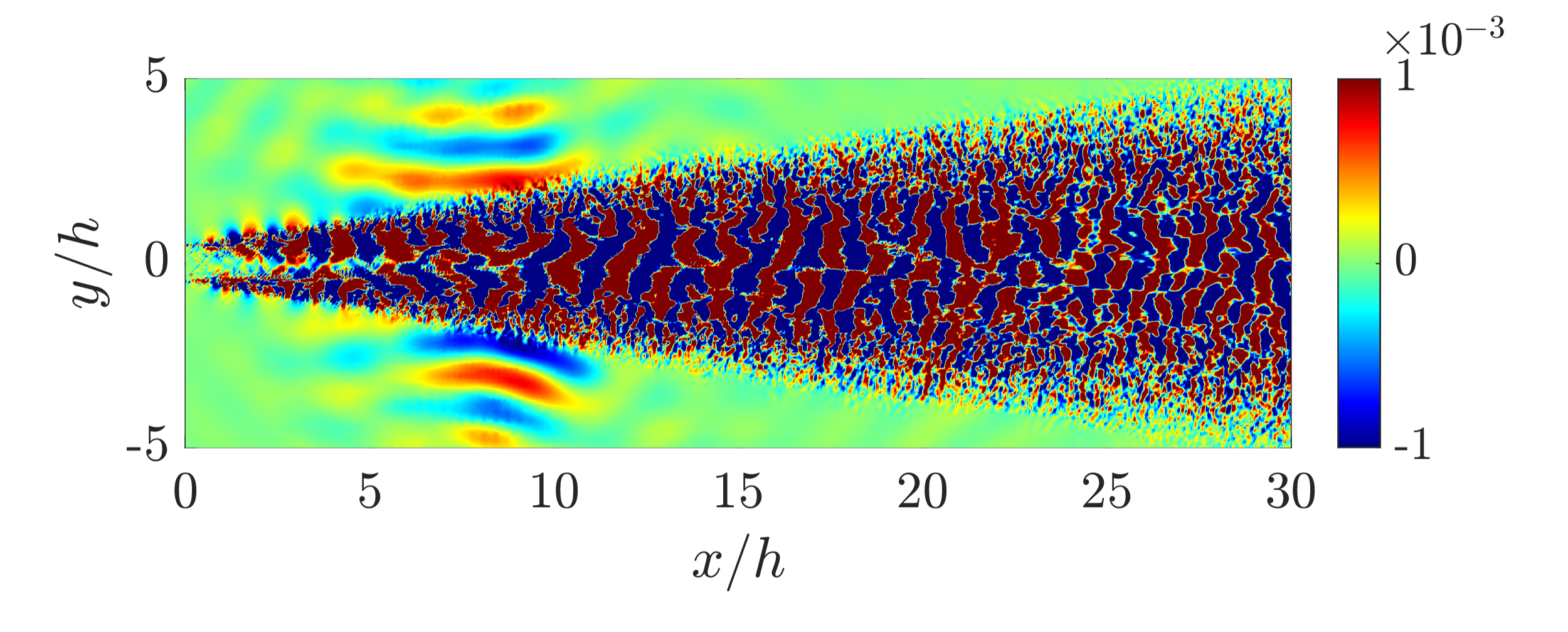} \\
  a) $St = 0.37$ & b) $St = 0.74$ \\
\end{tabular}
\caption{The leading SPOD modes visualized by the real part of the corresponding SPOD eigenfunctions computed by velocity fluctuations in $y$ for NPR = 3: (top) Jet 1 centered at $z/h$ = 1.75 and (bottom) Jet 2 centered at $z/h$ = -1.75.}
\label{fig:uy_spod_modes_xy}
\end{figure}

Figures~\ref{fig:p_spod_modes_xy} and~\ref{fig:uy_spod_modes_xy} show the leading SPOD modes visualized by the real part of the corresponding SPOD eigenfunctions using the pressure and transverse velocity fluctuations, respectively. At the fundamental frequency ($St$ = 0.37), the dominant coherent structures are mostly contained between $x/h$ = 0 and 20 in the streamwise direction for both cases. These modes consist of upstream- and downstream-travelling wave components as well as internal modes confined within the jet cores. The upstream radiation seems to originate between $x/h$ = 8 and 13. This location corresponds to the fifth or sixth shock-cells. Along the jet boundaries around this region, standing wave patterns~\cite{panda1997} are observed. The leading $p'$-SPOD modes are anti-symmetric with respect to the jet centerlines ($y/h$ = 0), whereas they are symmetric in the leading $v'$-SPOD modes. These patterns are reminiscent of the SPOD modes obtained for a single rectangular jet with an aspect ratio of 4:1, which also flaps in the minor direction~\cite{wu2020a}, and agree with the modes computed with a vortex-sheet model as well. From the comparison between the two jets (both in Figs.~\ref{fig:p_spod_modes_xy} and~\ref{fig:uy_spod_modes_xy}), we notice that they are out-of-phased with each other at $St$ = 0.37. 

At its harmonic ($St$ = 0.74), the modes are stretched much farther downstream, also peaking around somewhere between $x/h$ = 8 and 13. Unlike at the fundamental frequency, the two jets appear to be almost symmetric. In Fig.~\ref{fig:p_spod_modes_xy}, while the fundamental modes exhibit upstream-propagating wave components, the acoustic beams radiate in the sideline direction in this case. Yet, it is not aligned exactly at 90$^{\circ}$ as in a single rectangular jet~\cite{wu2020a}. Oscillations of the tails of jet plumes (15 < $x/h$ < 25) in the minor direction captured at $St$ = 0.37 (see Fig.~\ref{fig:uy_spod_modes_xy}) are less obvious at its harmonics.

To determine the coupling modes in the major axis, SPOD analysis can be employed onto the major axis plane ($y/h$ = 0) that contains both jets, using the pressure or transverse ($z$-) velocity components. However, due to the strong jet turbulence generated by the two closely placed nozzles that occupies most of the limited spatial domain in the cross-stream direction, the resulting energy spectra become broadened and the corresponding SPOD modes contain many misaligned lobes, which make it difficult to decide the coupling mode (not shown). Inter-nozzle region is also not resolved very well in this case.

\begin{figure}
\centering
\begin{tabular}{c}
  \includegraphics[width=0.45\textwidth]{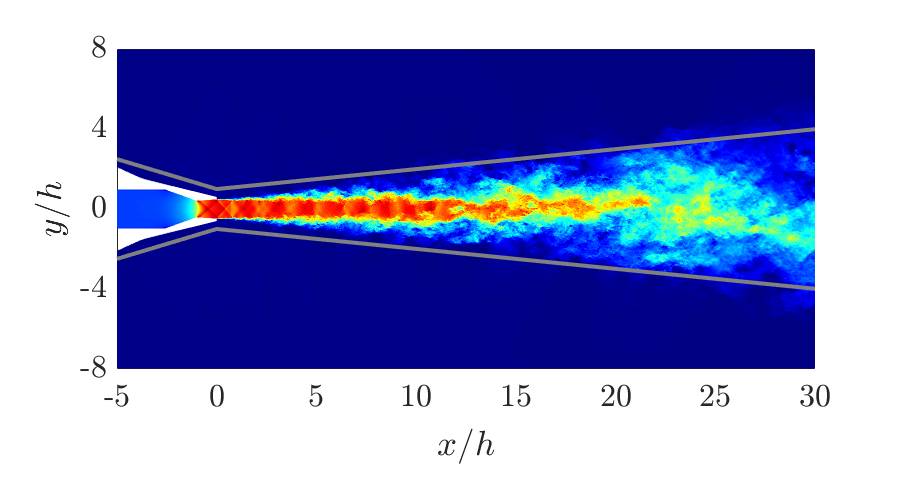} \vspace{-2mm}\\
\end{tabular}
\caption{Illustration of the additional probe planes inclined in the minor direction along the jet boundaries. The grids are distributed uniformly in both the streamwise and major directions. Denoted by gray lines, the two planes are placed symmetrically about the $x$-axis, i.e., above and below of the jet turbulence.}
\label{fig:probe_planes}
\end{figure}

Instead, we choose to place an additional pair of probe planes just outside of the jets as described in Fig.~\ref{fig:probe_planes}. Probe points are distributed uniformly in the streamwise and transverse directions on both planes, which are inclined along the nozzle walls in the upstream and jet boundaries in the downstream of the nozzle exits, respectively. Each plane extends from $x/h$ = -5 to 30 and from $z/h$ = -8 to 8, spanning larger areas compared to the previous SPOD analyses. One of the planes is located above the twin-jets ($y > 0$), and the other one is mirrored in the minor direction ($y < 0$). However, by utilizing the tilted probe planes, analysis based on the pressure or the minor axis velocities may be clouded by stronger jet screech and flapping motions in the minor axis rather than by the flow dynamics in the major axis. To counteract this, the major axis velocity fluctuations are used for SPOD analysis.

\begin{figure}
\centering
\begin{tabular}{cc}
  \includegraphics[width=0.4\textwidth]{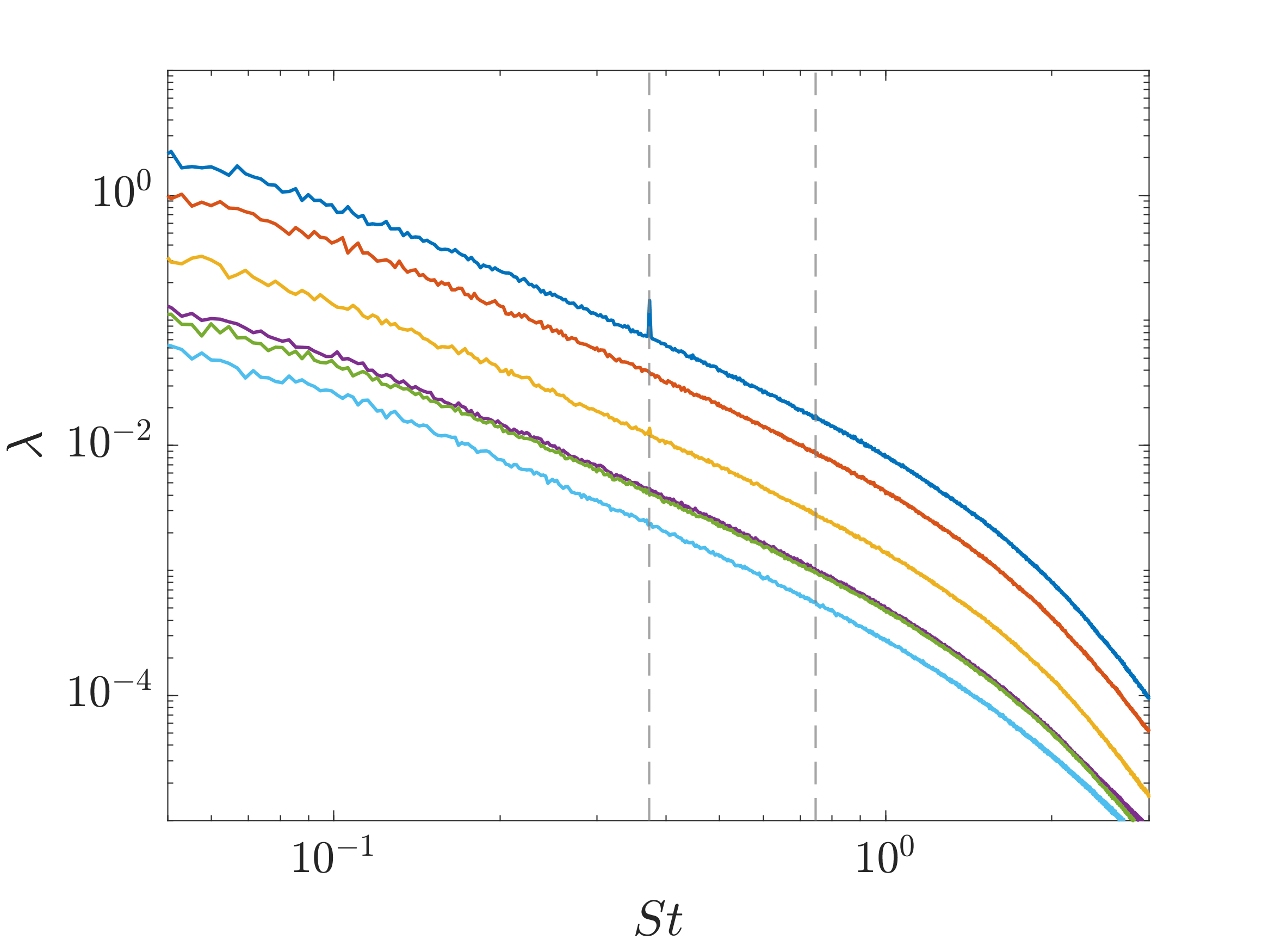} & \includegraphics[width=0.4\textwidth]{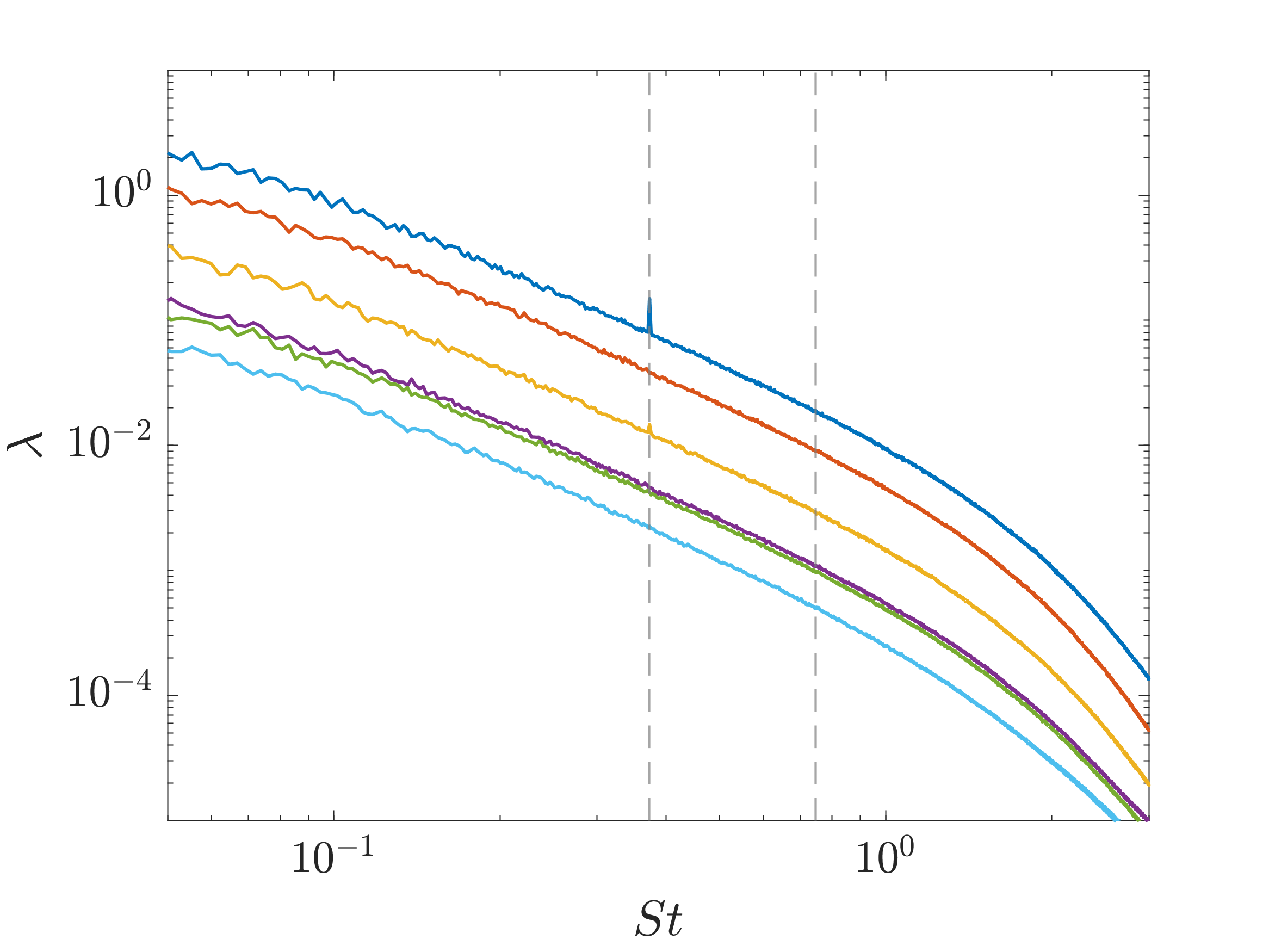} \\
  a) $y/h > 0$ & b) $y/h < 0$ \\
\end{tabular}
\caption{SPOD energy spectra based on velocity fluctuations in $z$ for NPR = 3, measured on the probe planes illustrated in Fig.~\ref{fig:probe_planes}.}
\label{fig:uz_spod_spectra_xz_upper_lower}
\end{figure}

The resulting SPOD spectra are shown in Fig.~\ref{fig:uz_spod_spectra_xz_upper_lower}. The leading SPOD modes capture clear peaks at $St$ = 0.37, but their relative amplitudes (to other spectral components or higher modes) are still weak compared to those associated with the jet screech and flapping in the minor axis. No peaks are found at other frequencies. It is not too surprising since, compared with the jet behaviors in the minor axis direction, the screech amplitudes are small and no visible jet flapping motions are detected in this direction. For the suboptimal modes, the SPOD eigenvalues gradually decrease for high frequencies without any spectral peaks. 

\begin{figure}
\centering
\begin{tabular}{cc}
  \includegraphics[width=.45\textwidth]{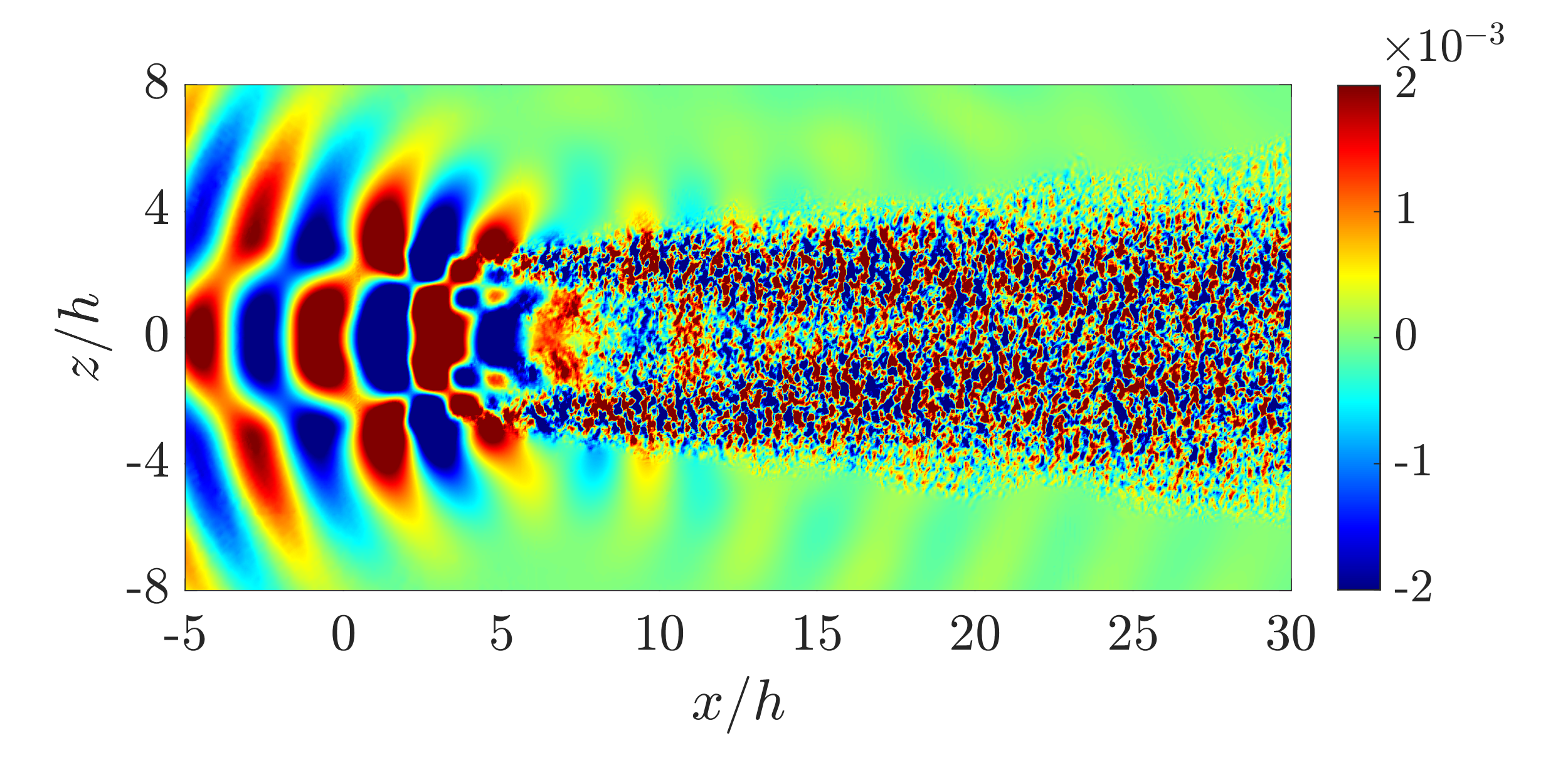} & \includegraphics[width=.45\textwidth]{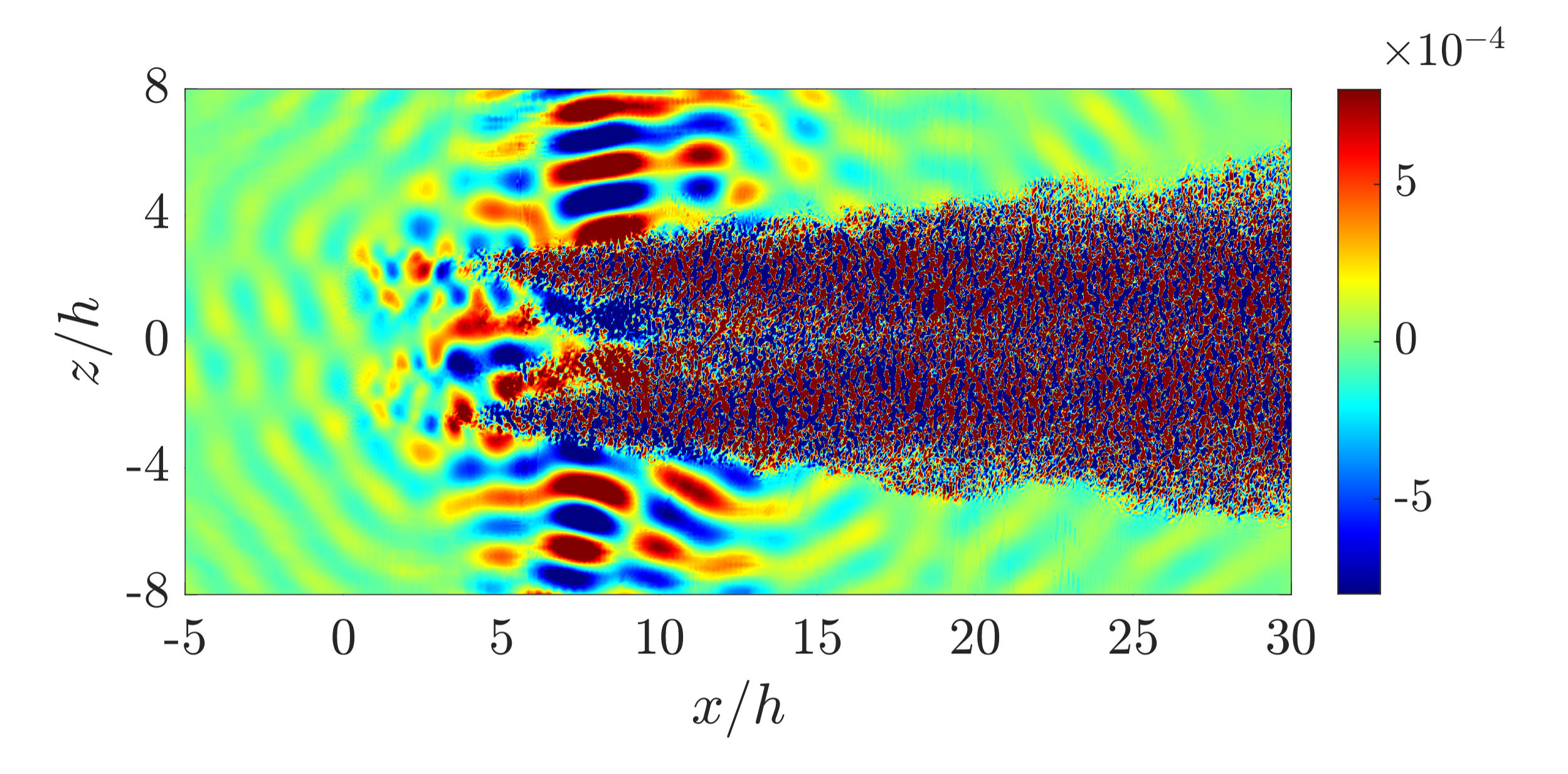} \vspace{-2mm}\\
  \includegraphics[width=.45\textwidth]{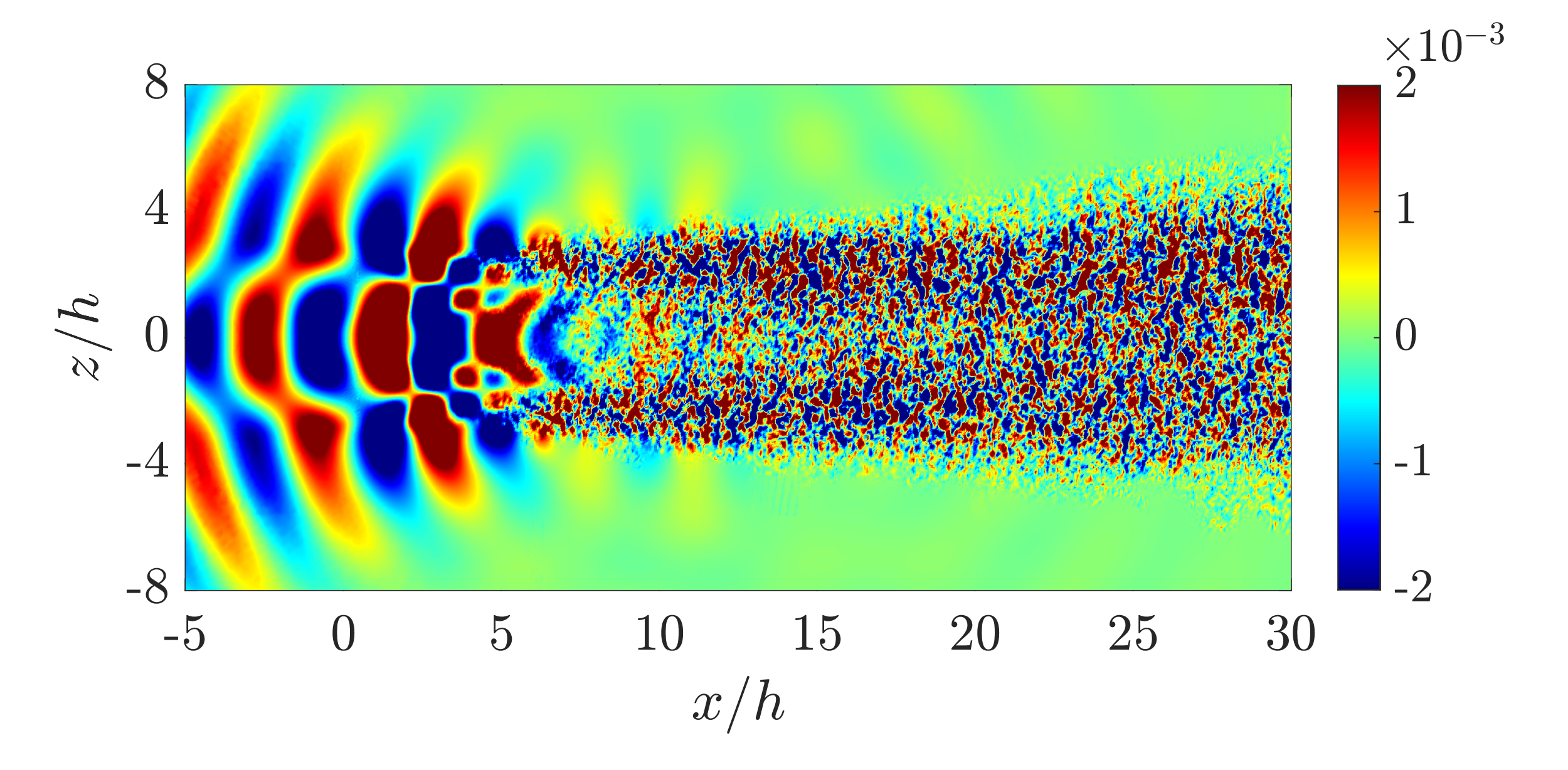} & \includegraphics[width=.45\textwidth]{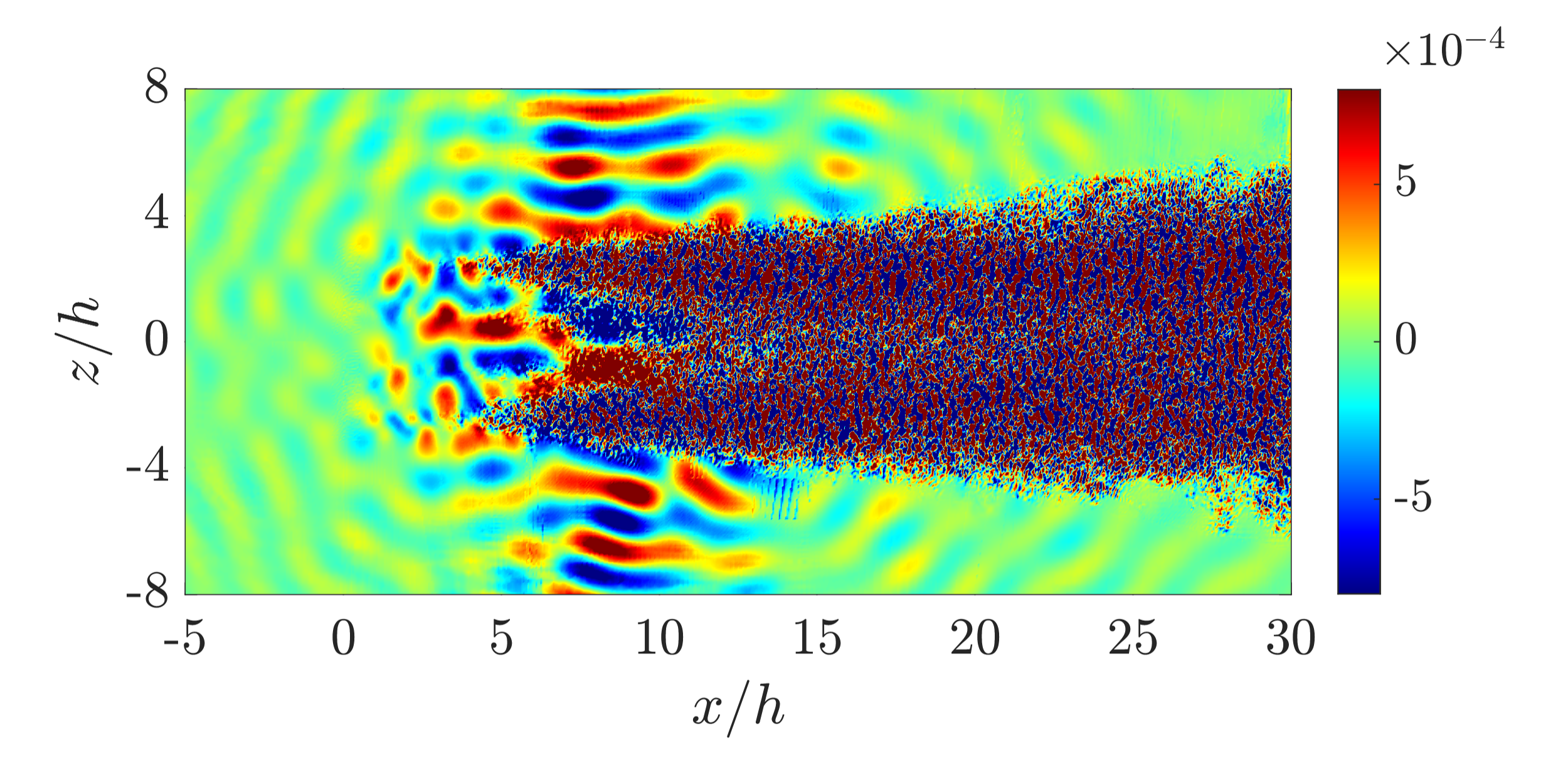} \\
  a) $St$ = 0.37 & b) $St$ = 0.74 \\
\end{tabular}
\caption{The leading SPOD modes at the fundamental screech frequency that are visualized by the real part of the corresponding SPOD eigenfunctions computed by velocity fluctuations in $z$ for NPR = 3: (top) probe plane with $y/h > 0$ and (bottom) probe plane with $y/h < 0$. The modal disturbance in the region just above or below the jets (in the probe plane) are quite weak compared to the upstream radiating signals. Nonetheless, there is an antisymmetric behavior between the top and bottom probe planes.}
\label{fig:uz_spod_modes_xz_upper_lower}
\end{figure}

Figure~\ref{fig:uz_spod_modes_xz_upper_lower} shows the corresponding leading SPOD modes at the fundamental screech frequency and its second harmonic. At $St$ = 0.37, the leading modes are described predominantly by the upstream-travelling components, and the two jets are coupled symmetrically in the major axis, as the experiments reported~\cite{karnam2021}. At $St$ = 0.74, the dominant structures are aligned perpendicular to the jet streamwise axis.

By examining the coupling modes in the minor and major axis planes simultaneously (Figs.~\ref{fig:p_spod_modes_xy},~\ref{fig:uy_spod_modes_xy}, and~\ref{fig:uz_spod_modes_xz_upper_lower}), an anti-symmetric flapping along the minor axis is the dominant coupling mode of the twin rectangular jets considered herein. Symmetric coupling of the two jets in the major axis direction is masked by much stronger anti-symmetric coupling in the minor axis. Oscillations in this direction are not as evident as in the minor axis direction. Also note that as discussed in Sec.~\ref{subsubsec:intermittency}, the screech tones are unsteady, and they are likely connected to the competition between the two coupling modes. More conclusive statement can be made by taking account of all these aspects and interpreting modes computed in the streamwise cross-sections.

\section{Conclusions and Future Work}
\label{sec:conclusion}
In this work the aeroacoustic characteristics of twin rectangular supersonic jets with an aspect ratio of 2:1 are explored using high-fidelity LES performed by a fully unstructured compressible flow solver, CharLES, developed by the Cascade Technologies. Across three jet operating conditions simulated, essential flow physics measured in the experiments are recovered by LES. Without inflow turbulence forcing, the nozzle-exit boundary layer remains laminar. Except some mismatch in the vicinity of the nozzle exit and around the end of the potential core, both the near-field flow statistics and the far-field acoustics predicted by LES match well with the experimental measurements. The fundamental screech tone and its harmonics are captured at multiple observer locations. The fundamental screech predominantly propagates upstream, while its harmonics mostly radiate into the sideline direction. Other noise components such as the broadband shock-associated noise and mixing noise are also recovered well. Particularly, LES estimates the experimental OASPL with 1-2 dB prediction accuracy at all inlet angles considered in the present study. Nevertheless, screech tones in the downstream inlet angle, which were clearly registered in the experiments, are missed. Such missing tones are probably due to unresolved fine scale turbulence in the current simulations. 

In the literature, the screech frequencies for single jets were predicted pretty accurately using the characteristic length scales of shock-cell systems and convection velocity of large-scale coherent structures. Interestingly, even for our twin jets, the screech frequency estimated by the analytic formula that was proposed for a single rectangular jet~\cite{tam1988} is found to be very close to those predicted by the LES and experiments. The screech amplitudes of the twin jets are however changed from those of a single jet~\cite{karnam2021}. These suggest that the same noise generation mechanisms of a single rectangular may be still valid for twin configurations, provided that the intensity variation due to jet-to-jet interaction is taken into account. The investigation of how the twin-jet coupling modifies screech intensity is on-going work for future publications. Concerning a single jet configuration with the same aspect ratio of 2 but equipped with a slightly different internal nozzle geometry, interested readers are referred to~\citep{viswanath2016,gojon2017,karnam2019}.

For the two over-expanded conditions, SPOD analysis of the near-field LES database captures strong spectral peaks at the screech frequencies. The leading SPOD modes at the fundamental screech frequency contain upstream-travelling components associated with the screech as well as downstream-travelling components that are similar to the Kelvin-Helmholtz instability wavepackets. The upstream radiation seems to originate around $8 < x/h < 13$, which corresponds to the fifth or sixth shock-cells. At the second harmonics, the leading modes show dominant directivity in the sideline direction. The SPOD energy at these frequencies turns out to be much weaker than that at the fundamental screech. As the time-frequency analysis indicates, the second harmonic signals appear to be much more irregular than the fundamental screech tone, and this might lead to considerably less energetic spectral peaks. 

Furthermore, by repeating SPOD analysis using other flow variables and onto planes with various orientations, out-of-phase (anti-symmetric) jet flapping motions/coupling are captured with respect to each other in the minor axis. In contrast, jets are coupled symmetrically in the major axis as the experiments reported. Also importantly, the leading SPOD modes are almost two orders of magnitude more energetic than the higher-order modes, implying strong potential value for its reduced-order models. 

The high-fidelity LES database and its SPOD analysis herein reported can provide valuable insights on the essential physics of the twin-jet screech. Still, much remains unanswered, including modeling of screech amplitude modification due to jet-to-jet interactions. When and how the presence of additional jet causes constructive/destructive coupling that yields the amplification/reduction of screech amplitudes are not clear at present as well. For the twin jets we consider, screech appears to be intermittent at over-expanded conditions but steady at under-expanded conditions~\cite{karnam2021,esfahani2021}. On the other hand, for twin round jets~\cite{bell2021} tones show intermittency at under-expanded conditions. \citet{bell2021} explained that this erratic behavior of screech tones were attributed to a competition between two global modes of the flow system, which are a function of the inter-nozzle spacing and NPR. Such imperfect phase-locking between jets must be considered for developing more accurate representation of the twin-jet screech, although modelling it may not be straightforward~\cite{jeun2021c}. Spatial linear stability analysis applied to round twin jets in a recent publication~\cite{nogueira2021} may be extended to our rectangular jets to assist deeper exploration on these matters.

Lastly, turbulent inflow forcing via surface roughness is being implemented to achieve the right nozzle-exit boundary layers. The implementation of turbulent inflow forcing may not affect the screech frequency selection of the twin jets. Nevertheless, the strength of twin-jet interactions and consequently their amplitudes could be affected, and some of missing tones may be recovered.

\section*{Acknowledgments}
This work is part of an on-going collaboration between teams from Stanford University and University of Cincinnati, supported by the Office of Naval Research under Grant No. N00014-18-1-2391 and Grant No. N00014-18-1-2582 respectively, and monitored by Dr. Steven Martens. Computational resources for large-eddy simulations were provided by the Extreme Science and Engineering Discovery Environment (XSEDE). The authors also acknowledge Cascade Technologies for granting us the access to their numerical software. Lastly, the authors want to offer special thanks to Dr. Guillaume Br{\`e}s at Cascade Technologies for his incredibly valuable advice and help on running the simulations.

\bibliography{references}

\end{document}